\documentclass[11pt]{article}
\usepackage[utf8]{inputenc}

\usepackage{latexsym}
\usepackage{amssymb,amsmath}
\usepackage{amsthm}
\usepackage{authblk}
\usepackage{graphicx}
\usepackage{sgame}
\usepackage{color}
\usepackage{graphicx}
\usepackage[hidelinks]{hyperref}
\usepackage{empheq}
\usepackage{blkarray}
\usepackage{cancel}
\usepackage{breqn}
\usepackage{comment}
\usepackage{makecell}
\usepackage{tikz} 
\usepackage{setspace}
\usepackage{appendix}

\usepackage{enumerate}

\usepackage{caption}
\usepackage{subcaption}

\usepackage{tocbasic}
\DeclareTOCStyleEntry[
  beforeskip=.2em plus 1pt,
  pagenumberformat=\textbf
]{tocline}{section}

\numberwithin{equation}{section}

\newcommand{\ZZ}{\mathbb{Z}}

\newcommand{\ds}{\displaystyle}
\newcommand{\mc}{\mathcal}

\newcommand{\bbm}{\begin{bmatrix}}
\newcommand{\bpm}{\begin{pmatrix}}
\newcommand{\ebm}{\end{bmatrix}}
\newcommand{\epm}{\end{pmatrix}}

 \newcommand{\del}[2]{\frac{\partial #1}{\partial #2}}
 \newcommand{\dsdel}[2]{\displaystyle\frac{\partial #1}{\partial #2}}

 \newcommand{\doubledelsame}[2]{\displaystyle\frac{\partial^2 #1}{\partial #2^2}}

\newcommand{\dsddt}[1]{\displaystyle\frac{d #1}{dt}}

\usepackage[numbers,sort&compress]{natbib}
\usepackage{algorithm}
\usepackage{algpseudocode}

\usepackage{authblk}

\title{Spatial Pattern Formation in Eco-Evolutionary Games with Environment-Driven Motion}
\author[1,2,*]{Tianyong Yao}
\affil[1]{Department of Mathematics, University of Michigan, Ann Arbor, MI, USA}
\author[2,3,*]{Daniel B. Cooney}
\affil[2]{Department of Mathematics, University of Illinois Urbana-Champaign, Urbana, IL, USA}
\affil[3]{Carl R. Woese Institute for Genomic Biology, University of Illinois Urbana-Champaign, Urbana, IL, USA}
\affil[*]{Correspondence to yuzutyao@umich.edu and dbcoone2@illinois.edu}
\date{\today}

\begin{document}

\newtheorem{definition}{Definition}[section]
\newtheorem{theorem}{Theorem}[section]
\newtheorem{lemma}[theorem]{Lemma}
\newtheorem{corollary}[theorem]{Corollary}
\newtheorem{claim}[theorem]{Claim}
\newtheorem{fact}[theorem]{Fact}
\newtheorem{proposition}[theorem]{Proposition}
\newtheorem{remark}[theorem]{Remark}
\newtheorem{observation}[theorem]{Observation}
\newtheorem{example}[theorem]{Example}

\maketitle
\begin{abstract}
The sustainable management of common resources often leads to a social dilemma known as the tragedy of the commons: individuals benefit from rapid extraction of resources, while communities as a whole benefit from more sustainable extraction strategies. Such a social dilemma can be further complicated by the role played by space for both resources and harvesters, where spatial diffusion of resources and directed motion of harvesters can potentially feature the emergence of clusters of environmental resource and sustainable harvesting strategies. In this paper, we explore a PDE model of evolutionary game theory with environmental feedback, describing how the spatial distribution of resource extraction strategies and environmental resources can change due to both local eco-evolutionary dynamics and environmental-driven directed motion of harvesters. Through linear stability analysis, we show that this biased motion towards higher-quality environments can lead to spatial patterns in the distribution of extraction strategies, creating local regions with improved environmental quality and increase payoff for resource extractors. However, by measuring the average payoff and environmental quality across the spatial domain, we see that this pattern-forming mechanism can actually decrease the overall success of the population relative to the equilibrium outcome in the absence of spatial motion. This suggests that environmental-driven motion can produce a spatial social dilemma, in which biased motion towards more beneficial regions can produce emergent patterns featuring a worse overall environment for the population.
\end{abstract}

{\hypersetup{linkbordercolor=black, linkcolor = blue}
\begin{spacing}{0.01}
\renewcommand{\baselinestretch}{0.1}\normalsize
\tableofcontents
\addtocontents{toc}{\protect\setcounter{tocdepth}{2}}
\end{spacing}
\singlespacing
}
\hypersetup{linkbordercolor=black, linkcolor = black}

\section{Introduction}

The sustainable management of common resources provides a clear example of a social dilemma, with a tension arising between the individual incentive to profitably extract a resource and the collective incentive to maintain the resource at a sustainable, mutually beneficial level. This misalignment between individual and collective incentives for use of shared resources is called the tragedy of the commons, and examples of such social dilemmas arise in a variety of natural settings ranging from competition for shared food in microbial populations \cite{maclean2006resource,maclean2008tragedy,schuster2017tragedy,de2019strong} to the use of natural resources by human populations \cite{schluter2016robustness,tavoni2012survival}. Spatial considerations can be an important factor in determining the ability to sustain common goods \cite{wilson1999scale} or to understand the role that local overharvesting can play on the spatial profile of environmental quality \cite{milne2022local}, and mathematical models of spatially explicit social-ecological systems can highlight the key role played by spatial heterogeneity and the movement of shared resources in shaping the long-time distribution of resources \cite{wulfing2024social}. This potential importance of such spatial factors in the management of common-pool resources motivates the goal of extending existing game-theoretic models of social-ecological systems to understand the spatial dynamics of the generation, movement, and extraction of natural resources.

The role of the spatial distribution of resources and harvesters has been frequently discussed in the context of the sustainability and management of the world's fisheries \cite{ciannelli2008spatial}, with spatial aggregation observed in both commercial ocean fishing \cite{mateo2017highly} and among recreational anglers based on the perceived yield of different fisheries \cite{post2012temporal}. A growing body of work has also explored spatial movement and directed motion towards greater abundance in agent-based models for harvesting natural resources \cite{perez2015effect,bousquet2004multi}, and fish dispersal has been used to understand the role of spatial closure in optimal management of fisheries \cite{white2014close,innes2024modeling}. Similar models have been applied to understand community-based management of logging resources, with spatial effects arising through both directed motion of loggers and seed dispersal across a forest landscape \cite{lapp2022new}. Agent-based models with directed motion towards neighboring grid cells with the greatest resource quality have also been used to described self-organized spatial assortment of extraction strategies \cite{pepper2000evolution,pepper2002mechanism,pepper2007simple} as well as the cultural coevolution of sustainable extraction strategies and social institutions favoring conservation practices \cite{waring2017coevolution}. More generally, spatial properties of ecological systems have been highlighted as important features for problems of biodiversity and conservation \cite{levin1974dispersion,levin2000multiple,LevinPacala1998,gibert2019laplacian,drechsler2022hitchhiker}, raising questions on optimal strategies for reserve design \cite{williams2005spatial} and the use of marine protected areas \cite{hastings1999equivalence,hastings2003comparing}. 

A framework for studying evolutionary game dynamics in which payoffs depend on an index of environmental quality was recently introduced by Weitz and coauthors \cite{weitz2016oscillating,paarporn2018optimal,lin2019spatial}, and a similar approach was used by Tilman and coauthors to describe ecological feedback motivated by models of renewable or exhaustible natural resources \cite{tilman2020evolutionary}. This modeling framework has been used to describe a range of scientific phenomena from microbial population dynamics to the coupled modeling of ecological conditions and the social dynamics of public sentiment regarding conservation efforts \cite{wang2020eco,thampi2018socio}. A variety of extensions of these models for environmental feedback have have been pursued, generalizing the class of game-theoretic interactions and social learning rules under consideration \cite{ito2024complete,li2025environmental}, exploring the role of ecological feedback with payoffs changing along the edges of networks \cite{betz2024evolutionary,hauert2019asymmetric,wang2025coevolutionary}, fixation probabilities of low-impact strategies in a finite population \cite{lv2024fixation}, and incorporating density-dependent effects to describe a growing population of harvesters with environmental feedback and logistic regulation \cite{das2021game}. Several of these extensions explored the role of spatial dynamics of environmental feedback with diffusion of resources, with Lin and coauthors demonstrating spatiotemporal oscillations in payoff and environmental quality \cite{lin2019spatial} and Yang and coauthors exploring how spatial heterogeneity of resources and overall resource abundance can impact the level of cooperation that can be supported by a population \cite{yang2024agent}. In a recent paper, Cheng and coauthors used a reaction-diffusion system of PDEs to describe the spatial dynamics of cooperation and ecological feedback from a public good to demonstrate spatial pattern formation through Turing and Turing-Hopf instabilities, showing how undirected spatial motion and diffusion of collective resource can help to promote localized regions featuring increased densities of cooperators and abundance of public goods \cite{cheng2024evolution}.

   Spatial models of evolutionary games have also incorporated directed motion in which individuals can either move towards increasing payoff levels \cite{helbing2009pattern,helbing2009outbreak,deforest2013spatial,xu2017strong,young2018fast} or to have biased motion towards cooperators or away from defectors \cite{funk2019directed}. These models build on prior work on reaction-diffusion equation models and other frameworks describing spatial dispersal in evolutionary game theory and theoretical ecology\cite{durrett1994importance,seri2012sustainability,cantrell2004deriving,wakano2009spatial,wakano2011pattern,arroyo2023local}, allowing for the exploration of how spatial movement rules and consideration of the spatial distribution of strategies can impact the strategic and payoff landscapes of a population. Prior work has also explored the emergence and stability of altruistic behaviors for the provision of diffusible public goods \cite{driscoll2010theory,scheuring2014diffusive,allen2013spatial,gerlee2019persistence}, with spatial clustering of public goods producers both providing an opportunity to maintain public goods and imposing a challenge that outward diffusion of shared goods can provide benefits to individuals in regions featuring lower levels of cooperation. In this paper, we will look to combine these ideas of directed motion towards greater payoffs and the harms of diffusion of common resources to understand how spatial motion of individuals and resource can impact the dynamics of spatial evolutionary games with environmental feedback. 

In this paper, we extend the model for evolutionary games with environmental feedback introduced by Tilman and coauthors, introducing a spatially explicit extension of their framework for evolutionary game dynamics coupled to the dynamics of an environmental quality variable. We consider the impact of spatial motion of both harvesters and the environmental quality variable, assuming that environmental quality can diffuse throughout the spatial domain and considering directed motion of harvesters based on directed motion towards regions of increasing environmental quality. We derive this model from an agent-based model in a finite population on a discrete spatial grid, showing that the spatial dynamics of the population can be described by a system of partial differential equations in the limit of continuous space and a hydrodynamic limit featuring a continuous description of population size. Our approach for considering PDE models for spatially explicit eco-evolutionary games builds on the approach introduced by Durrett and Levin in the context of reaction-diffusion spatial models of evolutionary games \cite{durrett1994importance}, and our model for directed motion climbing gradients of environmental quality takes the form of a chemotaxis-type term typically used to study spatial aggregation in cellular populations \cite{keller1970initiation,hillen2009user}. Our approach of considering harvesters who climb gradients of environmental quality builds on recent applications of chemotaxis-type models to study spatial pattern formation with directed motion in ecological systems featuring prey-taxis \cite{lee2009pattern,wang2021pattern} or spatial foraging \cite{tania2012role}, as well as applications to human social systems exploring questions including economic migration, gentrification, and residential segregation \cite{juchem2015capital,juchem2019returns,hasan2020transport,yizhaq2004mathematical,rodriguez2016exploring,short2008statistical,painter2019mathematical}.

We use this spatial model to explore how directed motion towards increasing levels of environmental quality can help to shape spatial patterns of the strategic composition of the population, the environmental state, and the resulting profile of payoffs for each strategy. We find that spatial patterns can emerge when individuals using a low-impact strategy display a greater sensitivity of directed motion towards increasing levels of environmental quality, in which case their more sustainable resource extraction allows for the creation of regions with improved environmental quality and increased payoffs in these locations. We demonstrate this pattern formation using a linear stability analysis around the coexistence equilibrium, and then we verify the prediction of our linear stability analysis using numerical simulations.

We further evaluate the impacts of this spatial pattern formation, exploring the spatial profiles of the distribution of strategies, the average payoff for low-effort and high-effort harvesters, and the environmental quality variable. We see that the mechanism of spatial pattern formation can help to produce local clusters featuring greater environmental quality, a greater proportion of low-effort harvesters, and greater average payoffs. However, we see that emergent spatial patterns can also feature negative aggregate properties, with the benefits of pattern formation not being shared across the full spatial environment. In particular, we find that it is possible the emergence of spatial patterns due to environmental-driven motion can decrease the total size of the population of harvesters, decrease the average payoffs of the population, and result in a lower average environmental quality across the spatial domain than would be achieved in a nonspatial model with well-mixed interactions. Interestingly, the behavior in this parameter regime suggests the possibility that directed motion by more sustainable harvesters towards regions of greater environmental quality can simultaneously produce beneficial local environments and a global degradation of environmental quality, indicating a potential social dilemma driven by environmental-driven motion. 

The remainder of the paper has the following organization. In Section \ref{sec:model}, we review the baseline model for evolutionary games with environmental feedback and present an extension of this model to describe spatial dynamics and environmental-driven motion. We then demonstrate the onset of spatial pattern formation using a linear stability analysis for our PDE model in Section \ref{sec:stabilityanalysis}, and we present further exploration of the emergent patterns and resulting collective payoff and environmental outcomes through numerical simulations in Section \ref{sec:numericalsimulations}. We then synthesize our results and present our outlook on future work in Section \ref{sec:discussion}, and we present additional derivations and analysis of our PDE model in the appendix.

\section{Model Description}
\label{sec:model}

In this section, we present our spatial model for evolutionary games with environmental feedback featuring directed motion towards regions of greater environmental quality. We first present a summary of the frequency-dependent model introduced by Tilman and coauthors that combines a replicator equation for harvester strategies with a model for how the strategic composition of the population impacts the level of environmental quality (Section \ref{sec:existingbackground}). We then extend this ODE model to allow for nonconstant population size for the total number of high-impact and low-impact harvesters (Section \ref{sec:nonspatialnonconstant}), combining reproduction based on payoff with logistic regulation limiting the local population density. Finally, we present our main PDE model eco-evolutionary games with environmental-driven motion, combining the reaction dynamics for our ODE model with non-constant population size with rules for spatial motion based on diffusion of environmental quality and directed motion of individuals toward regions with improved environmental quality (Section \ref{sec:fullspatialmodel}).

\subsection{Background on Existing Models of Games with Environmental Feedback}
\label{sec:existingbackground}

We first recall the model introduced by Tilman and coauthors \cite{tilman2020evolutionary} that describes fully frequency-independent eco-ecovolutionary game dynamics in a spatially well-mixed setting. Tilman and coauthors consider a population of individuals featuring two strategies for extraction of a renewable resource: low-effort harvesters (strategy $L$) and high-effort harvesters (strategy $H$) who exert efforts $e_L$ and $e_H$ for extracting a renewable resource. They describe the strategic composition of the population by the fraction $p$ of low-effort harvesters, with the remain fraction $1-p$ individuals in the population consisting of high-effort harvesters. To model the effects of environmental feedback on the payoffs received by individuals, Tilman and coauthors introduce a quantity $n(t) \in [0,1]$ described as a normalized environmental quality index, which can be derived from an ODE model for the dynamics of the harvesting of an renewable resource. It is assumed that the rate change for this environmental quality metric $n(t)$ depends on the fraction of individuals following the low-effort and high-effort strategies and the effort levels $e_L$ and $e_H$ used by the two harvesting strategies.

In this model, it is assumed that individuals play two-strategy, two-player games with the other members of the population and that the payoffs received in these games depend on the environmental quality metrix $n(t)$. For the case known as linear eco-evolutionary games, payoffs are modeled using a weighted payoff matrix given by
\begin{align} \label{eq:Payoffmatrix}
    \Pi(n) = (1-n) \begin{bmatrix}
    R_0 & S_0 \\
    T_0 & P_0
\end{bmatrix} + n \begin{bmatrix}
    R_1 & S_1 \\
    T_1 & P_1
\end{bmatrix},
\end{align}
in which payoffs are determined by a convex combination of the payoff matrix describing the payoffs in the case of an environment of minimal quality (corresponding to $\Pi(0)$) and the payoffs in the case of an environment of maximum quality (the payoffs when $n = 1$ (corresponding to $\Pi(1)$). The entries of these payoff matrices are named with the interpretation that $R_i$ is the reward for mutual sustainable effort, $P_i$ is the punishment for mutual aggressive harvesting, $T_i$ is the temptation to harvest aggressively when interacting with a sustainable harvester, and $S_i$ is a sucker payoff for harvesting sustainable when interacting with an aggressive harvester.

We typically place the following assumptions on the rankings of the payoffs
\begin{align}
    R_0 &> S_0 \quad T_0 > P_0 \quad R_1 > S_1 \quad T_1 > P_1 \\
    R_1 &> R_0 \quad S_1 > S_0 \quad T_1 > T_0 \quad P_1 > P_0,
\end{align}
which have the interpretation that individuals prefer to interact with low-effort harvesters than with high-effort harvesters and that each interaction in the maximum-quality environmental produces a better payoff than the correspond interaction in the minimum-quality environment. 

In this non-spatial model, we assume that individuals play the game against all other members of the population. In a population with a fraction $p$ of low-effort harvesters and an environmental quality $n$, we can describe the average payoffs received by low-effort and high-effort harvesters when playing the game against the members of the population by
\begin{align}
    f_L(p,n) &= (1-n) \left( R_0 p + S_0 \left( 1 - p \right) \right) + n \left( R_1 p + S_1 \left( 1 - p\right) \right)\\
    f_H(p,n) &= (1-n) \left( T_0 p + P_0 \left( 1 - p \right) \right) + n \left( T_1 p + P_1 \left( 1 -p \right) \right).
\end{align}

In this framework, the coupled dynamics the changing strategies of the population and the environmental impact of harvesting strategies is described by a coupled system of ODEs
\begin{subequations}
    \begin{align}
     \dsddt{p} &= \epsilon^{-1} p \left( 1 -p \right) (f_L\left(p,n\right) - f_H\left(p,n\right))  \label{eq:Tilmanreplicator} \\
   \dsddt{n} &= (r - a(e_L n + e_H (1-n))) \left( p - n \right), \label{eq:Tilmanresource}
\end{align}
\end{subequations}
where $\epsilon$ provides a relative time-scale of individual updating their strategies relative to the strength of environmental feedback, $r$ describes a background growth rate of the resource and $a$ describes a conversion efficiency of harvester effort to the extraction of the resource. Equation \eqref{eq:Tilmanreplicator} is a replicator equation, commonly used to describe the changing strategic composition of the population that can be derived from many individual-based rules for either social learning \cite{sandholm2010population} or natural selection \cite{hofbauer1998evolutionary,nowak2006evolutionary}. Equation \eqref{eq:Tilmanresource} characterizes how the environmental quality index changes in response to the harvesting efforts of the population, and was derived by Tilman and coauthors \cite{tilman2020evolutionary} based on a model of harvesting a renewable resource undergoing background logistic growth that is often used to study the management of common-pool resources \cite{gordon1954economic,munro1982fisheries,clark2010mathematical}.

Following the approach used by Tilman and coauthors, we will assume that the effort levels of the two strategies satisfy $e_L < e_H$ and that $e_H \in \left(0, \tfrac{r}{a} \right)$ \cite{tilman2020evolutionary}. By applying these assumptions to the dynamics of environmental quality in Equation \eqref{eq:Tilmanresource}, we see that these assumptions imply that the high-effort harvesters deplete environmental quality at a rate faster than the low-effort harvesters and that the level of environmental quality will remain positive under the long-time behavior of resource dynamics even in a population of harvesters consisting entirely of the high-impact strategy for all time (with $p(t) = 0$ for all $t$). This assumption on the effort levels and the assumptions we place on the payoff matrices from Equation \eqref{eq:Payoffmatrix} will allow us to determine equilibrium behavior of our nonspatial models of eco-evolutionary game, providing a baseline expectation that we can then apply to understand how spatial motion and environmental feedback interact.

\subsection{Nonspatial Model with Variable Population Size}
\label{sec:nonspatialnonconstant}

In order to describe spatial pattern formation, it will be helpful to consider a model of strategic dynamics that can allow for non-constant population size. To do this, we reformulate the model introduced by Tilman and coauthors \cite{tilman2020evolutionary} for evolutionary games with environmental feedback to explore how game-theoretic payoffs and density-dependent population regulation can determine the number of low-effort and high-effort harvesters present in a population. We adopt the approached used by Durrett and Levin that combines a frequency-dependent net birth rate based on the average payoff received by players of each strategy and a logistic regulation term based on the total density of individuals in the population \cite{durrett1994importance}.

In this version of our model, we describe the strategic composition of the population by the density of low-effort harvesters $u(t)$ and the density of high-effort harvesters $v(t)$. We describe the payoff-dependent dynamics based on the average payoffs $f_L(u,v,n)$ and $f_H(u,v,n)$ for $L$-strategists and $H$-strategists received when playing the game against the other members of the population, which are given by
\begin{align}
    f_L(u,v,n) &= (1-n) \left( R_0 \frac{u}{u+v} + S_0 \left( 1 - \frac{u}{u+v} \right) \right) + n \left( R_1 \frac{u}{u+v} + S_1 \left( 1 - \frac{u}{u+v} \right) \right) \\
    f_H(u,v,n) &= (1-n) \left( T_0 \frac{u}{u+v} + P_0 \left( 1 - \frac{u}{u+v} \right) \right) + n \left( T_1 \frac{u}{u+v} + P_1 \left( 1 - \frac{u}{u+v} \right) \right).
\end{align}
In particular, we note that $f_L(u,v,n)$ and $f_H(u,v,n)$ depend only on the numbers $u$ and $v$ of individuals through the fraction of individuals $p = \frac{u}{u+v}$ following the low-harvest strategy, which allows us to see that we can also write the average payoffs in terms of the payoff functions defined in Section \ref{sec:existingbackground} as $f_L(u,v,n) = f_L(p,n)$ and $f_H(u,v,n) = f_H(p,n)$.

We assume that average payoff individuals receive determines their payoff-dependent net birth rate, and that there is an additional density-dependent death rate with rate $\kappa (u + v)$, where the parameter $\kappa$ governs the strength of density-dependent competition in the population. We will further assume that the environmental quality index $n(t)$ experiences frequency-dependent feedback due to the strategic composition of the population that is analogous to the approach used in the model presented in Section \ref{sec:existingbackground}. With these assumptions, we can describe the coupled dynamics of population sizes of the harvester strategies and the environmental quality index by the following system of ODEs
\begin{subequations} 
    \begin{align}
   \dsddt{u}&=  \epsilon^{-1} u (f_L(u,v,n) - \kappa(u+v)) \\
    \dsddt{v} &= \epsilon^{-1} v (f_H(u,v,n) - \kappa(u+v)) \\
    \dsddt{n} &= (r - a(e_L n + e_H (1-n))) \left( \frac{u}{u+v} - n \right),
\end{align}
\end{subequations}
where the parameter $\epsilon$ now describes the relative time-scale of harvester population dynamics relative to the rate of environmental feedback. 

To analyze the dynamics of this system of ODEs, it is helpful to follow the approach used by Durrett and Levin \cite{durrett1994importance} by changing variables to describe the strategic dynamics in terms of the fraction $p = \frac{u}{u+v}$ of $L$-strategists and the total density $q = u + v$ of individuals in the population. With this change of variables, we can describe the coupled dynamics of strategy evolution and environmental feedback through the following system of ODEs
\begin{subequations} \label{eq:ODEafterCOV}
    \begin{align} 
\frac{dq}{dt} &= \epsilon^{-1} q \left[ \left(p f_L(p,n) + (1-p) f_H(p,n) \right)  - \kappa q \right] \label{eq:qODE} \\
\frac{dp}{dt} &= \epsilon^{-1}p(1-p)(f_L(p,n)-f_H(p,n)) \label{eq:pODE} \\
\frac{dn}{dt} &= (r - a(e_L n + e_H (1-n)))(p-n) \label{eq:nODE}
\end{align}
\end{subequations}
Notably, we see that Equations \eqref{eq:pODE} and \eqref{eq:nODE} are the same two ODEs that appear in the fully frequency-dependent model from Section \ref{sec:existingbackground} \cite{tilman2020evolutionary}, while Equation \eqref{eq:qODE} describes all of the density-dependent features of our model for eco-evolutionary games with non-constant population size. We will use this connection between these two ODE models to apply the results from Tilman and coauthors on how stability of equilibria depend on the payoff matrices and parameters for environmental feedback, providing us with baseline expectations for our model of spatial pattern formation in eco-evolutionary games.

\subsection{Full Spatial Model for Resource Dynamics with Environment-Driven Motion}
\label{sec:fullspatialmodel}

Now that we have described game-theoretic and environmental dynamics in the case of non-constant population size, we can incorporate these assumptions on population dynamics into a spatially explicit model for eco-evolutionary games featuring directed motion based on gradients in environmental quality. We consider the spatial dynamics of our eco-evolutionary game in a one-dimension domain of length $l$, and we describe the strategic composition of our system by the densities $u(t,x)$ and $v(t,x)$ of low-effort harvesters and high-effort harvesters at location $x \in [0,l]$ at time $t$. We describe the spatial profile of the environmental quality index by the density $n(t,x)$, and assume that environmental quality can diffuse throughout the spatial domain. We assume that the birth and death dynamics for the harvesters are the same as in our model for from Section \ref{sec:nonspatialnonconstant}, and we assume that individuals perform directed motion with a bias towards regions with increasing environmental quality.

Combining the local reaction terms described in Section \ref{sec:nonspatialnonconstant} and our assumptions on the spatial movement for harvesters and environmental quality, we can describe the dynamics of our spatial eco-evolutionary game using the following system of PDEs
\begin{subequations}\label{eq:PDEsystemuvn}
\begin{align}
\dsdel{u(t,x)}{t} &= D_u \doubledelsame{u(t,x)}{x} - \chi_u \dsdel{}{x} \left( u(t,x) \dsdel{n(t,x)}{x} \right)+ \epsilon^{-1} u (f_L\left(u,v,n\right) - \kappa(u+v)) \\
\dsdel{v(t,x)}{t} &= D_v \doubledelsame{v(t,x)}{x} - \chi_v \dsdel{}{x} \left( v(t,x) \dsdel{n(t,x)}{x} \right) + \epsilon^{-1} v (f_H\left(u,v,n\right) - \kappa(u+v)) \\
\dsdel{n(t,x)}{t} &= D_n \doubledelsame{n(t,x)}{x} + \left(r - a(e_L n + e_H (1-n))\right)\left(\frac{u}{u+v}-n\right)
\end{align}
\end{subequations}
for $x \in [0,l]$, which we will pair with zero-flux boundary conditions
\begin{subequations}
\begin{align}
\dsdel{u}{x} \bigg|_{x = 0} &= \dsdel{u}{x} \bigg|_{x = l} = 0 \\
\dsdel{v}{x} \bigg|_{x = 0} &= \dsdel{v}{x} \bigg|_{x = l} = 0 \\
\dsdel{n}{x} \bigg|_{x = 0} &= \dsdel{n}{x} \bigg|_{x = l} = 0
\end{align}
\end{subequations}
and arbitrary initial conditions 
\begin{subequations}
\begin{align}
u(0,x) &= u_0(x) \\
v(0,x) &= v_0(x) \\
n(0,x) &= n_0(x).
\end{align}
\end{subequations}

Here the parameters $D_u$, $D_v$, and $D_n$ are the diffusion coefficients for the two harvester strategies and the environmental quality index, $\chi_u$ and $\chi_v$ describe the sensitivity of directed motion for the low-effort and high-effort harvesters towards the gradient of environmental quality, and the parameter $\epsilon$ governs only the relative time-scale of the local population dynamics for the harvesters compared to the dynamics of environment feedback at a given spatial location. Here we model the biased directed motion based on environmental gradients as a chemotaxis-type term in which individuals move with directed motion with advection velocity proportional to the spatial derivative of environmental quality $\del{n(t,x)}{x}$. We present a derivation of this system of PDEs from a stochastic model that describes spatial motion as a biased random walk on a discrete grid, showing how the different sensitivity parameters $\chi_u$ and $\chi_v$ for the low-effort and high-effort harvesters can emerge from different biases towards choosing movement in directions with greater levels of environmental quality. 

Finally, it can be helpful to analyze spatial pattern formation in this PDE model by performing a change of variables to describe the spatial distribution of strategies in the population in terms of the fraction of $L$-players
\begin{subequations}
\begin{equation}
p(t,x) = \tfrac{u(t,x)}{u(t,x) + v(t,x)}
\end{equation}
and the total population density
\begin{equation}
q(t,x) = u(t,x) + v(t,x).
\end{equation}
\end{subequations} at a given point $x$ in space at time $t$. We show in Section \ref{sec:FrequencyDerivation} of the appendix that the one-dimensional version of our PDE can be written in terms of $p(t,x)$, $q(t,x)$, and $n(t,x)$ as
\begin{subequations} \label{eq:PDEsystemqpn}
\begin{align}
\dsdel{q(t,x)}{t} &= \left(D_u p + D_v (1-p) \right) \doubledelsame{q}{x} 
 + \left( D_u - D_v \right) q \doubledelsame{p}{x} -q \left[(1-p) \chi_v + p \chi_u \right] \doubledelsame{n}{x}\\
& + 2 \left( D_u - D_v \right) \dsdel{q}{x} \dsdel{p}{x} - q \left(\chi_u - \chi_v \right) \dsdel{p}{x} \dsdel{n}{x} - \left[ p \chi_u  + (1-p)\chi_v \right] \dsdel{q}{x} \dsdel{p}{x}  \nonumber \\
&+ \epsilon^{-1} q \left[ p f_L(p,n) + (1-p) f_H\left(p,n\right) - \kappa q \right],  \nonumber \\
\dsdel{p(t,x)}{t} &= \frac{1}{q} \left[ p(1-p) \left( D_u - D_v \right) \right] \doubledelsame{q}{x} + \left( (1-p) D_u + p D_v \right) \doubledelsame{p}{x} -p\left(1-p\right)\left(\chi_u-\chi_v\right)\doubledelsame{n}{x}\\
& + \frac{2}{q} \left[(1-p)D_u - p D_v \right] \dsdel{q}{x} \dsdel{p}{x}-\left[\left(1-p\right)\chi_u+p\chi_v\right]\dsdel{p}{x} \dsdel{n}{x} -\frac{1}{q}\left[p\left(1-p\right)\left(\chi_u-\chi_v\right)\right]\dsdel{q}{x} \dsdel{n}{x} \nonumber \\
&+ \epsilon^{-1} p \left( 1 - p \right) \left[ f_L(p,n) - f_H(p,n) \right],  \nonumber \\
\dsdel{n(t,x)}{t} &= D_n \doubledelsame{n}{x} + \left( r - a \left(e_L n + e_H \left(1-n\right) \right) \right) \left( p - n \right).
\end{align}
\end{subequations}
and this model inherits the zero-flux boundary conditions
\begin{subequations}
\begin{align}
\dsdel{q}{x} \bigg|_{x = 0} &= \dsdel{q}{x} \bigg|_{x = l} = 0 \\
\dsdel{p}{x} \bigg|_{x = 0} &= \dsdel{p}{x} \bigg|_{x = l} = 0 \\
\dsdel{n}{x} \bigg|_{x = 0} &= \dsdel{n}{x} \bigg|_{x = l} = 0
\end{align}
\end{subequations}
from our assumption on the boundary conditions for the quantities $u(t,x)$, $v(t,x)$, and $n(t,x)$. 

One benefit of using this representation of our PDE model is that all of the nonlinearities are quadratic in the quantities $q$, $p$, and $n$, while the original version of our PDE model represented the payoff functions as rational functions of the densities of strategies. In addition, this approach allows us to describe the stability of spatially uniform equilibria in terms of the conditions for stability in the original frequency-dependent model for eco-evolutionary games studied by Tilman and coauthors, helping us to facilitate the possibility of spatial instability introduced by environmental-driven motion. 

\section{Linear Stability Analysis and Onset of Spatial Patterns}
\label{sec:stabilityanalysis}

We now look to study the possibility of spatial pattern formation in our PDE model using a linear stability analysis. To do this, we first review in Section \ref{sec:LSAODE} the conditions for stability of our ODE model with non-constant population size from Equation \eqref{eq:ODEafterCOV}, emphasizing the conditions for stability of an equilibrium featuring coexistence of the low-effort and high-effort strategies. We then consider the possibility of instability with spatial perturbations from a uniform coexistence state in Section \ref{sec:LSAPDE}, showing how sufficient environmental sensitivity $\chi_u$ for the low-effort harvesters can allow for the formation of spatial patterns. 

\subsection{Stability Analysis of Equilibria for Nonspatial Dynamics}
\label{sec:LSAODE}

We first explore the possible equilibrium states of our nonspatial model described by the system of ODEs from Equation \eqref{eq:ODEafterCOV}. As these ODEs consist of the system of ODEs for the number of low-effort harvesters $p(t)$ and environmental quality index $n(t)$ introduced by Tilman and coauthors \cite{tilman2020evolutionary} 
\begin{subequations} \label{eq:TilmanCOV}
    \begin{align} 
\frac{dp}{dt} &= \epsilon^{-1}p(1-p)(f_L(p,n)-f_H(p,n)) \\
\frac{dn}{dt} &= (r - a(e_L n + e_H (1-n)))(p-n) 
\end{align}
and a separate ODE describing the total size of the harvester population
\begin{equation}
\frac{dq}{dt} = \epsilon^{-1} q \left[ \left(p f_L(p,n) + (1-p) f_H(p,n) \right)  - \kappa q \right],
\end{equation}
\end{subequations}
we see that equilibrium points $(q^*,p^*,n^*)$ will consist of points for which $(p^*,n^*)$ are equilibria for the model of Tilman and coauthors \cite{tilman2020evolutionary} and for which $q^*$ satisfies either $q^* = 0$ or 
\begin{equation}
q^* = \frac{p^* f_L(p^*) + (1-p^*) f_H(p^*)}{\kappa}.
\end{equation}
In other words, there are two possible equilibria of Equation \eqref{eq:ODEafterCOV} that correspond to each equilibrium state for the the $(p,n)$ subsystem, one featuring zero population size and one featuring a population size equal to ratio of the average payoff of harvesters in the equilibrium $(p^*,n^*)$ and the strength $\kappa$ of density-dependent regulation. 

Using the analysis of Tilman and coauthors for the $(p,n)$-subsystem, we can therefore deduce that there are six possible types of equilibrium points for our model from Equation \eqref{eq:ODEafterCOV}:
\begin{enumerate}
    \item $(q^*, p^*, n^*)  = \left(0, 0, 0\right)$
    \item $(q^*, p^*, n^*)  = \left(0, 1, 1\right)$
    \item $(q^*, p^*, n^*)  = \left(0, p^*, n^*\right)$ for $n^*$ satisfying $f_L(p^*,n^*)=f_H(p^*,n^*)$ and $p^*=n^*$
    \item $(q^*, p^*, n^*)  = \left(\frac{f_H(p^*)}{\kappa}, 0, 0\right)$ (all individuals play strategy $H$)
    \item $(q^*, p^*, n^*) = \left(\frac{f_L(p^*)}{\kappa}, 1, 1\right)$  (all individuals play strategy $L$)
    \item $(q^*, p^*, n^*)  = \left(\frac{p^* f_L(p^*) + (1-p^*) f_H(p^*) }{\kappa}, p^*, n^*\right)$ for $n^*$ satisfying $f_L(p^*,n^*)=f_H(p^*,n^*)$ and $p^*=n^*$.
\end{enumerate}
The equilibrium point of primary interest to our analysis is the final one, which describes a population with nonzero size and a positive fraction of both low-effort and high-effort harvesters.

We now explore the stability of equilibria for our ODE model with non-constant population size. We first note that the Jacobian matrix for the righthand side of the system of ODEs from Equation \eqref{eq:ODEafterCOV} is given by

\begin{align} \label{eq:ODE_jaco_C}
    C(p,q,n) = 
    \begin{pmatrix}
        \epsilon^{-1}[ pf_L + (1-p)f_H  - 2\kappa q] & c_{12}(q,p,n) & c_{13}(q,p,n)\\
        0 & c_{22}(p,n) & c_{23}(p,n)\\
        0 & c_{32}(p,n) & c_{33}(p,n)
    \end{pmatrix},
\end{align}
where
\begin{subequations}
\begin{align}
c_{12}(q,p,n) &= \epsilon^{-1} q \left[ f_L(p,n) - f_H(p,n)  + p \dsdel{f_L(p,n)}{p} + (1-p) \dsdel{f_H(p,n)}{p} \right] \\
c_{13}(q,p,n) &= \epsilon^{-1} q \left[ p \dsdel{f_L(p,n)}{n}  + (1-p) \dsdel{f_H(p,n)}{n} \right] \\
c_{22}(p,n) &= \epsilon^{-1} (1-2p) \left[f_L(p,n) - f_H(p,n) \right] + \epsilon^{-1} p (1-p) \left[\dsdel{f_L(p,n)}{p} - \dsdel{f_H(p,n)}{p} \right] \\
c_{23}(p,n) &= \epsilon^{-1} p(1-p) \left[\dsdel{f_L(p,n)}{n} - \dsdel{f_H(p,n)}{n}  \right] \\
c_{32}(p,n) &= r - a \left( e_L n + e_H (1-n) \right) \\
c_{33}(p,n) &= -a \left( e_L - e_H \right) (p-n) - \left( r - a\left( e_L n + e_H \left(1-n \right) \right) \right).
\end{align}
\end{subequations}

In particular, we notice that the two-by-two submatrix in the bottom-right corner is given by
\begin{equation}
\tilde{C}(p,n) = \bpm c_{22}(p,n) & c_{23}(p,n) \\ c_{32}(p,n) & c_{33}(p,n) \epm,
\end{equation}
which is the Jacobian matrix for the frequency-dependent model originally studied by Tilman and coauthors \cite{tilman2020evolutionary}. We can then see that, for any equilibrium point $(q^*,p^*,n^*)$ for the system from Equation \eqref{eq:ODEafterCOV}, the eigenvalues of the Jacobian $C(q^*,p^*,n^*)$ will satisfy
\begin{equation}
\det\left( C(q^*,p^*,n^*) - \lambda I \right) = \left( \epsilon^{-1}\left[ p^*f_L(p^*,n^*) + (1-p^*)f_H(p^*,n^*) - 2\kappa q^*\right] - \lambda  \right) \det\left(  \tilde{C}(p^*,n^*)-\lambda I \right),
\end{equation}
so the eigenvalues of $C(q^*,p^*,n^*)$ are given by \begin{equation}
    \lambda_1(q^*,p^*,n^*) = \epsilon^{-1}[(1-p^*)f_H(p^*,n^*) + p^*f_L(p^*,n^*) - 2\kappa q^*]
\end{equation}
and the eigenvalues of the matrix $\tilde{C}(p^*,n^*)$ corresponding to the Jacobian matrix for the model of Tilman and coauthors \cite{tilman2020evolutionary}. Because the equilibrium $(q^*,p^*,n^*)$ will be stable when all of the eigenvalues of $J(q^*,p^*,n^*)$ have negative real parts, we see that an equilibrium to the system of Equation \eqref{eq:ODEafterCOV} will be stable provided that
\begin{subequations}
\begin{align}
 \epsilon^{-1}[p^*f_L(p^*,n^*) +  (1-p^*)f_H(p^*,n^*) - 2\kappa q^*] &< 0 \\
Tr\left( \tilde{C}(p^*,n^*)\right) &< 0 \\
\det\left( \tilde{C}(p^*,n^*) \right) &> 0.
\end{align}
\end{subequations}
As equilibria to Equation \eqref{eq:ODEafterCOV} have total population size of either $q^* = 0$ or 
\begin{equation} q^* = \kappa^{-1}\left( p^* f_L(p^*,n^*) + (1-p^*) f_H(p^*,n^*) \right),\end{equation}  the eigenvalue $\lambda_1^*(q^*,p^*,n^*)$ is either given by
\begin{equation}
\lambda_1(0,p^*,n^*) = (1-p^*)f_H(p^*,n^*) + p^*f_L(p^*,n^*) 
\end{equation}
or
\begin{equation}
\lambda_1\left( \kappa^{-1}\left( p^* f_L(p^*,n^*) + (1-p^*) f_H(p^*,n^*) \right), p^*,n^*\right) = - \left[ (1-p^*)f_H(p^*,n^*) + p^*f_L(p^*,n^*)  \right].
\end{equation}
Therefore we can conclude that an equilibrium of the form $(q^*,p^*,n^*) = (0, p^*,p^*)$ is asymptotically stable if $(p^*,n*) = (p*,p^*)$ is an asymptotically stable equilibrium of the model by Tilman and coauthors and the average payoff at this equilibrium is negative, while equilibrium of the form $(q^*,p^*,n^*)$ with $q^* = \kappa^{-1}\left( p^* f_L(p^*,n^*) + (1-p^*) f_H(p^*,n^*) \right)$ and $p^* = n^*$ will be asymptotically stable if and only if $(p^*,n^*)$ is an asymptotically stable equilibrium of the model by Tilman and coauthors and the average payoff at this equilibrium is positive. 

For the equilibrium $(q^*,p^*,n^*) = (\kappa^{-1} f_L(p^*,p^*), p^*,p^*)$ satisfying $f_L(p^*,n^*) = f_H(p^*,n^*)$, we can then see that the stability will be determined by the following five quantities
\begin{subequations}
\begin{align}
c_{11}(q^*,p^*,n^*) &= \lambda_1(q^*,p^*,n^*) =  -\epsilon^{-1} f_L(p^*) \\
c_{22}(p^*,n^*) &= \epsilon^{-1} p^* (1-p^*) \left[ \dsdel{f_L}{p} - \dsdel{f_H}{p} \right] \bigg|_{\substack{p = p^* \\ n = n^* = p^*}} \\ 
c_{23}(p^*,n^*) &= \epsilon^{-1} p^* (1-p^*) \left[ \dsdel{f_L}{n} - \dsdel{f_H}{n} \right] \bigg|_{\substack{p = p^* \\ n = n^* =  p^*}} \\
c_{32}(p^*,n^*) &= r - a \left( e_L n^* + e_H (1-n^*) \right) \\
c_{33}(p^*,n^*) &= - \left[r - a \left( e_L n^* + e_H (1-n^*) \right)  \right].
\end{align}
\end{subequations}
From our assumptions on the effort levels that $e_L < e_H$ that $e_H \in \left(0, \frac{r}{a} \right)$, we can therefore deduce that $c_{32}(p^*,n^*) > 0$ and $c_{33}(p^*,n^*) < 0$ for the coexistence equilibrium. However, the process of determining the signs of $c_{22}(p^*,n^*)$ and $c_{23}(p^*,n^*)$ is more subtle, requiring us to consider the relative values of the payoff parameters for the game-theoretic interactions. In Remark \ref{rem:payoffstability},we provide further discussion on results derived by Tilman and coauthors \cite{tilman2020evolutionary} of the role of payoff parameters on the stability of equilibria featuring coexistence of the two harvesting strategies. 

\begin{remark}
\label{rem:payoffstability}
The existence and stability of such coexistence equilibria $(p^*,n^*)$ with $f_L(p^*,n^*) = f_H(p^*,n^*)$ was examined in detail by Tilman and coauthors \cite{tilman2020evolutionary} for the model with constant population size. They showed that the coexistence equilibria took the following form
\begin{equation}
p^*_{\pm} = n^*_{\pm} = \frac{2 \left( S_0 - P_0\right) + P_1 - S_1 + T_0 - R_0 \pm \sqrt{\left( P_0 - S_0 + R_1 - T_1 \right)^2 + 4 \left( T_1 - R_1\right) \left(S_0 - P_0 \right)}}{2 \left[ \left( R_1 - S_1 - T_1 + P_1 \right) - \left( R_0 - S_0 - T_0 + P_0 \right) \right]},
\end{equation}
and that only the equilibrium $(p^*_{-},n^*_-)$ corresponding to the negative root will be biologically feasible provided that $T_1 \geq R_1$ and $S_0 > P_0$. In addition, the equilibrium $(p^*_{-},n^*_-)$ will be asymptotically stable if at least one of the following two conditions is satisfied by the payoff parameters:
\begin{itemize}
    \item $P_1 - S_1 + R_0 - S_0 < 0$
    \item $ (P_1 - S_1) (R_0 - S_0) < (T_1 - R_1) (S_0 - P_0)$.
\end{itemize}
In addition, Tilman and couathors used linear stability analysis to show that the edge equilibria $(p,n) = (1,1)$ and $(0,0)$ are unstable when $T_1 > R_1$ and $S_0 > P_0$ \cite{tilman2020evolutionary}. 

For the spatial dynamics of our PDE model, we will typically consider parameter regimes for our payoff matrices for which there is a unique coexistence equilibrium $(q^*, p^*,n^*)$ with $p^* \in (0,1)$ and for which this equilibrium is locally stable. The choice of local stability for the coexistence equilibrium will allow us to perform linear stability analysis around a spatially uniform state that is stable to spatially uniform perturbations, while the assumption of a unique spatially stable equilibrium can be important for understanding the role of instability of a uniform state in the formation of spatial patterns (as it has been shown that linear instability of a uniform state can be insufficient for the achievement of spatial patterns in the case of multistability of the reaction dynamics \cite{krause2024turing}). 
 \end{remark}

\subsection{Linear Stability Analysis of Uniform State with Spatial Perturbations}
\label{sec:LSAPDE}

In this section, we consider the case of payoff and ecological parameters for which the nonspatial ODE dynamics will have a stable equilibrium point $(q_0,p_0,n_0)$ featuring nonzero population size $q_0 > 0$ and the coexistence of the low-effort and high-effort harvesters corresponding to $p_0 \in (0,1)$. This means that the spatially uniform state $(q(x),p(x),n(x)) = (q_0,p_0,n_0)$ will be an equilibrium for our PDE model of Equation \eqref{eq:PDEsystemqpn}, and we will perform a linear stability analysis to see whether choosing certain parameters for the rules of spatial motion can allow for the formation of spatial clustering of strategies and environmental quality. We will perform a linear stability analysis of our PDE model from Equation \eqref{eq:PDEsystemqpn} around this uniform equilibrium, looking to determine the threshold strengths of environmental-driven motion $\chi_u$ of low-effort harvesting that will allow for spatial pattern formation in our spatial model.  

We then consider a small parameter $\delta$ and spatially varying functions $\tilde{q}(t,x)$, $\tilde{p}(t,x)$, and $\tilde{n}(t,x)$, allowing us to consider solutions to consider solutions of our PDE model of the form
\begin{subequations}
\begin{align}
    q(t,x) &= q_0 + \delta \tilde{q}(t,x) \\
    p(t,x) &= p_0 + \delta \tilde{p}(t,x)\\
    n(t,x) &= n_0 + \delta \tilde{n}(t,x),
\end{align}
\end{subequations}
which constitute a small perturbation from our spatially uniform equilibrium state. Substituting these expressions for $q(t,x)$, $p(t,x)$, $n(t,x)$ into the nonlinear system of PDEs from Equation \eqref{eq:PDEsystemqpn}, we can linearize our system of PDEs around the uniform state $(q(x),p(x),n(x)) = (q_0,p_0,n_0)$ by neglecting terms of order $\mc{O}\left(\delta^2\right)$ as $\delta \to 0$ and using the fact that $(q_0,p_0,n_0)$ is an equilibrium of the ODE reaction system of Equation \eqref{eq:ODEafterCOV}. This allows us to obtain the following system of PDEs to obtain the following linearized system of PDEs for the perturbation functions $\tilde{q}(t,x)$, $\tilde{p}(t,x)$, and $\tilde{n}(t,x)$:
\begin{subequations} \label{eq:linearizedsystem}
   \begin{align}
   \dsdel{\tilde{q}}{t} &= \left(D_u p_0 + D_v (1-p_0) \right) \doubledelsame{\tilde{q}}{x} 
 + \left( D_u - D_v \right) q_0 \doubledelsame{\tilde{p}}{x} \\
 &-q_0 \left[(1-p_0) \chi_v + p_0 \chi_u \right] \doubledelsame{\tilde{n}}{x}+ c_{11} \tilde{q} + c_{12} \tilde{p} + c_{13} \tilde{n} \nonumber\\
   \frac{\partial \tilde{p}}{\partial t} &= \frac{1}{q_0} \left[ p_0(1-p_0) \left( D_u - D_v \right) \right] \doubledelsame{\tilde{q}}{x} + \left( (1-p_0) D_u + p_0 D_v \right) \doubledelsame{\tilde{p}}{x} \\
   &-p_0\left(1-p_0\right)\left(\chi_u-\chi_v\right)\doubledelsame{\tilde{n}}{x}+ c_{22} \tilde{p} + c_{23} \tilde{n} \nonumber \\
    \frac{\partial \tilde{n}}{\partial t} &= D_n \frac{\partial^2 \tilde{n}}{\partial x^2} + c_{32} \tilde{p} + c_{33} \tilde{n},
\end{align}
\end{subequations}
where the terms of the form $c_{ij}$ correspond to the $(i,j)$th entry of the Jacobian matrix $C$ for the ODE system given by Equation \eqref{eq:ODE_jaco_C}, evaluated at the coexistence equilibrium point $(q_0,p_0,n_0)$. For completeness, we present the full derivation of this linearized system of PDEs in Section \ref{sec:App_linearization} of the appendix.

We can then write our linearized system of PDEs in the following matrix form 
\begin{subequations} \label{eq:linearizedoperator}
\begin{equation}
\dsdel{}{t} \bpm \tilde{q} \\ \tilde{p} \\ \tilde{n} \epm = \mc{L} \bpm \tilde{q} \\ \tilde{p} \\ \tilde{n} \epm, 
\end{equation}
where $\mc{L}$ is the linearized differential operator 
\begin{equation}
\resizebox{\linewidth}{!}{
$\mc{L} := \begin{pmatrix}
(D_u p_0 + D_v (1 - p_0)) \partial_{xx} + c_{11} & (D_u - D_v) q_0\partial_{xx} + c_{12} & - (\chi_u p_0 q_0 + \chi_v q_0 (1 - p_0)) \partial_{xx} + c_{13} \\
q_0^{-1}(1-p_0)p_0 (D_u - D_v) \partial_{xx}  & \left(D_u (1-p_0) + p_0 D_v \right) \partial_{xx} + c_{22} & - p_0(1-p_0) (\chi_u - \chi_v) \partial_{xx} + c_{23} \\
0 & c_{32} & D_n \partial_{xx} + c_{33}
\end{pmatrix}
$}.
\end{equation}
\end{subequations}

We can then assume that our small perturbation to the uniform state $(\tilde{q},\tilde{p},\tilde{n})^T$ takes the following form
\begin{align}
    \tilde{q} (t, x) &=  k_{\tilde{q}} e^{\sigma t} \cos\left(\frac{m \pi x}{l}\right) \\
    \tilde{p} (t, x) &= k_{\tilde{p}}e^{\sigma t} \cos\left(\frac{m \pi x}{l}\right) \\
    \tilde{n} (t, x) &=k_{\tilde{n}} e^{\sigma t} \cos\left(\frac{m \pi x}{l}\right)
\end{align}
for some constants $k_{\tilde{q}}$, $k_{\tilde{p}}$, $k_{\tilde{n}}$,
and we substitute the expressions for our perturbation into the linearized system from Equation \eqref{eq:linearizedoperator} to see that the growth rate $\sigma$ of these perturbations will satisfy the following eigenvalue problem in terms of our linearized system for sinusoidal perturbations with wavenumber $m$:

\begin{equation}
\sigma \begin{pmatrix}
\tilde{q}(t,x) \\
\tilde{p}(t,x) \\
\tilde{n}(t,x)
\end{pmatrix}  = A(m) \begin{pmatrix}
\tilde{q}(t,x) \\
\tilde{p}(t,x) \\
\tilde{n}(t,x)
\end{pmatrix},
\end{equation}
where the matrix $A(m)$ is given by
\begin{equation} 
\resizebox{\textwidth}{!}{$
    A(m) = \begin{pmatrix} \label{eq:PDE_Jaco_A}
c_{11} - \left(\frac{m \pi}{l}\right)^2 (D_u p_0 + D_v (1 - p_0)) & c_{12} - \left(\frac{m \pi}{l}\right)^2 (D_u - D_v) q_0 & c_{13} + \left(\frac{m \pi}{l}\right)^2 (\chi_u p_0 q_0 + \chi_v q_0 (1 - p_0)) \\
-q_0^{-1} \left(\frac{m \pi}{l}\right)^2 (1-p_0) p_0(D_u -D_v)  & c_{22} - \left(\frac{m \pi}{l}\right)^2 (D_u (1-p_0) + p_0 D_v) & c_{23} + \left(\frac{m \pi}{l}\right)^2 p_0(1-p_0) (\chi_u - \chi_v) \\
0 & c_{32} & -\left(\frac{m \pi}{l}\right)^2 D_n + c_{33}
\end{pmatrix}
$}.
\end{equation}
As these perturbations will grow in time if any of the eigenvalues $\sigma$ have positive real part, we see that the uniform equilibrium $(q_0,p_0,n_0)$ will only be stable to a perturbation with wavenumber $m$ if each of the eigenvalues of $A(m)$ has a nonpositive real part. 

We will now further explore the properties of our linearized system to determine the conditions required for the instability of the uniform state to spatial perturbations. We first focus on the case of equal diffusivities $D_u = D_v$ for the low-effort and high-effort harvesters in Section \ref{sec:LSAequalDuDv}, which will allow a more straightforward analysis of the strength of environmental-driven motion $\chi_u$ for the low-effort harvester required to generate spatial patterns. We then consider the more general case in Section \ref{sec:LSAunequalDuDv} in which the diffusion coefficients $D_u$ and $D_v$ can be unequal, studying how these diffusion rates can also impact the onset of pattern formation.

\subsubsection{Linear Stability Analysis for Case of Identical Diffusion Coefficients}
\label{sec:LSAequalDuDv}

For the case of equal diffusivities of the two strategies, we see that the condition $D_u = D_v$ allows for substantial simplification of the form of the Jacobian matrix $A(m)$, resulting in the linearization matrix taking the form  %
\begin{equation} \label{eq:Amsimplified}
    A(m) = \begin{pmatrix}
c_{11} - \left(\frac{m \pi}{l}\right)^2 D_u & c_{12} & c_{13} + \left(\frac{m \pi}{l}\right)^2 (\chi_u p_0 q_0 + \chi_v q_0 (1 - p_0)) \\
0 & c_{22} - \left(\frac{m \pi}{l}\right)^2D_u & c_{23} + \left(\frac{m \pi}{l}\right)^2 p_0(1-p_0) (\chi_u - \chi_v) \\
0 & c_{32} & -\left(\frac{m \pi}{l}\right)^2 D_n + c_{33}.
\end{pmatrix}.
\end{equation}

The homogeneous steady state will be unstable to the spatial perturbation we consider provided that at least one of the eigenvalues of the linearization matrix $A$ has positive real part. Noting that the characteristic polynomial for $A(m)$ is given by
\begin{align}
    P_A(\lambda_A)=\det(A(m)-\lambda_A I) = \left(c_{11} - \left(\frac{m \pi}{l}\right)^2 D_u - \lambda_A\right) \det(\tilde{A}(m) - \lambda_A I),
\end{align}
where $\tilde{A}$ is the submatrix given by
\begin{equation} \label{eq:Atildesubmatrix}
   \tilde{A}(m) = \begin{pmatrix}
         c_{22} - \left(\frac{m \pi}{l}\right)^2D_u &c_{23} + \left(\frac{m \pi}{l}\right)^2 p_0(1-p_0) (\chi_u - \chi_v) \\
         c_{32} & -\left(\frac{m \pi}{l}\right)^2 D_n + c_{33}
    \end{pmatrix},
\end{equation}
so the eigenvalues of the matrix $A(m)$ are $\lambda_1 = c_{11} - \left(\frac{m \pi}{l}\right)^2 D_u$ and the two eigenvalues of the submatrix $\tilde{A}(m)$ from Equation \eqref{eq:Atildesubmatrix}. As we have assumed that $D_u \geq 0$ and we have shown in Section \ref{sec:LSAODE} that $c_{11} < 0$, we can deduce that $\lambda_1 < 0$ and therefore the two possible ways for our matrix $A$ to have an eigenvalue with positive real part are to have either $\det(\tilde{A}) < 0$ or $\mathrm{Tr}(\tilde{A}) > 0$. 
We can then use the non-negativity of the diffusion coefficients and the fact that $\mathrm{Tr}(\tilde{C}) = c_{22} + c_{33} < 0$ due to the assumption that the uniform equilibrium $(q_0,p_0,n_0)$ in stable under our nonspatial ODE model to deduce that the sign of $\mathrm{Tr}(A)$s satisfies
\begin{align}
    \operatorname{Tr}(\tilde{A}(m)) =c_{22} - \left(\frac{m \pi}{l}\right)^2 D_u - \left(\frac{m \pi}{l}\right)^2 D_n + c_{33} = Tr(\tilde{C}) - \left( D_u + D_n \right)  \left(\frac{m \pi}{l}\right)^2 < 0.
\end{align}
As $\mathrm{Tr}(\tilde{A}) < 0$, the only way for $A$ to have an eigenvalue with positive real part is for $\det{\tilde{A}} < 0$, which means that our condition for the spatial pattern formation through the instability of the uniform state to spatial perturbations is given by
\begin{equation} \label{eq:det(A)}
\begin{aligned}
    \det(\tilde{A}(m)) &= \left(c_{22} - \left(\frac{m \pi}{l}\right)^2 D_u\right)\left(-\left(\frac{m \pi}{l}\right)^2 D_n + c_{33}\right) - c_{32}\left(c_{23} + \left(\frac{m \pi}{l}\right)^2 p_0(1-p_0) (\chi_u - \chi_v)\right) \\
    &= \underbrace{c_{22} c_{33} - c_{32} c_{23}}_{= \det{\tilde{C}}} \\ &+ \left( \frac{m \pi}{l} \right)^4 D_u D_n - c_{22} \left( \frac{m \pi}{l} \right)^2 D_n - c_{33} \left( \frac{m \pi}{l} \right)^2 D_u - c_{32} \left( \frac{m \pi}{l}\right)^2 p_0 (1-p_0) \left( \chi_u - \chi_v \right) \\ &< 0.
\end{aligned}
\end{equation}

We can then rewrite the condition from Equation \eqref{eq:det(A)} to  see that it will be possible to achieve instability of the uniform state to a perturbation with wavenumber $m$ provided that the parameter $\chi_u$ for the strength of environmental-driven motion of the low-effort harvesters satisfies
\begin{align}
    \chi_u > \chi_u^*(m) :=\frac{\left( \left(\ds\frac{m \pi}{l}\right)^2 D_u - c_{22}\right) \left(\ds\frac{m \pi}{l}\right)^2 D_n + \left(c_{22} - \left(\ds\frac{m \pi}{l}\right)^2 D_u\right) c_{33} - c_{32} c_{23}}{c_{32} \left(\ds\frac{m \pi}{l}\right)^2 p_0 (1 - p_0)} + \chi_v, 
\end{align}
where $\chi_u^*(m)$ denotes the threshold value of $\chi_u$ for given wavenumber $m$. We can further write our threshold quantity $\chi_u^*(m)$ in the following form
\begin{align} \label{eq:chistarum}
    \chi_u^*(m) = b \left(\frac{m \pi}{l}\right)^2  + \frac{c}{\left(\frac{m \pi}{l}\right)^2} + d,
\end{align}
where we have defined the quantities $b$, $c$, and $d$ as
\begin{align} \label{eq:bcdpattern}
    b = \frac{D_u D_n}{c_{32} p_0 (1 - p_0)}, \quad
    c = \frac{c_{22} c_{33} - c_{32} c_{23}}{c_{32} p_0 (1 - p_0)}, \quad
    d = \chi_v - \frac{c_{22} D_n + D_u c_{33}}{c_{32} p_0 (1 - p_0)}.
\end{align}

Given the conditions mentioned in Section \ref{sec:LSAODE} for stability of the equilibrium point $(q_0,p_0,n_0)$ under the ODE system, we must have that $c_{22} + c_{23} < 0$ and $c_{32} > 0$, which means that the assumption of a stable equilibrium to the reaction system allows us to deduce that $b > 0$ and $c > 0$. Therefore the minimum value of $\chi_u^*(m)$ is $2\sqrt{bc} + d$, occurring at $m^* = \tfrac{l}{\pi} \left(\tfrac{c}{b}\right)^{1/4}$ if we interpret the wavenumber $m$ as a real number.  %
However, because we consider our spatial dynamics on the interval $[0,l]$ and assume zero-flux boundary conditions, we can only consider perturbations with integer wavenumber $m$, so we see that the minimum strength of environmental-driven motion by low-effort harvesters will be given by
\begin{equation}
\chi^*_u := \min_{m \in \ZZ_{\geq 0}} \chi^*_u(m),
\end{equation}
the minimum of the instability thresholds across all possible non-negative integer wavenumbers $m$. We will therefore expect to see spatial pattern formation for strengths of environmental-driven motion $\chi_u$ for which $\chi_u > \chi_u^*$. 

\begin{figure}[!ht]
    \centering
\includegraphics[width = 0.45\textwidth]{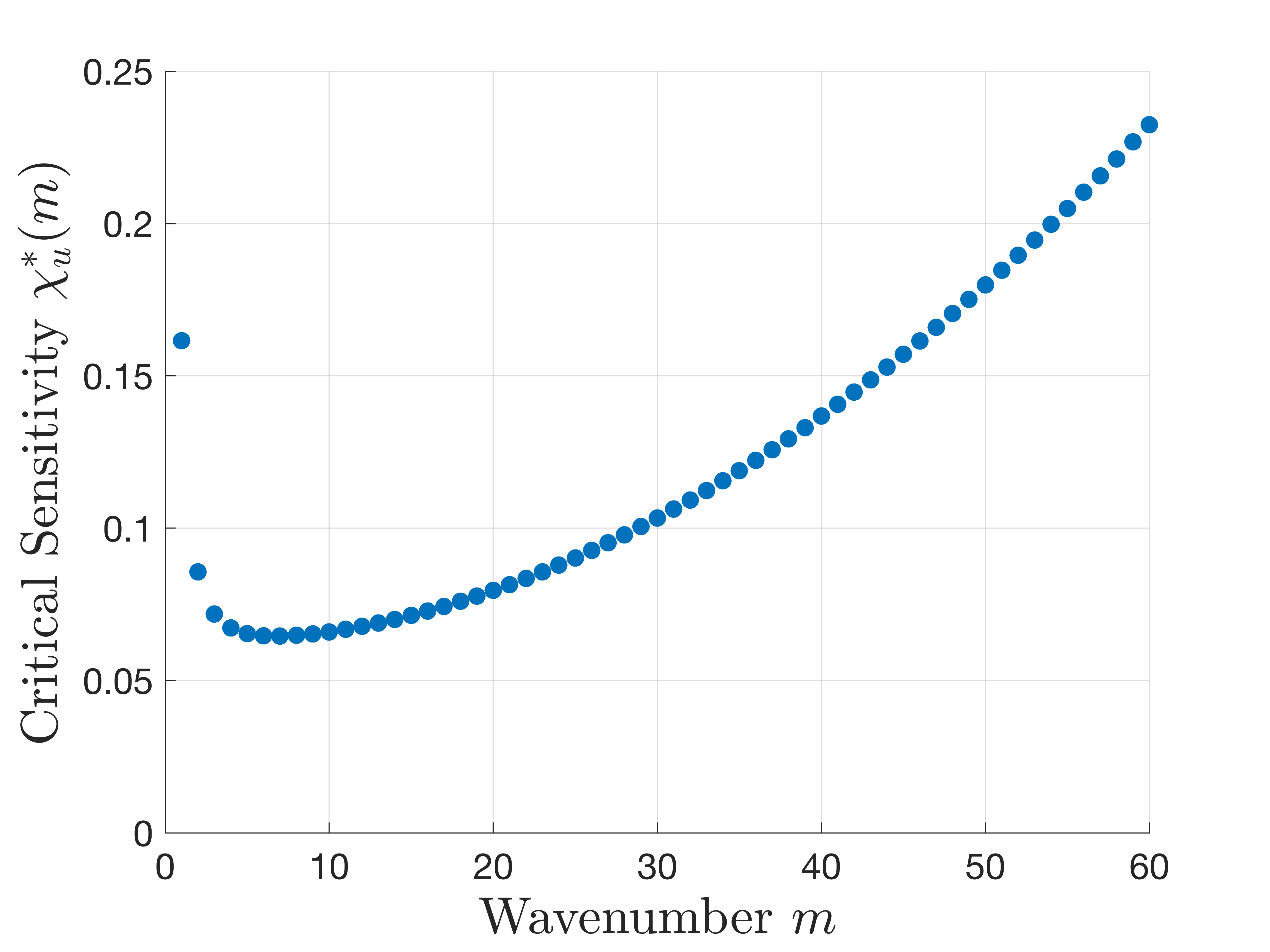}
\includegraphics[width = 0.45\textwidth]{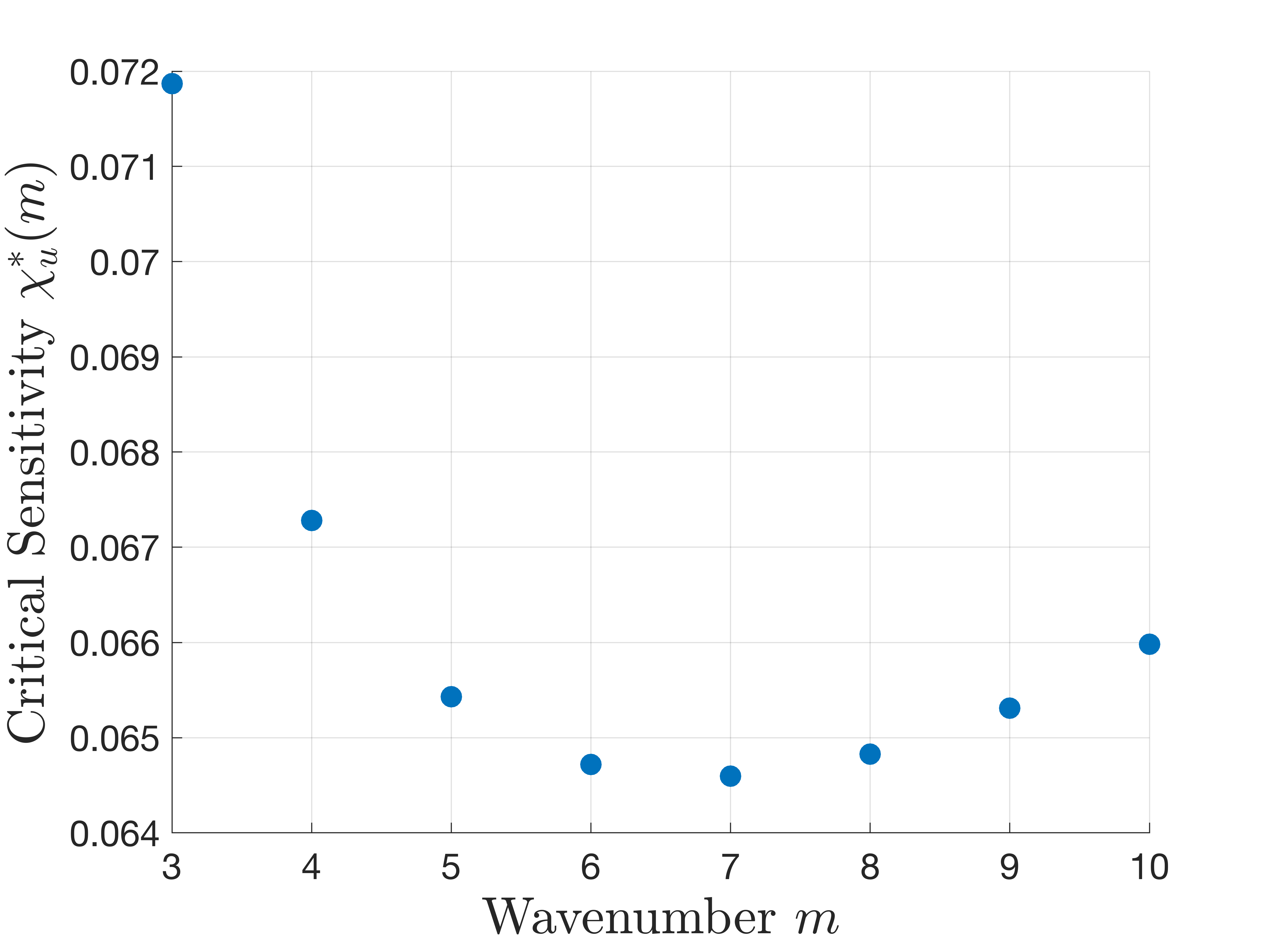}
   \caption{Plots showing the critical value of $\chi_u^*(m)$ as a function of the mode $m$ of the sinusoidal perturbation for the case of equal diffusion coefficients $D_u = D_v$ for the low-effort and high-effort harvesters. The values of $\chi_u^*(m)$ are presented for integer wavenumbers $m$ between 1 and 40 for the left panel and between 1 and 10 for the right panel, with the right panel serving to highlight the wavenumber $m$ that is most conducive for patter formation. The minimum value of $\chi_u^*(m)$ is $0.0646$, which is achieved by the wavenumber $m^* = 7$. The initial conditions for the simulation are set as: $p_0 = 0.5$ and $n_0 = 0.5$. The domain length is $l = 10$, and the key parameters include diffusion coefficients $D_u = 0.01$ and $D_n = 0.01$, along with the sensitivity of environment-driven motion $\chi_v = 0.02$. Additional game-theoretic and ecological parameters include: $r = 1$, $a = 0.5$, $e_L = 0.2$, $e_H = 0.5$, $R_0 = 0.35$, $S_0 = 0.3$, $R_1 = 0.6$, $S_1 = 0.35$, $T_0 = 0.5$, $P_0 = 0.2$, $T_1 = 0.6$, and $P_1 = 0.3$. These parameters dictate the critical behavior of $\chi_u^*(m)$ across different perturbation modes. %
}
    \label{fig:chi_u_m}
\end{figure}

In particular, when the constant $d$ defined in Equations \eqref{eq:chistarum} and \eqref{eq:bcdpattern} is positive, we expect that there will be a range of positive strengths $\chi_u$ of environmental-driven motion for the low-effort harvesters for which small perturbations from the uniform state will not allow for the emergence of spatial patterns.  We further illustrate this threshold behavior Figure \ref{fig:chi_u_m}, displaying the relationship between the threshold  strength environmental-driven motion $\chi_u^*(m)$ for a range of perturbation wavenumbers $m$ for an example set of payoff and ecological parameters for which there is a unique stable equilibrium featuring coexistence of low-effort and high-effort harvester. This allows us to explore the overall threshold $\chi_u^*$ required to promote spatial pattern formation, which is achieved by the minimum point achieved for an integer wavenumber $m$. We see for this choice of parameters that the the integer wavenumber with minimal threshold $\chi^*_u(m)$ is $m = 7$, which constitutes the most unstable wavenumber in our linearization and provides the wavenumber we predict for emergent patterns when $\chi_u$ is sufficiently close to the threshold $\chi^*_u$. 

However, the constant $d$ from Equation \eqref{eq:bcdpattern} can also be negative if $c_{22} > 0$ and $D_n$ is sufficiently large. This means that it may be possible to achieve spatial pattern formation even for the case in which $\chi_u \leq \chi_v$ and high-effort harvester is at least as efficient as low-effort harvesters in terms of climbing gradients of environmental quality. In any event, it is possible to see that, regardless of the sign of $c_{22}$, it is possible to require $\chi_u > \chi_v$ as a necessary condition for the achievement of spatial pattern formation provided that the diffusion of the environmental quality index is sufficiently weak (corresponding to $D_n$ sufficiently small). Therefore we see that there may be multiple pattern-forming mechanisms that allow for spatial instability of the uniform state for our PDE model, but that our mechanism of low-effort harvesters having an advantage under environmental-driven motion will be necessary for spatial-driven motion for the case of weak diffusion of environmental quality and equal diffusivity of the low-effort and high-effort harvesters.  

\begin{remark}
For the case in which $c_{22} > 0$ and $\chi_u = \chi_v$, the determinant of our Jacobian matrix becomes
\begin{equation}
\det(\tilde{A}) = \underbrace{c_{22} c_{33} - c_{32} c_{23}}_{= \det{\tilde{C}}} + \left( \frac{m \pi}{l} \right)^4 D_u D_n - c_{22} \left( \frac{m \pi}{l} \right)^2 D_n - c_{33} \left( \frac{m \pi}{l} \right)^2 D_u,
\end{equation}
which bears some resemblance to the linearization for a Turing instability. When $c_{22} > 0$ and $c_{23} < 0$, we see that the linearization of the $(p,n)$ subsystem has the sign pattern
\begin{equation}
\tilde{C} = \bpm c_{22} & c_{23} \\ c_{32} & c_{33} \epm = \bpm + & - \\ + & - \epm,
\end{equation}
which is the activator-inhibitor sign pattern that allows for Turing instability \cite{gierer1972theory,murray2007mathematical} (with low-effort harvesters serving as the activator and environmental quality serving as the inhibitor). A similar activator-inhibitor sign pattern for Turing instability was found in the model by Cheng and coauthors \cite{cheng2024evolution}, who considered a reaction-diffusion model for evolutionary games with environmental feedback as a system of two PDEs for the density of low-effort harvesters and an environmental quality index. 
\end{remark}

\subsubsection{Linear Stability Analysis for Case of Non-Identical Diffusion Coefficients}
\label{sec:LSAunequalDuDv}

Next, we consider the possibility that the diffusion coefficients are unequal between the two strategies. This will allow us to explore how changing the relative diffusivity of the low-effort harvesters will impact the threshold sensitivity $\chi_u^*$ for environmental-driven motion required to produce spatial pattern formation. For the case in which $D_u \ne D_v$, we do not obtain the same simplification of the linearization matrix we see in Section \ref{sec:LSAequalDuDv}, and instead we need to study the stability of the uniform state using the full linearization matrix $A(m)$ from Equation \eqref{eq:PDE_Jaco_A}. One can use the Routh-Hurwitz criteria for the eigenvalues of a $3\times3$ matrix to analyze the stability in this case, which is often applied to study pattern formation in spatial models featuring three interacting species \cite{satnoianu2000turing,tania2012role,parshad2014turing,piskovsky2025turing,regueira2025network}. We explain how to use the Routh-Hurwitz criteria to establish conditions for instability in Section \ref{sec:app_LSA_unequal}, and the explicit expressions for this threshold $\chi_u^*$ are provided in supplementary Mathematica notebooks. %

\begin{figure}[!ht]
    \centering
    \includegraphics[width = 0.45\textwidth]
    {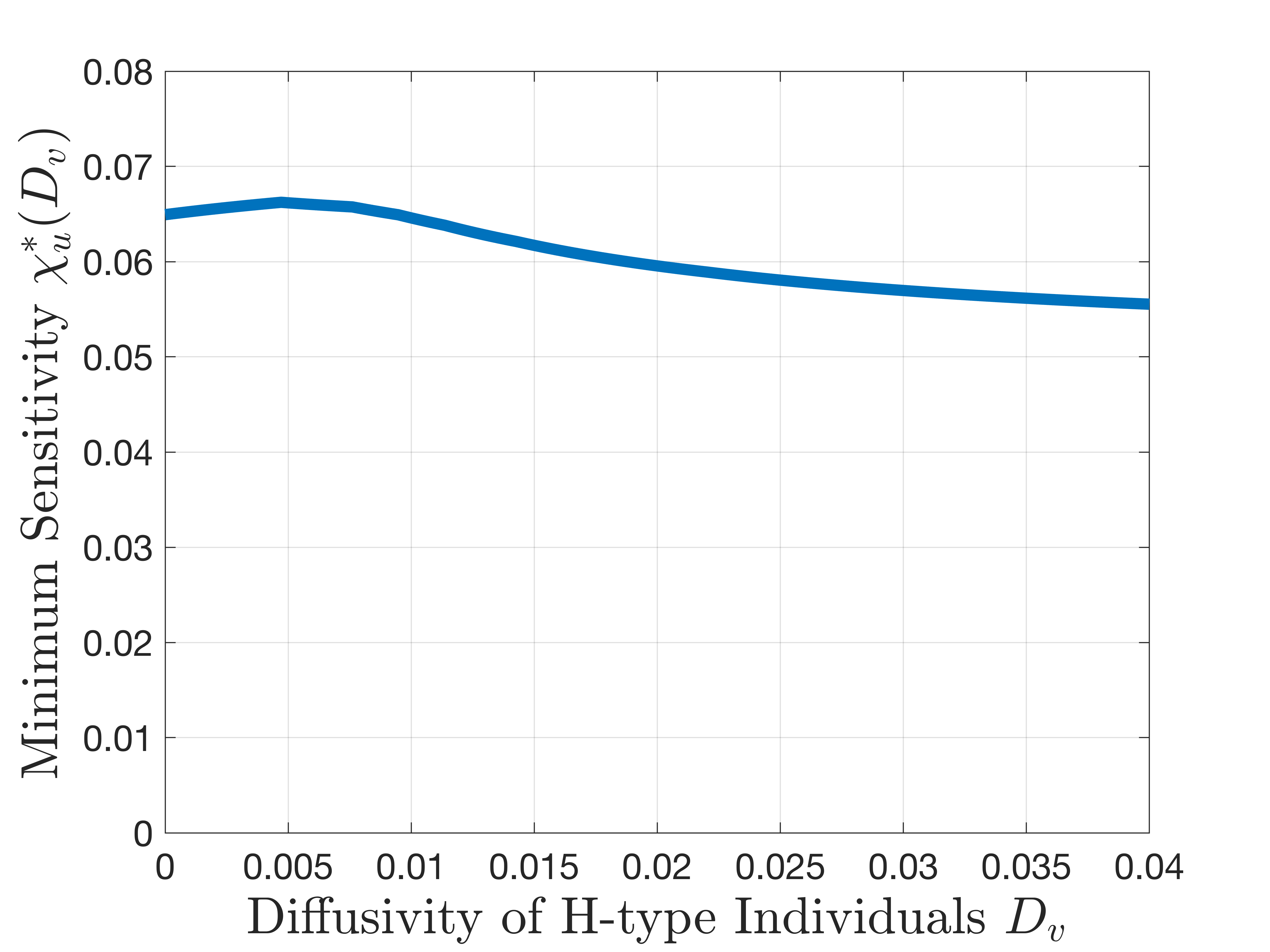}
     \includegraphics[width = 0.45\textwidth]
    {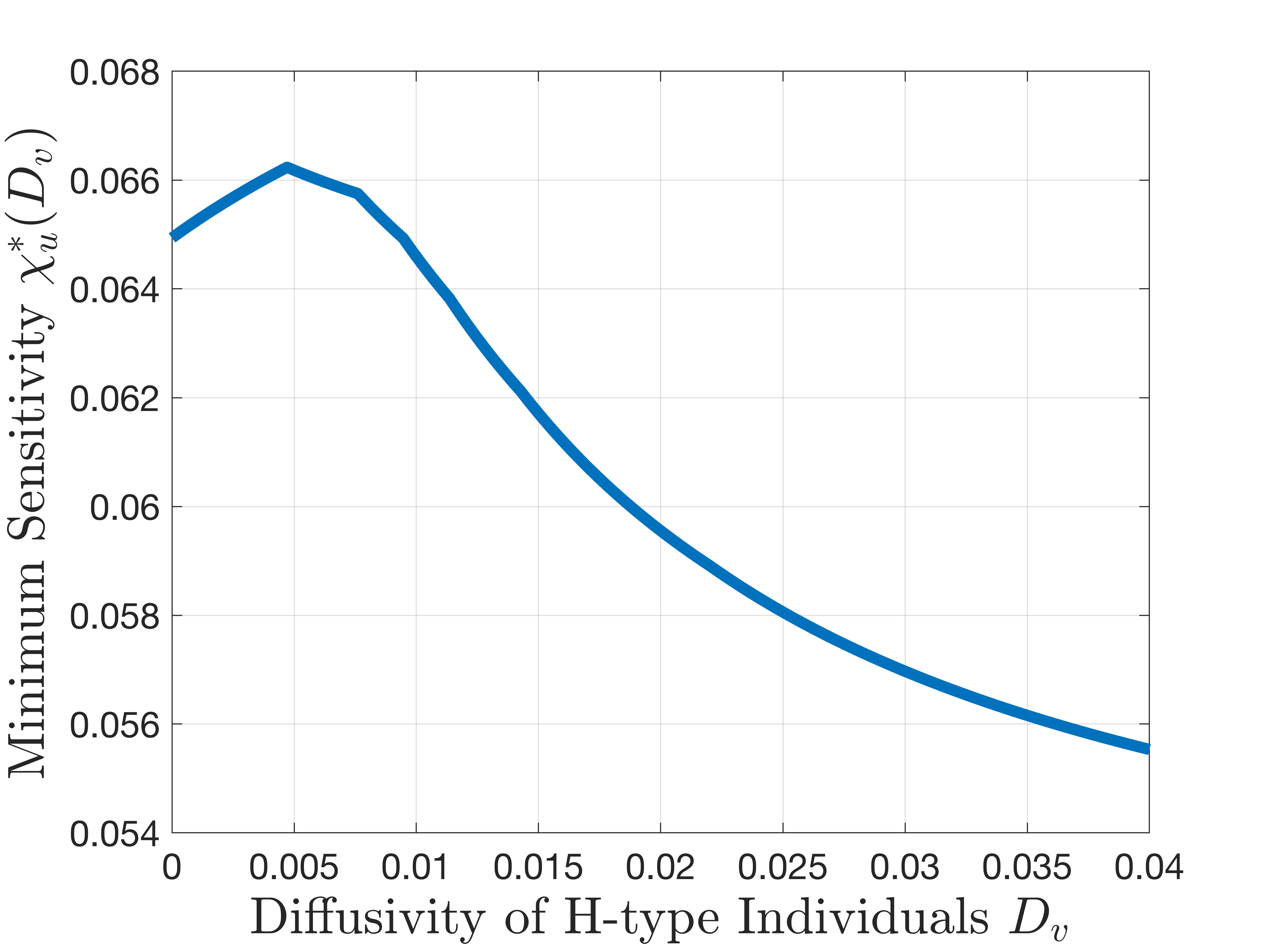}

   \caption{Plots of critical $\chi_u^*$ as a function of the varying diffusion coefficient $D_v$ for the high-effort harvester and fixed diffusivity $D_u = 0.1$ for the low-effort harvester. The two panels depict the same curve for $\chi_u^*$ with different ranges displayed for the vertical axis, with the right panel zooming in on the values close to the curve to highlight the non-monotonic dependence of the critical strength of environmental-driven motion $\chi_u^*$ on the diffusivity $D_v$ of high-impact harversters. All other parameters for spatial movement (including the diffusivity of the environmental quality metric $D_n$, sensitivities of environment-driven motion $\chi_u$ and $\chi_v$) and all game-theoretic and ecological parameters are consistent with those defined in Figure \ref{fig:chi_u_m}. 
   }
    \label{fig:Chi_u_D_v}
\end{figure}

We can then use the expression for $\chi_u^*$ derived from the Routh-Hurwitz criteria to explore how the pattern-forming threshold on the strength of directed motion $\chi_u^*$ will depend on the unequal diffusion coefficients $D_u$ and $D_v$ for the two harvesting strategies. In Figure \ref{fig:Chi_u_D_v}, we present the threshold strength of environmental-driven motion $\chi_u^*$ obtained by varying the diffusivity $D_v$ of the high-effort harvester and holding fixed the diffusivity $D_u$ of the low-effort harvester. We see that this pattern-forming threshold $\chi_u^*$ has a non-monotonic dependence on $D_v$, initially increasing as the high-effort diffusivity increases before reaching a maximum around $D_v = 0.0047$ and then decreasing with $D_v$ for all subsequent values of $D_v$ considered. In particular, this maximum threshold is achieved when the diffusivity of the high-effort harvesters is less than the diffusivity of the low-effort harvesters, but considering either $D_v$ close to 0 or large $D_v$ can be more conducive to the formation of spatial patterns by environmental-driven motion than the case of equal diffusivity $D_u = D_v$. From a biological standpoint, it is possible that this non-monotonic behavior is due to the fact that low and high values of high-effort diffusivity $D_v$ correspond respectively to slower overall motion and to rapid movement towards a uniform spatial distribution, both of which could potentially hinder the ability to move towards regions featuring greater environmental quality. 

\section{Transient and Long-Time Behavior of Numerical Simulations of PDE Model}
\label{sec:numericalsimulations}

Having established the possibility of spatial pattern formation when environmental-driven motion of low-impact harvesters is sufficiently strong, we can now look to examine the emergence of spatial patterns using numerical simulations of our PDE models starting from initial conditions close to the uniform equilibrium state. We first perform numerical simulations of our model for values of $\chi_u$ slightly above the the threshold $\chi_u^*$ required for pattern formation, allowing us to compare our numerical results with the predictions of linear stability analysis (Section \ref{sec:numericaldynamicsandlinear}). We then illustrate the patterns achieved for different example values of the strength $\chi_u$ of environmental-driven motion for the $L$-strategies (Section \ref{sec:numericalsteady}) and compare key quantities of our numerical simulations for as a function of $\chi_u$ (Section \ref{sec:numericalchiucomparison}), showing how increasing the level of environmental-driven motion can have detrimental effects on steady-state payoff and environmental quality. Finally, we present examples time-dependent numerical solutions for $\chi_u$ further away from $\chi_u^*$ illustrating transient dynamics of the merging and splitting of local regions with improved environmental quality (Section \ref{sec:numericaltransient}).

We use \textsc{Matlab} to simulate our system of partial differential equations (PDEs) with the built-in \texttt{pdepe} function, which implements numerical solutions to the PDE model using the method of lines, a Petrov-Galerkin spatial discretization, and a variable-step, variable-order solver in time\cite{PDEPEdocumentation,skeel1990method}. We consider initial conditions that consist of densities that are close to the spatially uniform equilibrium states, taking the form
\begin{equation} \label{eq:initialcondtionsrandom}
\begin{aligned}
 u(0,x_i) &= u_0 + \alpha_0 \: \mc{U}(x_i) \\
 v(0,x_i) &= v_0 + \beta_0 \: \mc{V}(x_i) \\
 n(0,x_i) &= n_0 + \gamma_0 \: \mc{N}(x_i),
\end{aligned}
\end{equation}
where $\mc{U}(x_i)$, $\mc{V}(x_i)$, and $\mc{N}(x_i)$ are each drawn independently from a uniform distribution on $[-1,1]$ for each grid-point $x_i$ and $\alpha_i$, $\beta_i$, and $\gamma_i$ give a constant range for the random perturbation drawn at each grid-point. We draw the random variables independently for each grid-point $x_i$ in the discretization of our spatial domain, allowing us to explore the possible emergence of spatial patterns from a small, random perturbation from a uniform equilibrium state. Because we will consider the effect of different parameter values of the behavior of emergent patterns, we use the Mersenne Twister algorithm to generate random seeds for our initial distributions that will allow us to consider the same set of randomly drawn initial conditions across simulations with different parameter values. 

\subsection{Pattern-Forming Dynamics and Comparison with Linear Stability Analysis}
\label{sec:numericaldynamicsandlinear}

\begin{figure}[!ht]
    \centering
    \subfloat[Heatmap of $u$]{
        \includegraphics[width=0.45\textwidth]{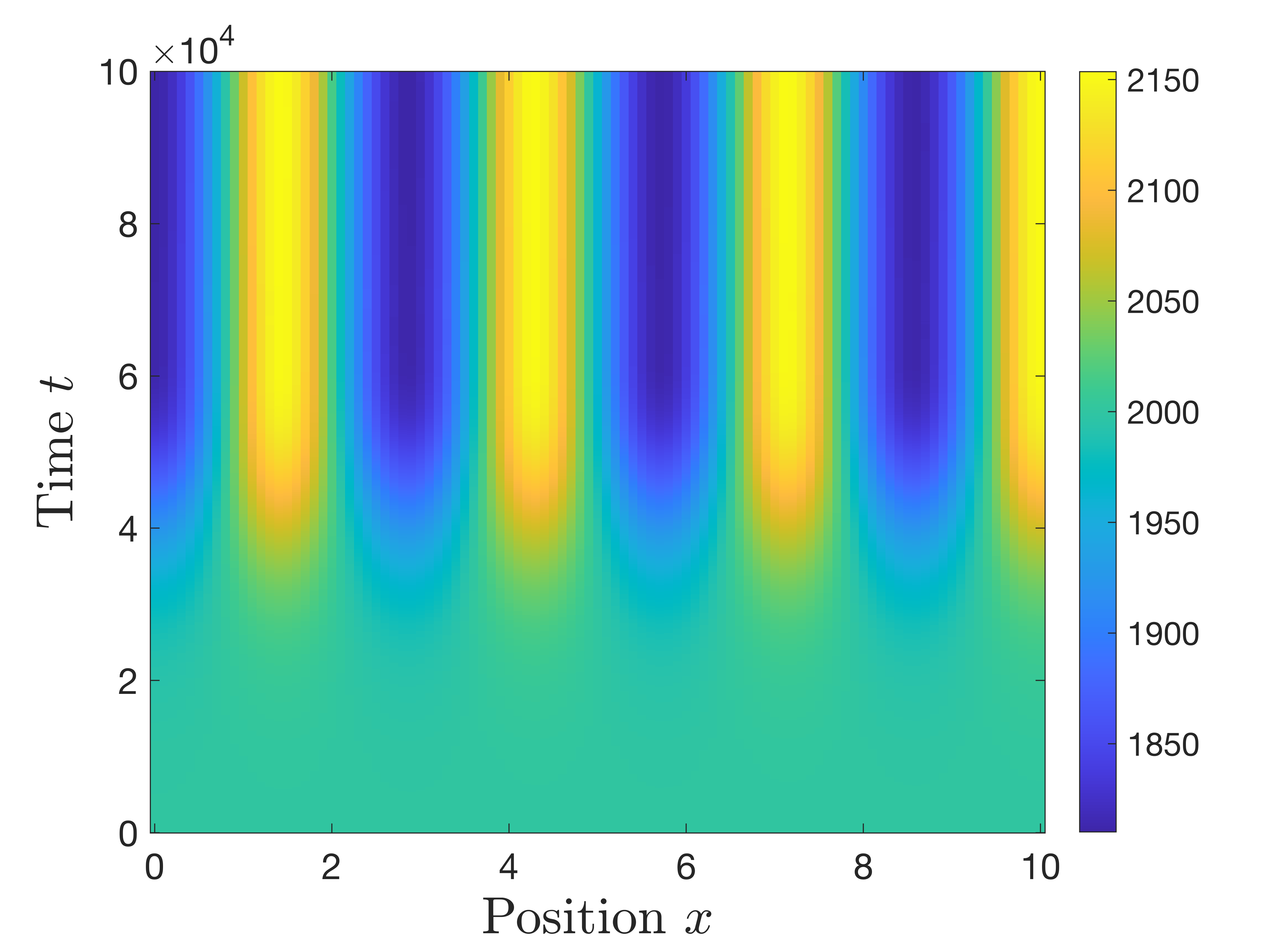}
        \label{fig:heatmap_u}
    }\hspace{5mm}
    \subfloat[Heatmap of $v$]{
        \includegraphics[width=0.45\textwidth]{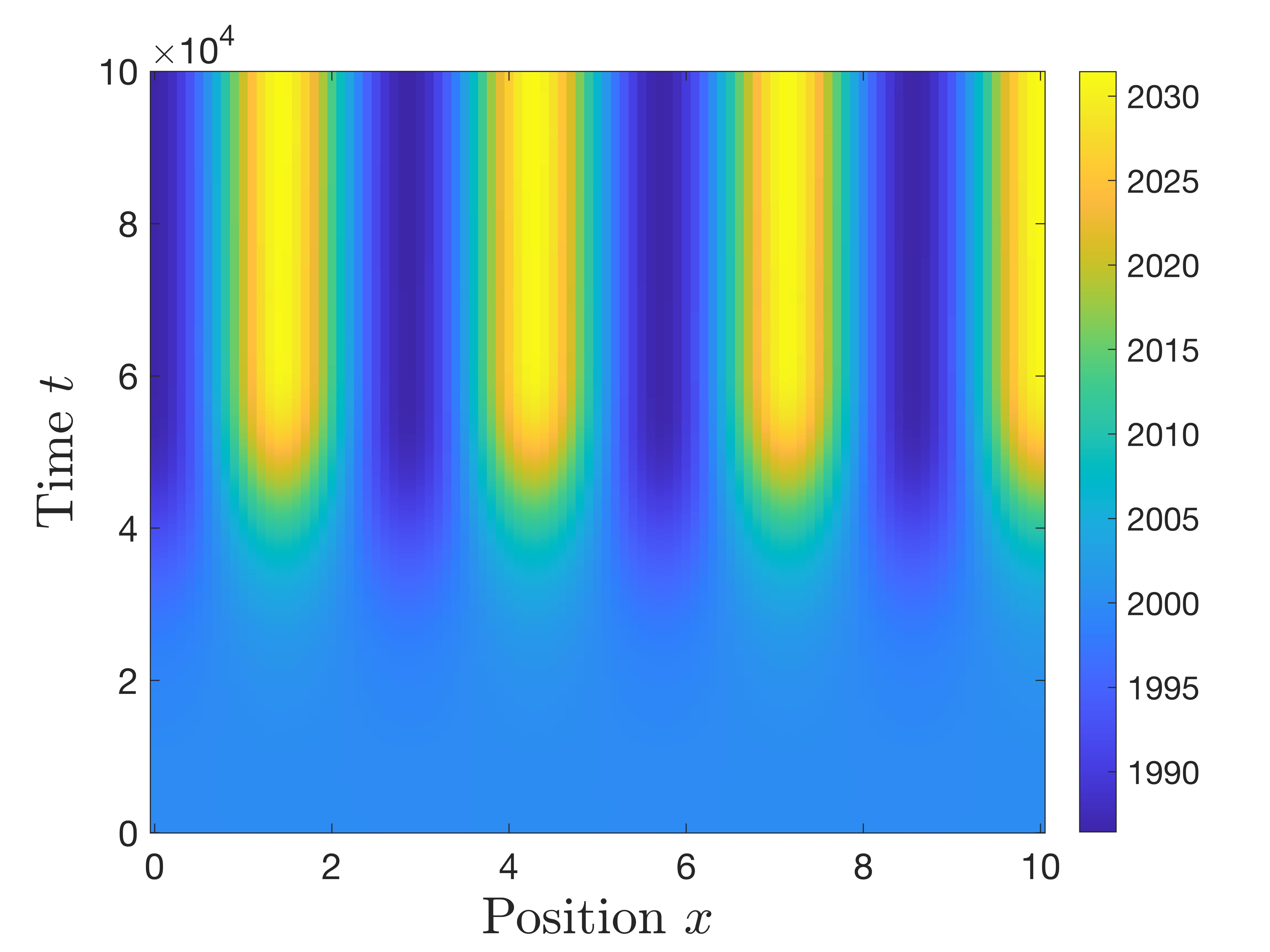}
        \label{fig:heatmap_v}
    }\hspace{5mm}
    \subfloat[Heatmap of $f_L$]{
        \includegraphics[width=0.45\textwidth]{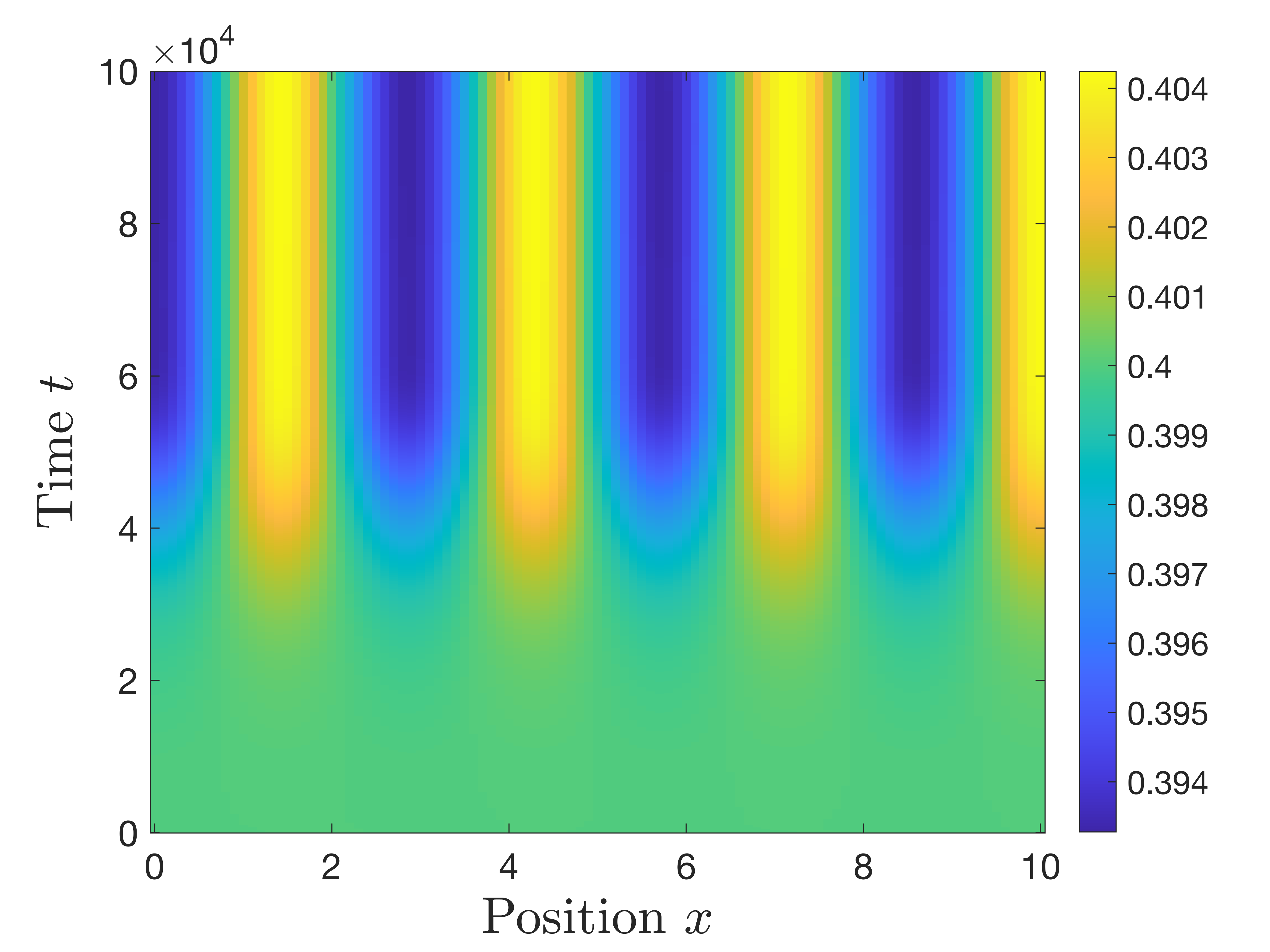}
        \label{fig:heatmap_pi_L}
    }\hspace{5mm}
    \subfloat[Heatmap of $f_H$]{
        \includegraphics[width=0.45\textwidth]{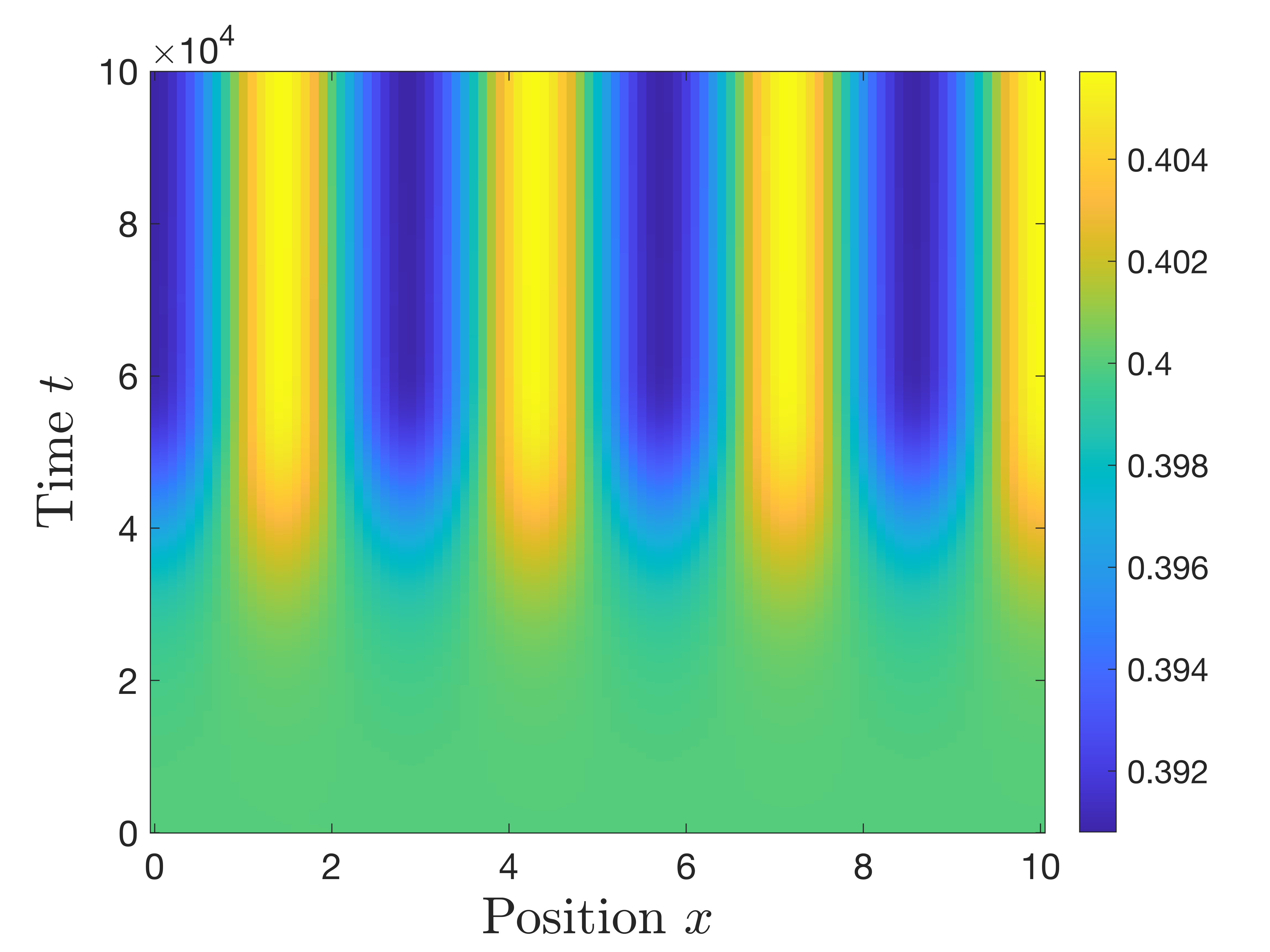}
        \label{fig:heatmap_pi_H}
    }
    \caption{Heatmaps of populations $u$, $v$, and payoffs $f_L$, $f_H$ plotted in terms of the spatial position $x$ and time $t$. Simulations are conducted with sensitivities of environmental-driven motion of $\chi_u = 0.0653$ as $\chi_v = 0.02$ and with diffusion coefficients $D_u = D_v = D_n = 0.01$. The initial conditions were given by Equation \eqref{eq:initialcondtionsrandom}, which consist of perturbations of the uniform state $(u_0,v_0,n_0) = (2000,2000,0.5)$ with uniform random variables drawn at each spatial grid-point and with the range of initial perturbations characterized by $\alpha_i = 0.1$, $\beta_i = 0.1$, and $\gamma_i = 0.001$. The game-theoretic and ecological parameters were the same as used to produce Figure \ref{fig:chi_u_m}.}
    \label{fig:heatmap}
\end{figure}

We start our numerical exploration by considering how the densities of strategies $u(t,x)$ and $v(t,x)$ and the resource distribution $n(t,x)$ change in time starting from an initial condition close to the uniform equilibrium state. We consider a case of strength of environmental-driven motion $\chi_u$ for the $L$-strategist that is close to the threshold value $\chi_u^*$ for the pattern forming instability, exploring how the nonlinear dynamics of our PDE model affect the long-time behavior for a regime in which linear stability analysis predict spatial pattern formation.  In Figure \ref{fig:heatmap}, we present heat-maps of the values of the strategy distributions $u(t,x)$ and $v(t,x)$ and the spatial profiles of payoffs $f_L(u(t,x),v(t,x))$, and $f_H(u(t,x),v(t,x))$ as functions of time, representing spatial location on the horizontal axis and representing time on the vertical axis. Starting from initial conditions featuring small uniform noise around the uniform equilibrium state of the nonspatial ODE system, we see that noticeable patterns start to form after around $t = 4\times 10^4$, appearing to equilibrate around a steady-state density featuring a periodic distribution in both the densities of strategies and the densities of the average payoffs in space. 

As we are considering the case of $\chi_u$ very close to the threshold for pattern formation, we can also use the form of the patterns observed in Figure \ref{fig:heatmap} as a point of comparison with the prediction made by our linear stability analysis. From our linear stability analysis, we characterized conditions for perturbations proportional to $\cos\left( \frac{m \pi}{l}\right)$ to grow in time when the strength of environmental-driven motion for $L$-strategists satisfies $\chi_u > \chi_u^*(m)$. For the choice of parameters chosen in Figure \ref{fig:heatmap}, we see from Figure \ref{fig:chi_u_m} that we expect to see the most unstable perturbation of the uniform state to be proportional to be the perturbation with index $m = 7$, and therefore we anticipate that the spatial profiles of patterns will have the wavenumber $k = \frac{m \pi}{l} = \frac{7 \pi}{10}$ for numerical simulations on a domain of length $l = 10$ when $\chi_u$ is slightly above the critical threshold $\chi_u^*$. From the numerical simulations presented in Figure \ref{fig:heatmap}, we see that the spatial profiles settle around a distribution featuring approximately $3.5$ wave peaks over the spatial domain $l = 10$, corresponding to a wavelength $\lambda = \frac{20}{7}$. This means that the wavenumber of the patterns obtained from our numerical simulation is $k = \frac{2 \pi}{\lambda} = \frac{7 \pi}{10}$, agreeing with the prediction we found from our linear stability analysis. %

\subsection{Example Long-Time Outcomes Achieved Under Numerical Simulations}
\label{sec:numericalsteady}

Next, we examine the emergent spatial patterns achieved under our PDE model both for a strength of environmental-driven motion $\chi_u$ close to the pattern-forming threshold $\chi_u^*$ and for a strength further from the threshold value. To examine the patterns, we run our simulations until time $t = 10^6$ and then plot the resulting spatial densities of the low-effort and high-effort strategies $u(t,x)$ and $v(t,x)$, the spatial profiles of the payoffs $f_L(u(t,x),v(t,x))$ and $f_H(u(t,x),v(t,x))$ for each strategy, and the spatial profile of the environmental quality index $n(t,x)$ at this time $t$. 

\begin{figure}[!ht]
    \centering
       \subfloat[Spatial Distribution of Populations $u$ and $v$]{
        \includegraphics[width=0.45\textwidth]{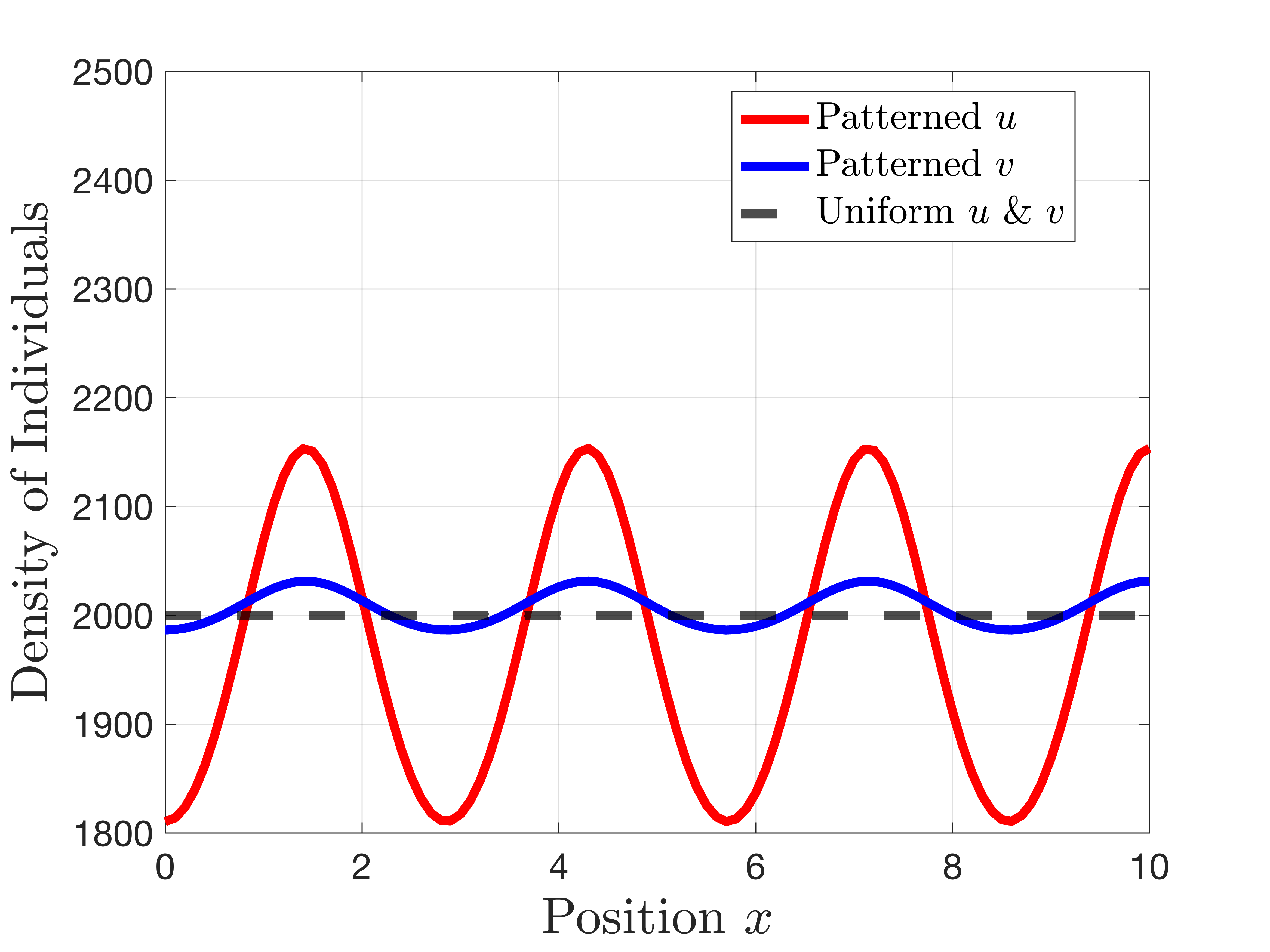}
        \label{fig:uv_pi_close}
    }\hspace{5mm}
    \subfloat[Spatial Distribution of Environmental Factor $n$]{
        \includegraphics[width=0.45\textwidth]{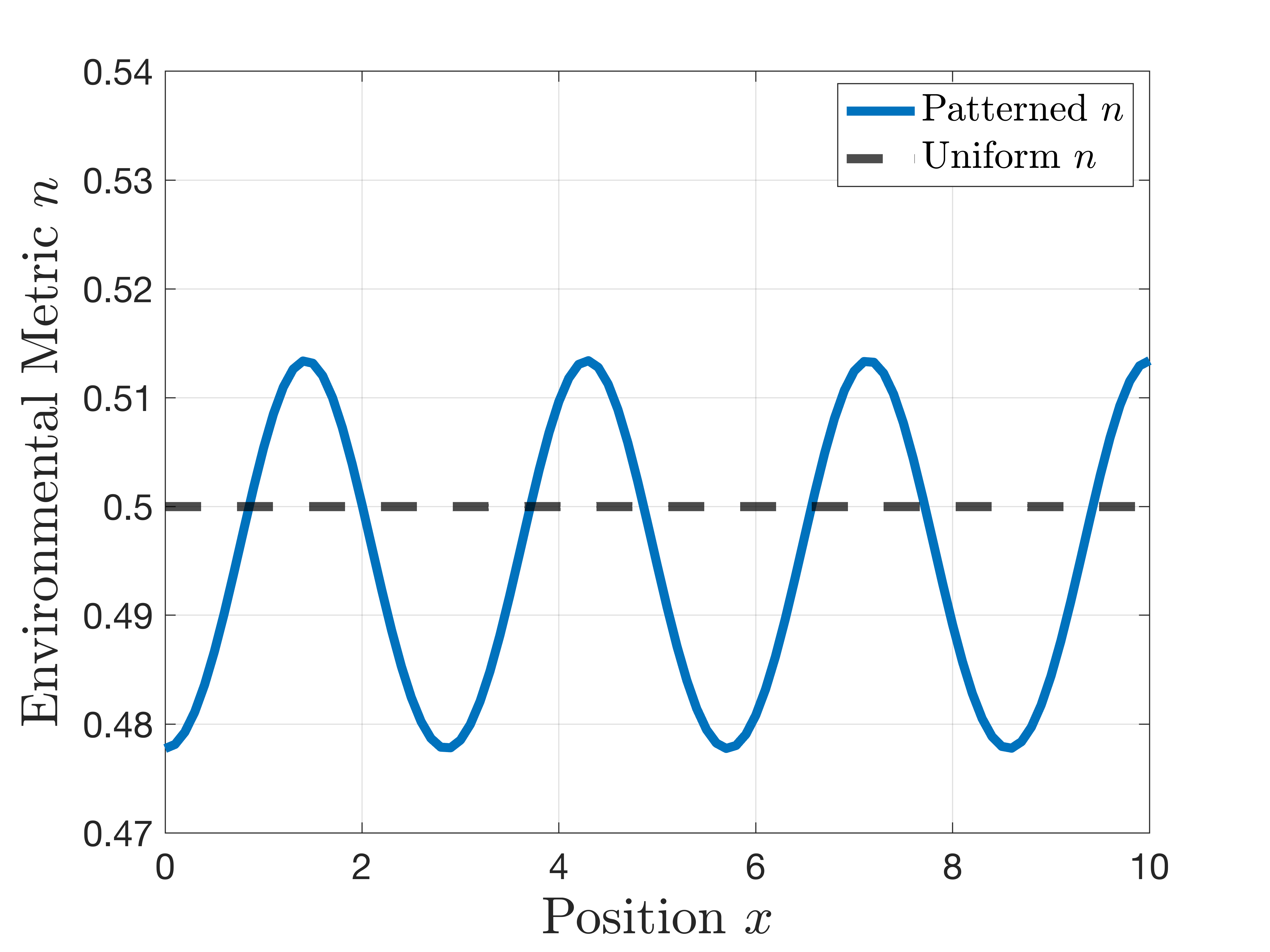}
        \label{fig:n_pi_close}
    }\hspace{5mm}
    \subfloat[Payoffs $f_L$ and $f_H$ Across the Spatial Domain]{
        \includegraphics[width=0.45\textwidth]{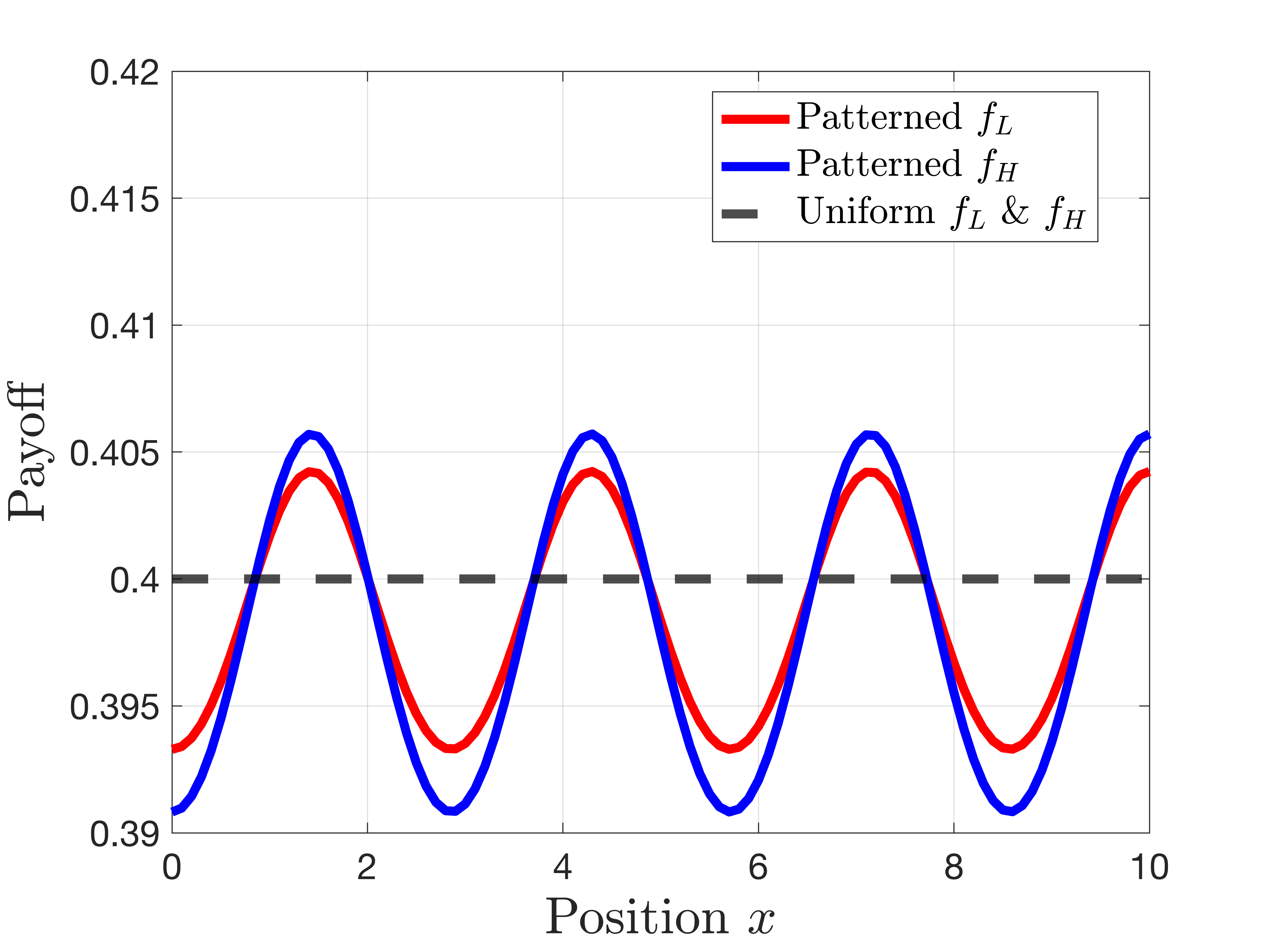}
        \label{fig:pi_LH_close}
    }
    \caption{Payoff distributions $f_L(t,x)$ and $f_H(t,x)$ for low-impact and high-impact harvesters at time $t = 10^6$, with $\chi_u = 0.0647$ and $\chi_v = 0.02$. Spatial distributions of populations $u$, $v$, and environmental factor $n$ at $t = 10^6$, with $\chi_u = 0.0647$ and $\chi_v = 0.02$. The game-theoretic and ecological parameters are consistent with Figure \ref{fig:chi_u_m}, while we draw our initial conditions using the form and parameters considered in Figure \ref{fig:heatmap} featuring small perturbations of the uniform equilibrium $(u_0,v_0,n_0) = (2000,2000,0.5)$ at each spatial gridpoint. Note that the dashed lines in each panel correspond to the values of each of the values taken for the plotted quantities achieved in the stable spatially uniform equilibrium for our eco-evolutionary game. }
    \label{fig:pi_uvnLH_close}
\end{figure}

In Figure \ref{fig:pi_uvnLH_close}, we display the spatial patterns that arise for the case of a strength of environmental-driven motion $\chi_u$ for the low-effort harvester close to the threshold for our pattern-forming instability. We see that the numerical solutions produce sinusoidal patterns with the wavenumber predicted from linear stability analysis for the densities of strategies, the payoffs achieved by each strategy, and the distribution of environmental quality. In particular, we see that there exist regions in space corresponding to peaks of the pattens featuring clustering of greater population density and payoffs for both strategies as well as an improved environmental quality. Correspondingly, there are correlated troughs of these patterns corresponding to lower population density, lower payoffs for each strategy, and degraded environmental quality. Furthermore, the regions of high payoff and high environmental quality feature a majority of individuals following the low-impact strategy, while the regions of lower payoff and lower environmental quality feature a majority of individuals following a high-impact strategy. We find that the payoff and environmental quality at the peaks of the patterns exceed the equilibrium value achieved in the spatially-uniform state and fall below the spatially-uniform equilibrium value at the troughs of the patterns,  suggesting that the improved ability of the low-impact harvesters to climb gradients of environmental quality promote the formation of high-payoff, abundant-environment regions at the expense of leaving behind many high-effort harvesters to form a region with low payoffs and environmental quality. 

\begin{figure}[!ht]
    \centering
    \subfloat[Spatial Distribution of Populations $u$ and $v$]{
        \includegraphics[width=0.45\textwidth]{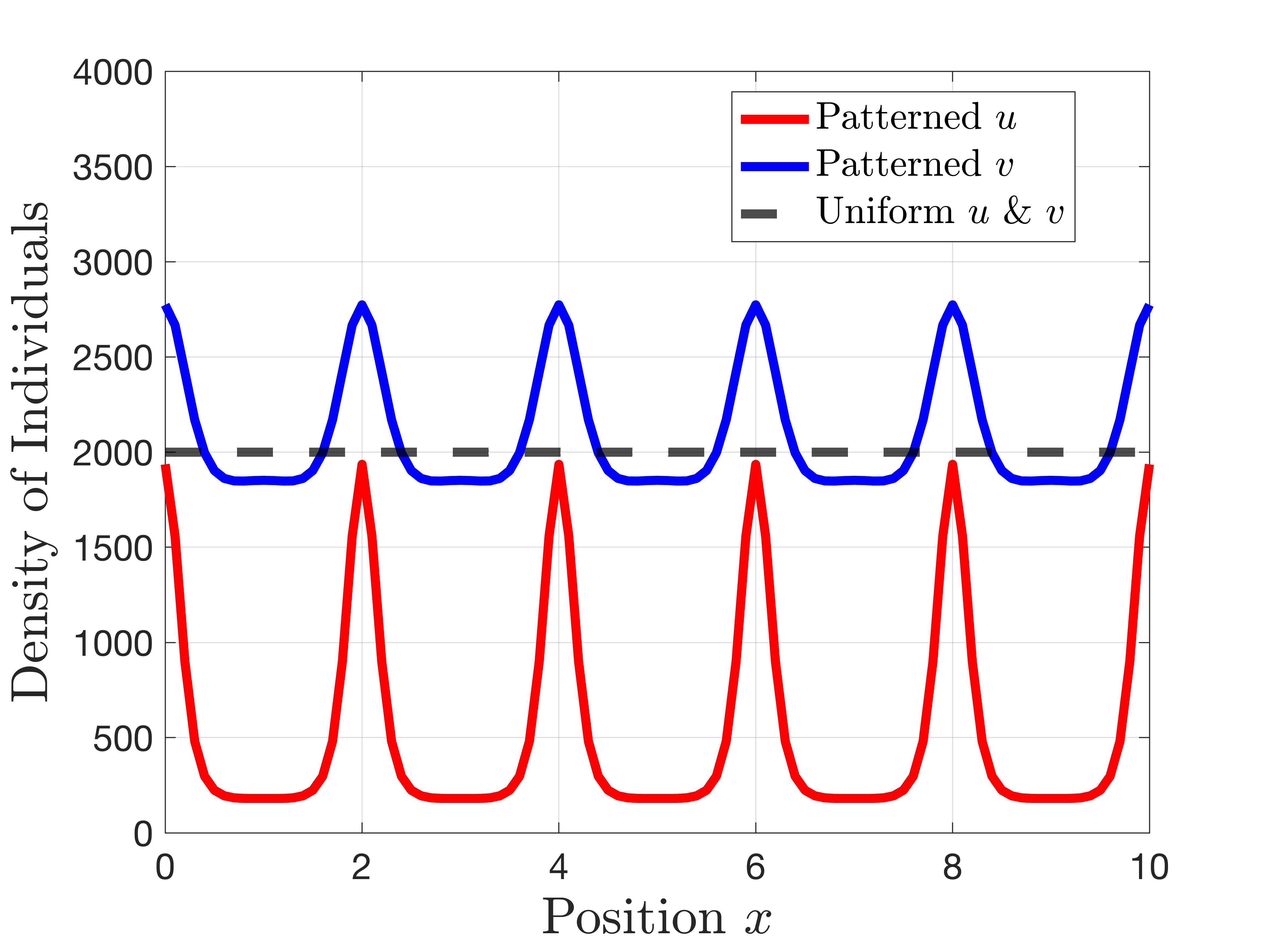}
        \label{fig:uv_pi_far}
    }\hspace{5mm}
    \subfloat[Spatial Distribution of Environmental Factor $n$]{
        \includegraphics[width=0.45\textwidth]{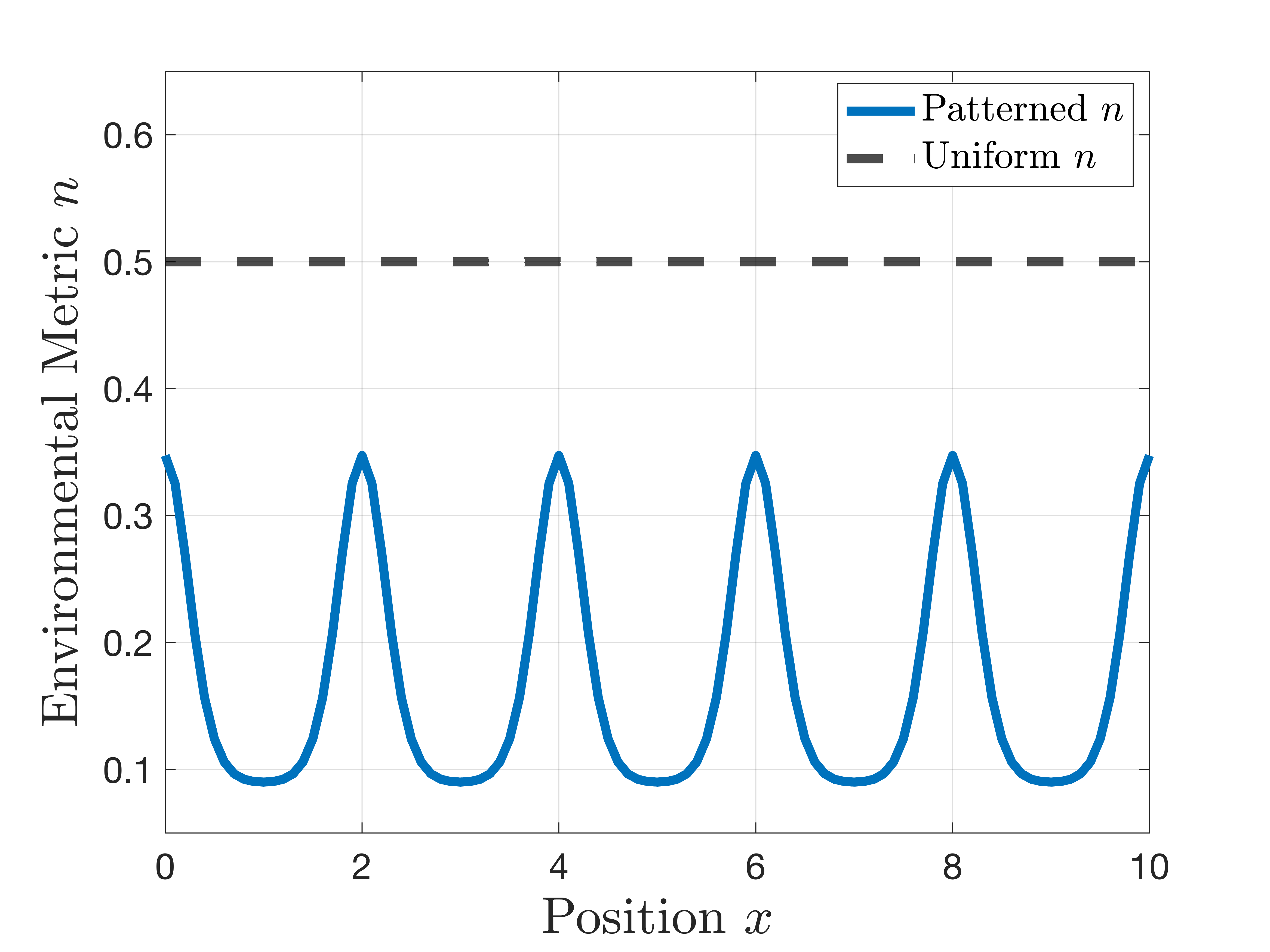}
        \label{fig:n_pi_far}
    }\hspace{5mm}
    \subfloat[Payoff dynamics for $f_L$ and $f_H$]{
        \includegraphics[width=0.45\textwidth]{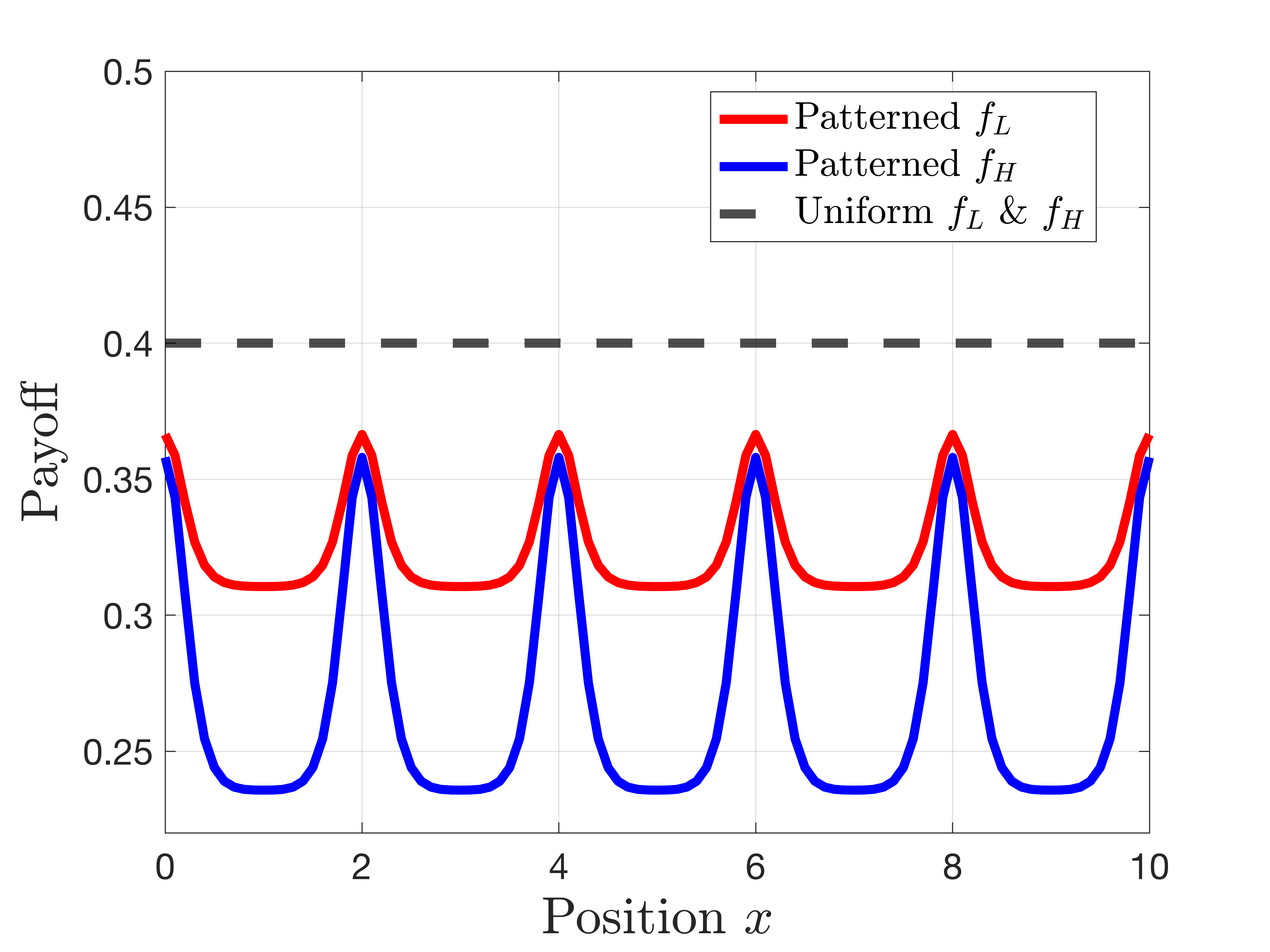}
        \label{fig:pi_LH_far}
    }
    \caption{Payoff dynamics for $L$-type $f_L$ and $H$-type $f_H$ individuals at $t = 10^6$, with $\chi_u = 0.1$ and $\chi_v = 0.02$. Spatial distributions of populations $u$, $v$, and environmental factor $n$ at $t = 10^6$, with $\chi_u = 0.1$ and $\chi_v = 0.02$. The game-theoretic and ecological parameters are consistent with Figure \ref{fig:chi_u_m}, while the initial condition was drawn in the same manner as in Figure \ref{fig:heatmap}. The dashed lines in each panel correspond to the values taken for the quantities of interest for the spatially uniform equilibrium for the eco-evolutionary game.}
    \label{fig:pi_uvnLH_far}
\end{figure}

We then study the case of stronger environmental-driven motion in Figure \ref{fig:pi_uvnLH_far}, exploring the states achieved at time $t= 10^6$ by numerical simulations of our PDE model for a strength of environmental-driven motion $\chi_u$ further from the threshold value $\chi_u^*$. As with the case of $\chi_u$ close to the threshold, we see that the emergent spatial patterns feature regions of greater relative population density, increased average payoff for each strategy, and an improved relative level of environmental quality.Unlike the case seen in Figure \ref{fig:pi_uvnLH_close}, the spatial patterns we observe have more of a spike shape as compared to the sinuosoidal patterns we observed close to the instability threshold. In addition, we see that the average payoffs of both strategies (Figure \ref{fig:pi_LH_far}) and the level of environmental quality (Figure \ref{fig:n_pi_far}) achieves lower values at all points in the spatial domain when compared to the uniform equilibrium values achieved in the absence of any spatial movement. Therefore no individuals end up receiving a higher payoff or living in an improved environment by following the scheme of environmental-driven motion relative to the outcomes they would experience in a nonspatial setting, suggesting that environmental-driven motion with sufficient strength $\chi_u$ of directed motion for low-effort harvested can result in both an individually and collectively worse outcome for the population than the payoff and environment achieved under nonspatial evolutionary games with environmental feedback.

\subsection{Impact of Environmental Sensitivity on Populations and Average Payoff}
\label{sec:numericalchiucomparison}

\begin{figure}[!ht]
     \centering
     \subfloat[Payoff dynamics as a function of $\chi_u$]{
      \includegraphics[width=0.45\textwidth]{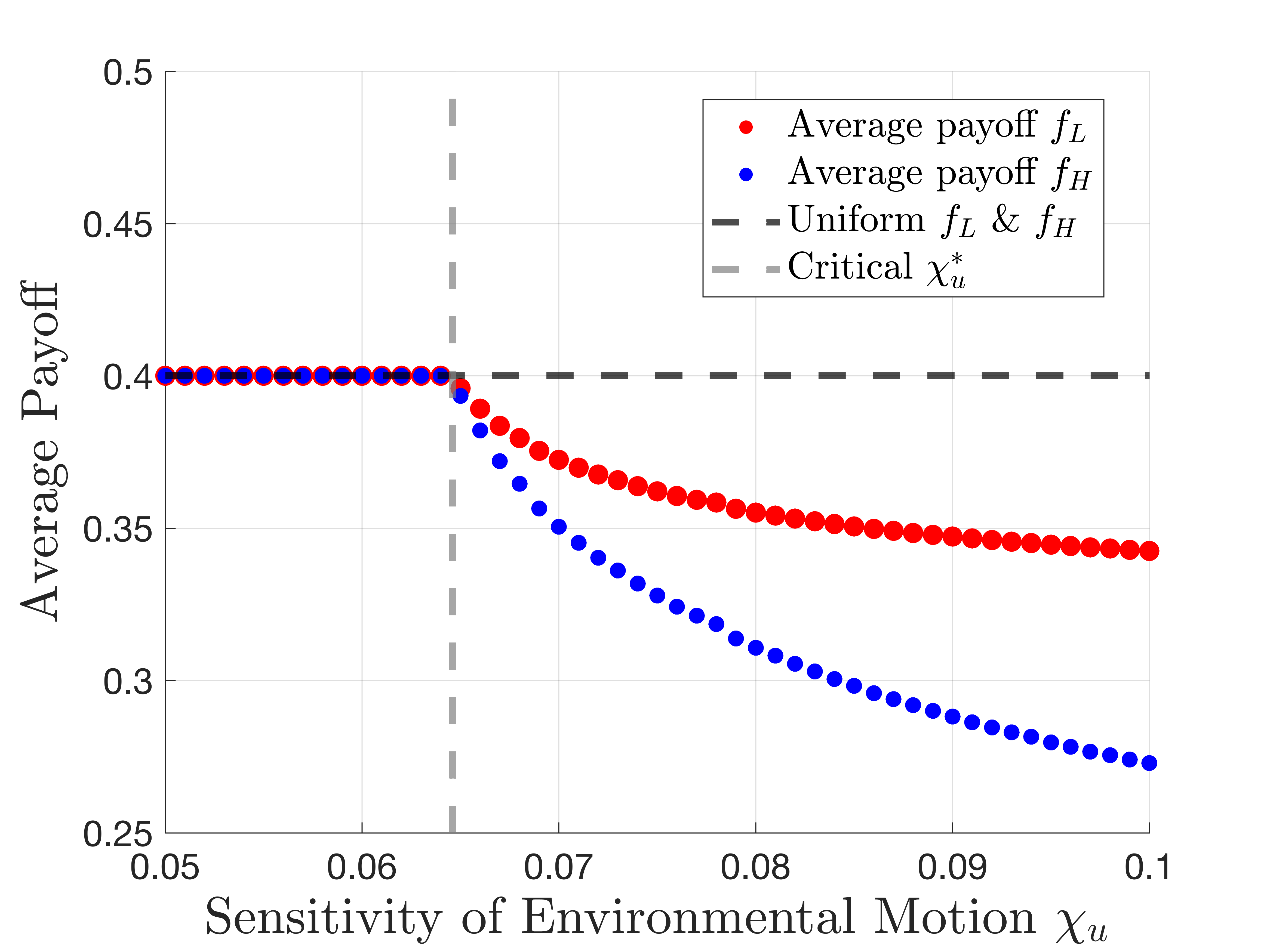}
      \label{fig:pi_chi_u}
     }
         \subfloat[Population dynamics as a function of $\chi_u$]{
      \includegraphics[width=0.45\textwidth]{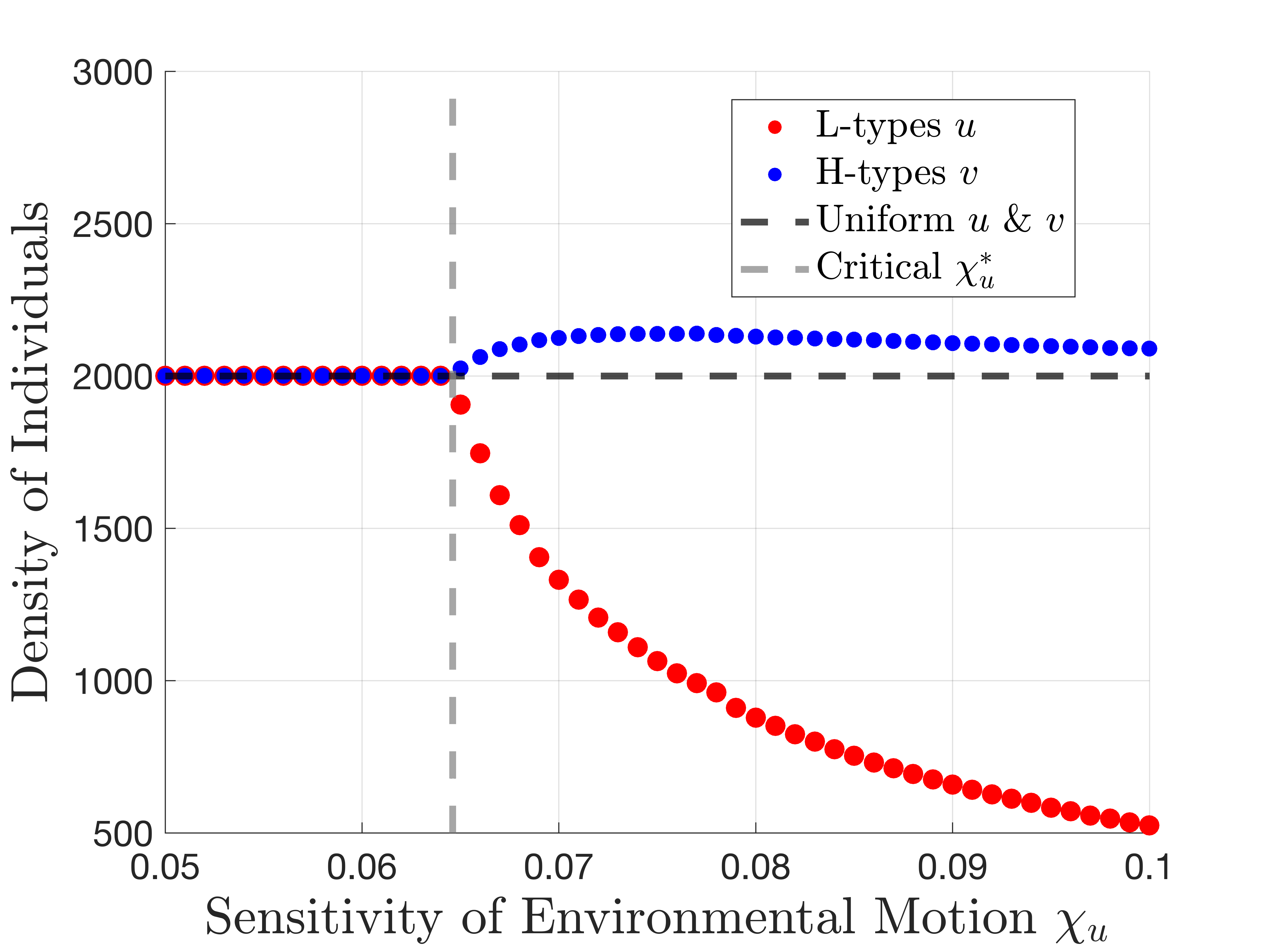}
      \label{fig:population_chi_u}
     }\hspace{5mm}
     \subfloat[Environmental factor $n$ response to $\chi_u$]{
      \includegraphics[width=0.45\textwidth]{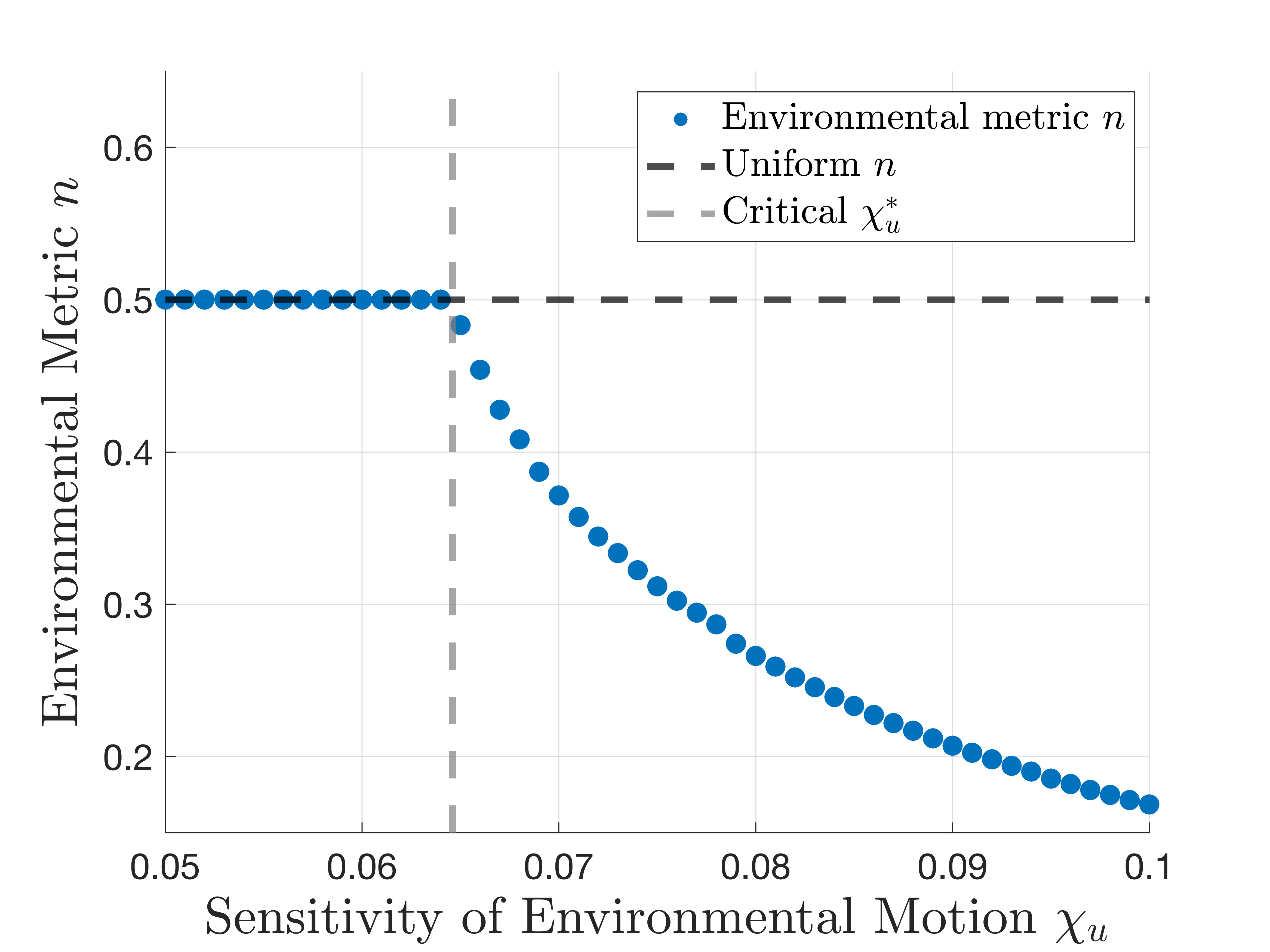}
      \label{fig:environment_chi_u}
     }
     \caption{Effects of increasing environmental sensitivity $\chi_u$ for L-type individuals on average payoff, average population sizes, and average environmental quality across the spatial domain. 
     Figure \ref{fig:pi_chi_u} shows a higher sensitivity to environmental conditions $\chi_u$ leads to increased movement of L-type individuals toward resource-rich areas, resulting in a decline in average payoff for both L-type and H-type individuals. Figure \ref{fig:population_chi_u} shows the average of low-impact and high-impact harvesters as $\chi_u$ increases, with a noticeable decrease in L-type individuals and an increase in H-type individuals. Figure \ref{fig:environment_chi_u} depicts the corresponding decrease in the environmental metric $n$ as $\chi_u$ rises.  Simulation parameters include $D_u = 0.01$, $D_v = 0.01$, $D_n = 0.01$, and $\chi_v = 0.02$, with $\chi_u$ increasing by $0.001$ from $0.05$ to $0.1$. All other parameters are consistent with those used in  Figure \ref{fig:chi_u_m}. Each dot in the plots represents the average value of a given quantity averaged across fifty sets of randomized initial conditions drawn from the distribution described in Equation \eqref{eq:initialcondtionsrandom}.  %
     }
     \label{fig:uvnLH_chi_u}
\end{figure}

Now that we have seen that the strength of environmental-driven motion can impact the spatial distribution of payoffs of environmental quality, we look to understand the overall impact of this form of directed motion on the collective outcomes achieved by the population of harvesters and the environment across the spatial domain. In particular, we look to explore how the average population densities, payoff, and environmental quality change as we increase the $L$-strategist's strength of directed motion $\chi_u$, which we describe in Figure \ref{fig:uvnLH_chi_u}. To do this, we use numerical integration to evaluate the average values of these quantities of the spatial domain, calculating the average payoff of the two strategies as
\begin{subequations}
\begin{align}
\langle f_L \rangle(t) &= \frac{\int_0^{l} f_L\left(u(t,x),v(t,x) \right) u(t,x) dx}{\int_0^l u(t,x) dx} \\
\langle f_H \rangle(t) &= \frac{\int_0^{l} f_H\left(u(t,x),v(t,x) \right) v(t,x) dx}{\int_0^l v(t,x) dx}
\end{align}
\end{subequations}
and calculating the average densities of the two strategies and the environmental quality index as
\begin{subequations}
\begin{align}
\langle u \rangle(t) &= \frac{1}{l} \int_0^l u(t,x) dx \\
\langle v \rangle(t) &= \frac{1}{l} \int_0^l v(t,x) dx \\
\langle n \rangle(t) &= \frac{1}{l} \int_0^l n(t,x) dx.
\end{align}
\end{subequations}
We then compare these quantities achieved in the patterned states with the expected averages achieved in the uniform equilibrium state, for which $\langle f_L \rangle(t) = \langle f_H \rangle(t) = 0.4$, $\langle u \rangle(t) = \langle v \rangle = 2,000$, and $\langle n \rangle(t) = 0.5$ for the game-theoretic and ecological parameters we consider in the previous numerical simulations. 

In Figure \ref{fig:uvnLH_chi_u}, we compare the average population densities $\langle u \rangle(t)$ and $\langle v \rangle(t)$ of each strategy, calculating the average payoffs $\langle f_L \rangle(t)$ and $\langle f_H \rangle(t)$ achieved by each strategy, and the average level of environmental quality $\langle n \rangle(t)$. We calculate the averages of these quantities for a shared set of 50 initial conditions for each strength of environmental-driven motion $\chi_u$, and then average the spatial averages across our set of 50 simulations to determine the quantities we plot in Figure \ref{fig:uvnLH_chi_u}. In Figure \ref{fig:pi_chi_u}, we show that increasing the environmental sensitivity $\chi_u$ for L-type individuals results in a decreasing average payoffs for both $L$-type and $H$-type individuals across the spatial domain. While we have seen that the spatial pattern formation induced by environmental-driven motion can result in clusters with higher population density, greater environmental quality, and higher payoff for both strategies, we see that, overall, the presence of these spatial patterns actually results in a situation that decreases the collective outcome from the viewpoint of average payoff. However, the average payoff for low-effort harvesters is higher than the average payoff of the high-effort harvesters, so the greater ability to perform environmental driven-motion results in a better relative outcome for $L$-strategists than for $H$-strategists. %

We also consider the total population size of each strategy (Figure \ref{fig:uvnLH_chi_u}) and the average environmental quality (Figure \ref{fig:uvnLH_chi_u}c) achieved after many time-steps across the spatial domain. We see that the number of $L$-strategists decreases with an increase in their level of sensitivity for environmental gradients $\chi_u$, while the number of $H$-strategists increases with $\chi_u$. The average environmental quality also decreases with $\chi_u$, so the biased motion of sustainable harvesters towards areas of greater environmental quality can result in a long-time degradation of the overall environmental quality even though it can help produce localized pockets of abundant resource and payoff. Taken together, these results regarding how payoff, population density, and environmental quality respond to the strength of environmental-driven motion of low-effort harvester suggest that this environmental-driven directed motion can produce a spatial social dilemma, in which local benefits of environmental-driven motion can be individually and locally beneficial while producing a negative overall consequence in the long-run and across the whole spatial domain. 

\begin{figure}[!ht]
     \centering
     \subfloat[Maximal payoff variation across extended $\chi_u$ range]{
      \includegraphics[width=0.45\textwidth]{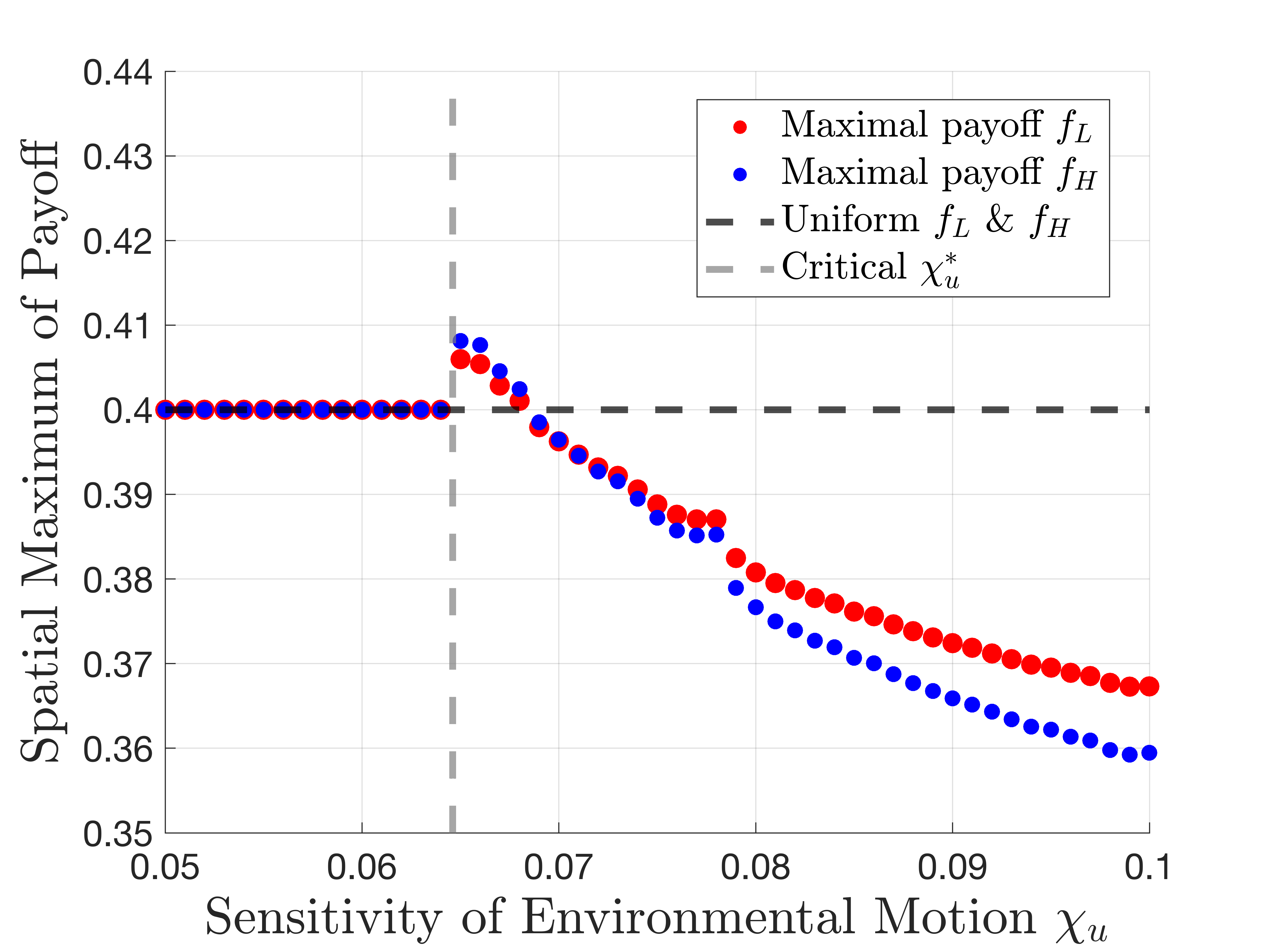}
      \label{fig:max_pi_chi_u}
     }
         \subfloat[Maximal population density dynamics as a function of $\chi_u$]{
      \includegraphics[width=0.45\textwidth]{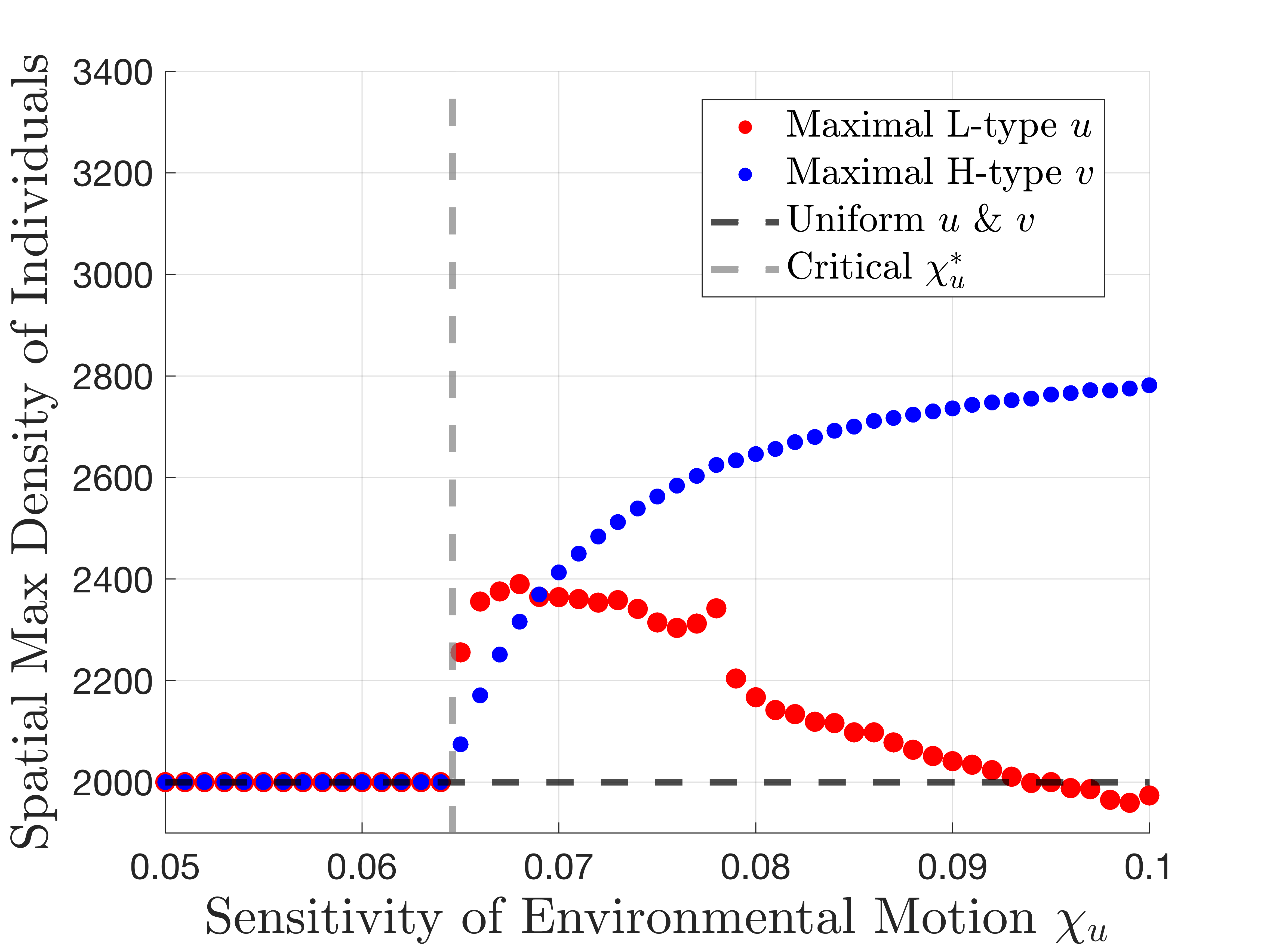}
      \label{fig:max_population_chi_u}
     }\hspace{5mm}
     \subfloat[Maximal environmental metric $n$ response to varying $\chi_u$]{
      \includegraphics[width=0.45\textwidth]{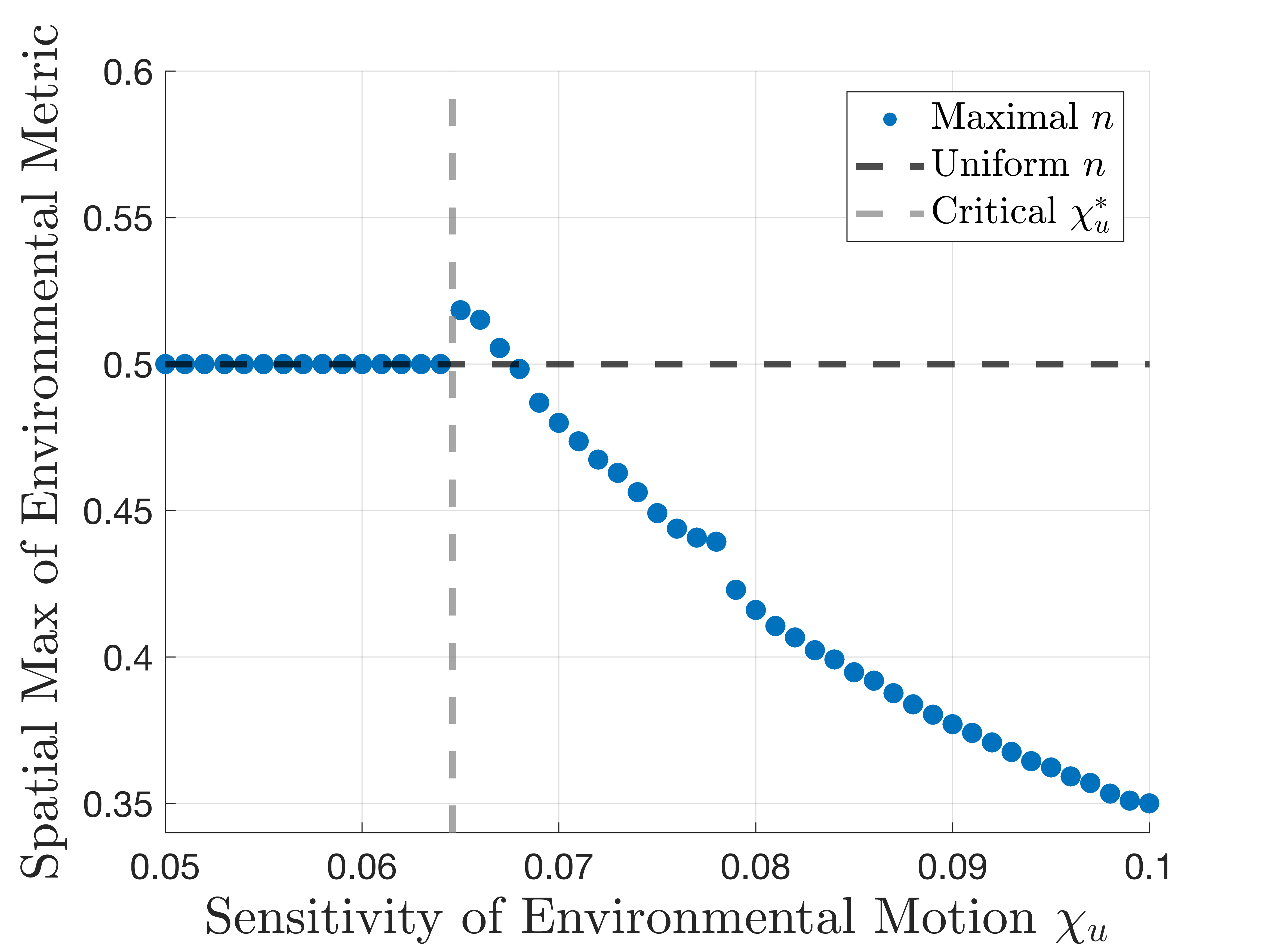}
      \label{fig:max_environment_chi_u}
     }\hspace{5mm}
     \subfloat[Maximal payoff behavior within narrow $\chi_u$ range]{
      \includegraphics[width=0.45\textwidth]{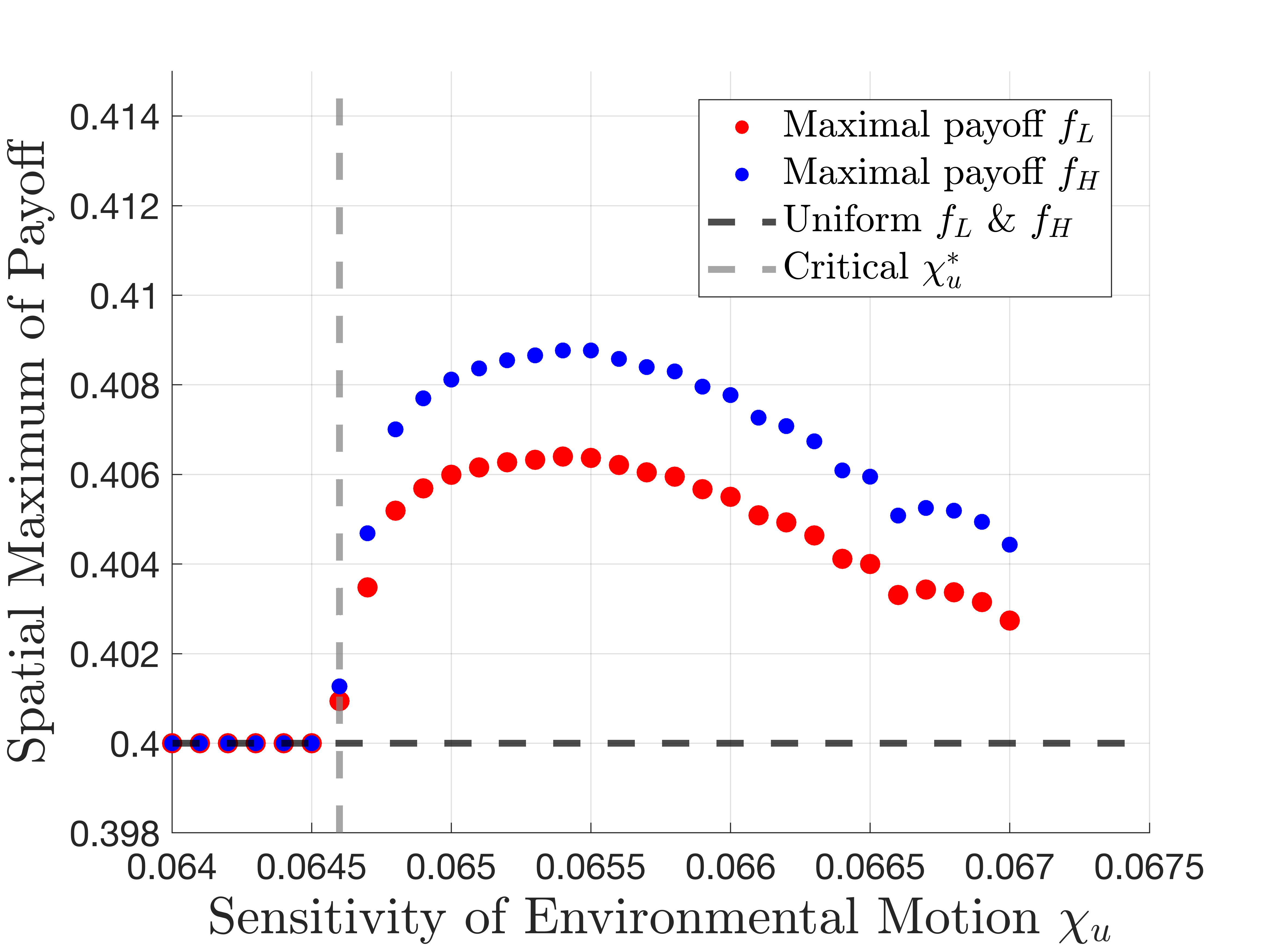}
      \label{fig:max_pi_chi_u_small}
     }
    \caption{Plots of the effects of the peak values across the spatial domain after many time-steps for payoff for each strategy (panel (a)), the population density of each strategy (panel (b)), and the environmental quality index (panel (c)), all plotted as a function of the strength $\chi_u$ of environmental-driven motion for the low-impact harvester. We also plot the maximum value achieved by the average payoff for each strategy across the spatial domain for a range of values of $\chi_u$ zoomed in upon the critical threshold $\chi_u^*$ required for pattern formation (panel (d)), which allows us to see that there appears to be a rapid, but continuous, increase in these peak quantities as $\chi_u$ increases past the threshold value. Each point plotted represents the average value of the given quantity achieved after time $t = 10^6$ starting from a set of 50 randomized initial conditions using the description described by Equation \eqref{eq:initialcondtionsrandom}.  All game-theoretic, ecological, and movement parameters and the range of $\chi_u$ variation are consistent with Figure \ref{fig:uvnLH_chi_u}. %
    }
     \label{fig:max_uvnLH_chi_u}
\end{figure}

In addition to studying the overall collective outcome achieved in spatially patterned states, we can try to explore the maximal values of population densities of each strategy, the average payoff of each strategy, and the environmental quality achieved in our spatially patterned states. In Figure \ref{fig:max_pi_chi_u}, we plot the average maximum value of each of these quantities achieved across simulations with our set of 50 initial conditions. We see that, as $\chi_u$ increases above the pattern-forming threshold $\chi_u^*$, the maximum payoffs initially experience a rapid increase before decreases to peak payoffs that are below the average payoffs achieved in a spatially uniform equilibrium state. In particular, we notice that the maximal payoff achieved by low-effort harvesters exceeds the maximum payoff achieved by low-effort harvesters when the peak payoffs exceed the payoff of the uniform state, meaning that increased strength of directed motion $\chi_u$ for low-effort harvesters can allow for some low-effort harvesters to achieve better payoffs than any high-effort harvesters in the population. We also explore the maximum payoffs achieved by the two strategies in Figure \ref{fig:max_pi_chi_u_small} for the case of strength of directed motion $\chi_u$ close to the threshold $\chi_u^*$, which suggests that this peak payoff rapidly increases right above the threshold for pattern formation, but that this rapid increase potentially occurs in a continuous manner from the payoff achieved in the uniform state. 

In Figure \ref{fig:max_population_chi_u}, we show the maximum population densities achieved by the two strategies as we increase the strength of environmental-driven motion. We see that the maximal density of each strategy initially increases, with the low-effort harvesters achieving higher peaks than high-effort harvesters when $\chi_u$ is relatively close to the pattern-forming threshold. However, the peak densities of high-effort harvesters continue to increase with $\chi_u$, while the peak abundance of low-effort harvesters eventually decreases and passes below the density of low-effort harvesters in the spatially uniform equilibrium. In a similar manner, we show in Figure \ref{fig:max_environment_chi_u} that the peak level of the environmental quality index first increases with $\chi_u$ close to the pattern-forming threshold, but that the peak environmental eventually begins to decrease with $\chi_u$ and falls below the quality of the spatially uniform equilibrium for sufficiently large $\chi_u$. This reveals a particular strength of the spatial social dilemma we observe, as the directed spatial motion of low-impact harvesters towards regions of improved environmental quality can actually result in degraded environmental quality at all spatial locations within our domain.

\subsection{Time-Dependent Behavior of Emerging Patterns}
\label{sec:numericaltransient}

So far, we have primarily focused on the time-dependent behavior of numerical solutions to our model. In this section, we examine the time-dependent behavior of patterned states within our system, seeing how the aggregates of low-effort harvesters and resulting regions of higher-quality environments change in time. While we saw heatmaps of the time-dependent solution for $\chi_u$ near the pattern formation threshold resulted in stationary patterns with a wavenumber predictable from linear stability analysis, we are curious to explore whether more complicated transient dynamics are possible when low-effort harvesters experience stronger environmental-driven motion and there are many possible unstable wavenumbers that could potentially characterize the form of emergent spatial patterns. This approach allows us to explore the transient dynamics of pattern formation, showing coarsening of the number aggregates of population and environmental quality over time. 

\begin{figure}[!ht]
    \centering
   
    \subfloat[Heatmap of $u$ illustrating merging behavior]{
        \includegraphics[width=0.45\textwidth]{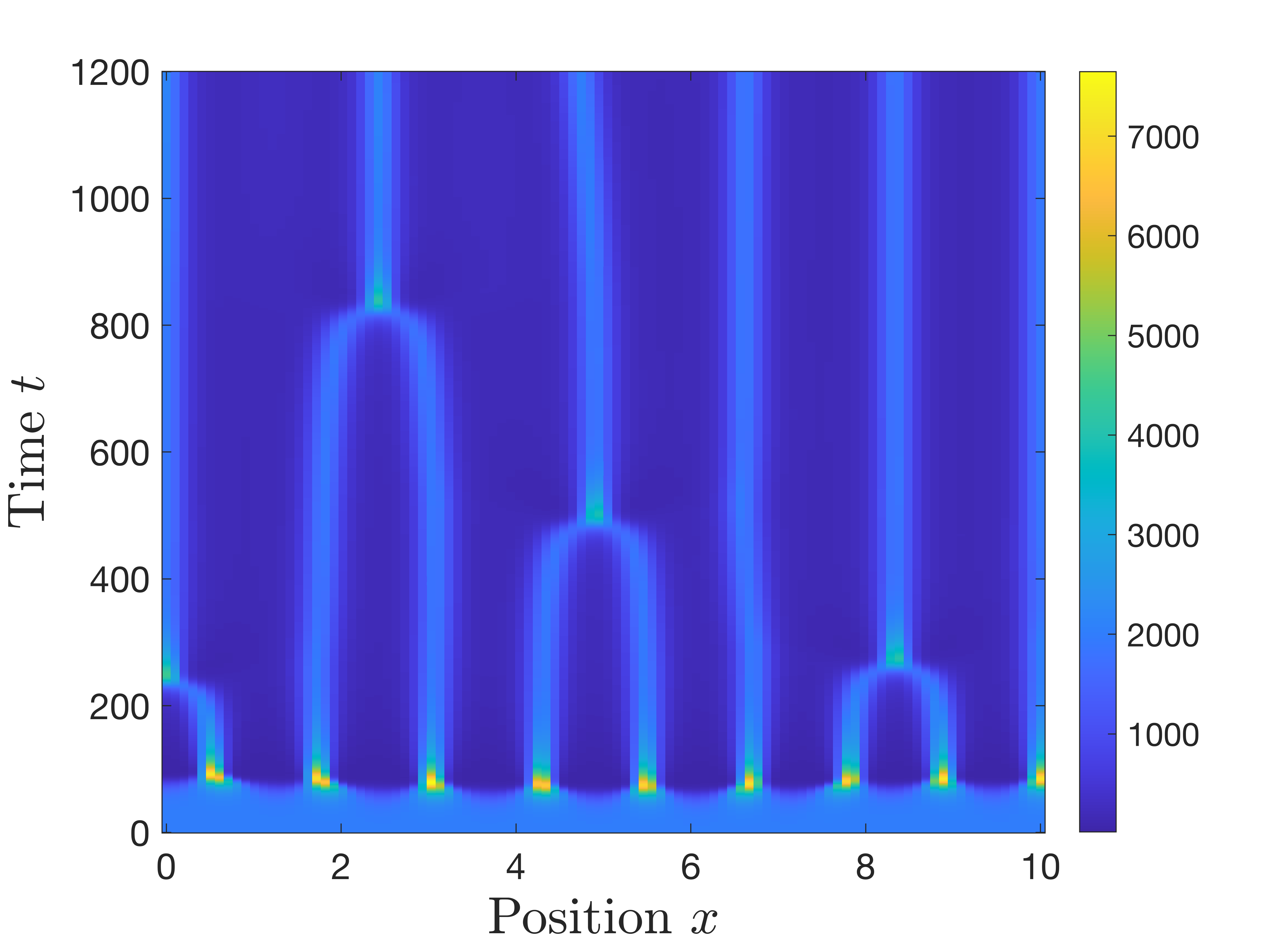}
        \label{fig:meta_heatmap}
    }
         \subfloat[Average payoff oscillations over time]{
        \includegraphics[width=0.45\textwidth]{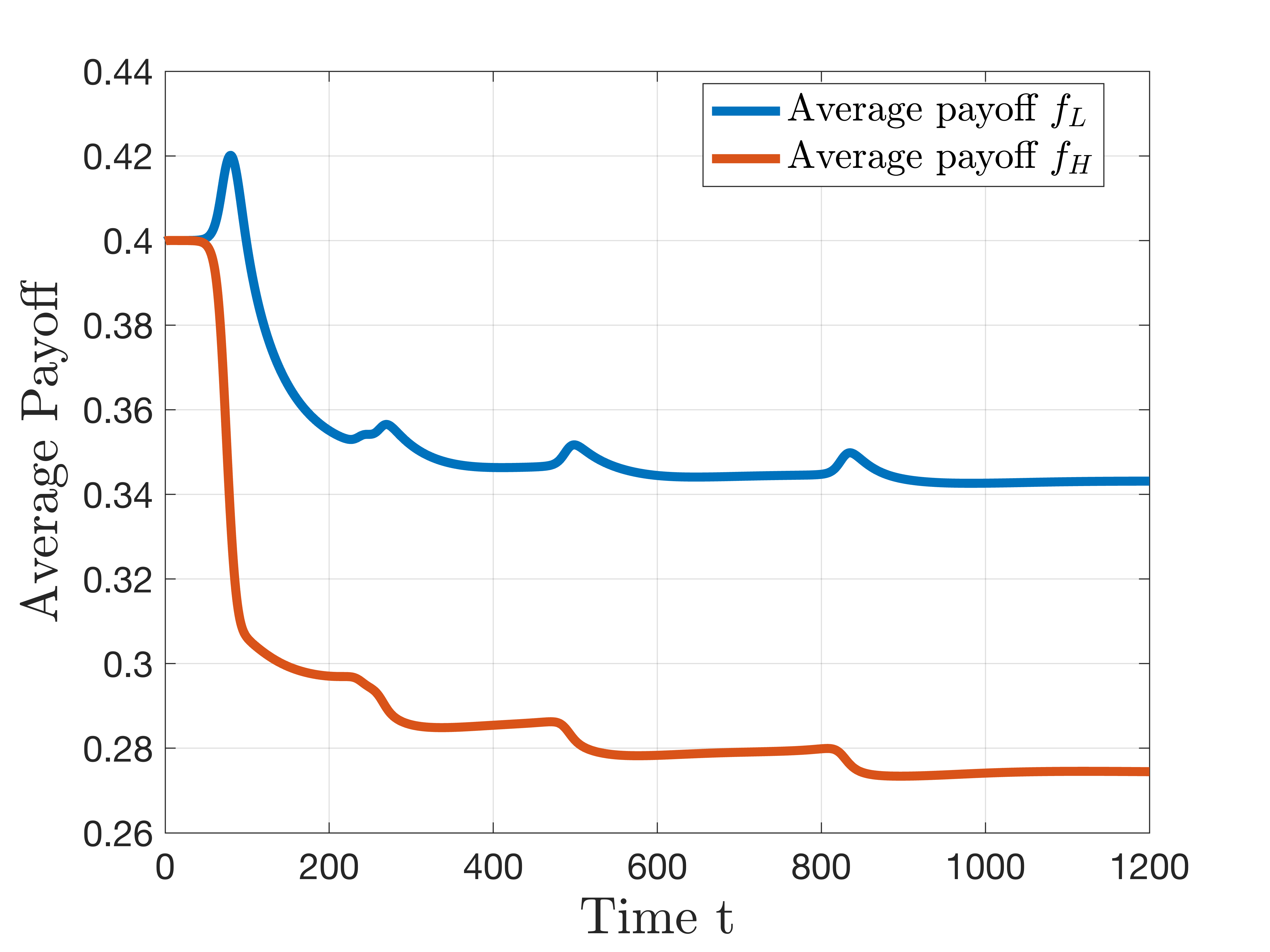}
        \label{fig:meta_pi}
    }

   \caption{Illustration of the time-dependent behavior solutions to the PDE model. Panel (a) shows the heatmap of population $u$ merges into advantageous environments during oscillation onset, with corresponding impacts on payoffs. Panel (b) presents the average payoff of the population as a function of time. The sensitivities for environment-driven motion are $\chi_u = 0.1$ and $\chi_v = 0.02$. All other parameters and boundary conditions are identical to those in Figure \ref{fig:chi_u_m}.}
    \label{fig:meta}
\end{figure}

In Figure \ref{fig:meta_heatmap}, we plot heatmaps of the density $u(t,x)$ of low-effort harvesters as a function of space and time, showing that, after an initial phase when the population is distributed near the uniform state, about nine peaks form around time $t = 100$. As time progresses, we see quick windows of time in which the pairs of clusters approach one another and merge to form a larger cluster, resulting in a reduction to six noticeable clusters over the course of our numerical simulation. As the locations of spatial aggregates seem to move relatively slowly away from these merging events, the dynamical behavior bears some resemblance to metastable dynamics seen in the formation and maintenance of spatial patterns for PDE models with chemotactic behavior \cite{potapov2005metastability,tse2016hotspot}. This coarsening of the spatial pattern over time raises questions about the long-time behavior of the clusters, suggesting future exploration of whether the patterned dynamics for $\chi_u$ far above $\chi_u^*$ will ultimately result in convergence upon a single spike of greater population size and increased environmental quality, or whether two or more clusters can stably coexist in the long-time limit.  

We can also explore the underlying behavior related to the merging dynamics of clusters by considering the temporal dynamics of the mean payoff received by individuals across the spatial domain. In
Figure \ref{fig:meta_heatmap}, we plot the average payoff $f_H(u(t,x),v(t,x))$ and $f_L(u(t,x),v(t,x))$ of low-effort and high-effort harvesters over the spatial domain as a function of time, showing that the low-effort harvesters experience an initial increase in average payoff as they move to form the initial patterns, but that both low-effort and high-effort harvesters experience a rapid decrease in average payoff around the time at which spatial patterns become apparent in the heatmap from Figure \ref{fig:meta_heatmap} (at approximately $t = 100$). We then see that that average payoff of low-effort harvesters features several regions of quick increase around the points in time at which we see regions of greater environmental quality merge in Figure \ref{fig:meta_heatmap} (at approximate times $ t = 310 $, $ t = 630 $, and $ t = 800 $), while the payoffs of both strategies then quickly decline after than point before settling into a period of slow change in payoff until the next event in which clusters merge. This increase in average payoff seen by the low-effort harvesters in the process of merging clusters may be due to the fact that individuals near the boundaries of each cluster may see improved environmental quality nearby as two clusters approach, resulting in greater payoffs and a greater tendency to move towards the approaching cluster with directed motion towards increasing environmental quality. 

\section{Discussion}
\label{sec:discussion}

In this paper, we extended the framework of evolutionary games with environmental feedback to include spatial strategic and resource dynamics, exploring the effects diffusion of environmental quality and directed motion of individuals towards regions of increasing environmental quality. We formulated a PDE model of eco-evolutionary games with environmental-driven motion modeled as a chemotaxis-type term, and used a mix of linear stability analysis and numerical simulations to explore how environmental-driven motion can produce regions featuring clusters of lower-impact resource extraction and high levels of environmental quality. We further explored qualitative properties of the emergent patterns, seeing how the average payoff and average levels of population is impacted by these rules for spatial motion. We found that spatial pattern formation can occur when more sustainable low-effort harvesters are more effective at climbing gradients of environmental quality than high-effort harvesters who also deplete the environment, but this ability for more sustainable individuals to create pockets of higher-quality environments may not actually improve the overall collective outcome of the population in terms of payoff or environmental quality. 

Notably, it appears that directed motion towards increasing levels of environmental quality actually decreases the long-run average payoff of the population and the overall environmental quality in the spatial domain relative to the average payoffs and environment achieved in the absence of spatial movement. However, close to the pattern-forming threshold, we saw that the individuals who were able to successfully migrate to locations of higher-quality environment in the patterned state were able to achieve higher personal payoffs than would be achieved in the uniform state, with individuals remaining in lower-quality environments receiving lower payoffs than in the uniform equilibrium. This could potentially be interpreted as an additional social dilemma in which the collective payoff of the population decreases even though the rules for spatial movement aim to increase individual payoff in the short run. In addition, an even more extreme scenario was seen when the strength $\chi_u$ of environmental-driven motion was further above the pattern-forming threshold, in which the collective payoff and environmental quality in the patterned state achieved lower values than the corresponding values in the uniform equilibrium state at all points in the spatial domain. These results suggest that environmental-driven motion of sustainably-minded individuals may actually harm the ability to produce sustainable environments and abundant payoffs for a population of harvesters.

As we saw that directed motion towards increasing concentrations of environmental resources tended to decrease the long-time collective payoff of the population, it may be of interest to explore how such spatial movement rules interact with models for mechanisms that help to induce the sustainable management of common-pool resources \cite{schluter2016robustness,tavoni2012survival,tilman2017maintaining,tilman2018revenue,tilman2023public}. Recent work has explored the role of cultural evolution of collective management of resources \cite{andrews2024cultural} and a model of localized prosocial preferences have shown the ability to sustain the provision of common-pool resources, so it could also be important to explore how spatially localized mechanisms for cooperative management can interact with spatial diffusion of resources or directed motion towards more sustainable locations. In general, our observation of spatial social and environmental inequality motivates further work in understanding the role of spatial and temporal factors in the sustainable management of shared resources \cite{levin2000multiple}.

One feature of existing PDE models for evolutionary games with payoff-driven motion is that directed motion towards greater feedback features a negative diffusion term corresponding to a self-taxis behavior with a population climbing the gradient of its own local population density. This behavior results in a short-wave instability in which there are infinitely many unstable wavenumbers when payoff-driven motion is sufficiently strong \cite{helbing2009pattern}, which raises questions of what characteristic pattern size can be achieved when there is no finite dominant wavenumber for the PDE's linearization around the spatially uniform equilibrium. Because directed motion in our model is based on individuals climbing the gradient of the environmental quality variable, the diffusion of environmental quality loosens the feedback for directed motion and allows for the achievement of a finite wavenumber that maximizes the growth rate of spatial patterns. For this reason, our model may be more amenable to further analytical study than models of payoff-driven motion in evolutionary games, and therefore this model of spatial evolutionary dynamics with environmental feedback may also serve as a useful model problem for the dynamics of cooperation in continuous space. 

While the analysis in this paper focused on game-theoretic payoffs featuring coexistence of the low-impact and high-impact strategists in the nonspatial strategic dynamics, prior work on eco-evolutionary games has shown the possibility of more complex nonspatial dynamics featuring oscillatory long-time behavior \cite{tilman2020evolutionary,weitz2016oscillating}. An interesting question for future research would be to explore how the incorporation of spatial movement rules including diffusion and environmentally-driven directed motion would impact the eco-evolutionary dynamics for those payoff scenarios. Prior work on reaction-diffusion models for ecological public goods games have demonstrated behaviors including diffusion-driven coexistence of strategies and spatiotemporal chaos for cases in which the respective nonspatial replicator equations predicted exclusion of a cooperative strategy and long-time cycling of the cooperator and defector strategies \cite{wakano2009spatial,wakano2011pattern}, so a natural question is whether similar behaviors could be seen when incorporating spatial diffusion for evolutionary games with environmental feedback. Combined with our exploration of environment-driven motion in this paper, these questions of complex dynamics of spatial games with environmental feedback provide an abundant setting for studying spatial eco-evolutionary dynamics and PDE models for the sustainable management of natural resources. 

Even for our setting of studying a spatial instability of a uniform coexistence equilibrium of low-effort and high-effort harvesters, there are many more questions that can be explored regarding the dynamics of pattern formation. While our analysis focused on the case of pattern formation in one spatial dimension, it would be interesting to explore the emergence and classification of spatial patterns in two dimensions or to account for a heterogeneous spatial domain based on differing intrinsic resource levels or levels of environmental quality. The transient behavior featuring merging and splitting of spatial clusters of harvesters and environment also raises question about predicting the long-time number of peaks in the emergent spatial patterns at strengths of biased motion far from the threshold value. In particular, substantial work has explored the number and stability of coexisting spike or plateau states in the Keller-Segel model and related PDE models with directed spatial motion \cite{kang2007stability,kong2024existence,tse2016hotspot,buttenschoen2020cops}, and it could be interesting to explore whether a similar analysis could reveal possible long-time behavior in our model for eco-evolutionary games.

The formation of spatial patterns results in the self-organized creation of regions with greater environmental quality, a greater fraction of low-effort harvesters, and higher levels of payoff. This bottom-up creation of regions with sustainable extraction bears some similarity to game-theoretic models that show that competition over fishing territory can result in the emergence of marine protected areas as a Nash equilibrium between rival fishers on the high seas \cite{herrera2016high} or as the result of an optimal control policy for maximizing the yield of a renewable resource that can diffuse in space \cite{upmann2021optimal}. The question of self-organized and top-down creation of conservation areas has also been explored in the context of conversation efforts in land use policies \cite{albers2008patterns}, and therefore it may an interesting direction for future research to understand how models of environmental-driven individual behavior interact with questions of central planning and institutional management of shared resources.

\renewcommand{\abstractname}{Acknowledgments}
\begin{abstract}
Tianyong Yao received a Research Support Grant from the Office of Undergraduate Research at the University of Illinois Urbana-Champaign.
\end{abstract}

\renewcommand{\abstractname}{Statement on Code Availability}
\begin{abstract}
\sloppy{All code used to generate figures is archived on GitHub (see \href{https://github.com/YuzuruTesla/Spatial_Evolutionary_Games_with_Environmental_Feedback}{GitHub Repository}) and licensed for reuse, with appropriate attribution/citation,
under a BSD 3-Clause Revised License. This repository contains the Matlab code used to run numerical simulations of the PDE model and to generate the figures in the paper, as well as two Mathematica notebooks used to find and present analytical expressions for the conditions required for pattern formation for the case of unequal diffusion coefficients (as discussed in Sections \ref{sec:LSAequalDuDv} and \ref{sec:app_LSA_unequal}).}
\end{abstract}

\bibliographystyle{unsrt}
\bibliography{references}

\appendix
\appendixpage
\addtocontents{toc}{\protect\setcounter{tocdepth}{1}}

\section{Derivation of the PDE Model from an Individual-Based Model with Biased Random Walk Influenced by Environmental Quality}

To provide intuition on the biological interpretation of the relative values of the diffusivities $D_u$, $D_v$, and $D_n$ and the parameters $\chi_u$ and $\chi_v$ describing the strength of environmental-driven motion, we describe how it is possible to derive our PDE model from Equation \eqref{eq:PDEsystemuvn} from a discrete-time, discrete-space model for the eco-evolutionary game with spatial motion of harvesters characterized by a biased random walk towards neighboring lattice sites with featuring greater levels of environmental quality.

In this section, we present a way to derive our PDE model for the spatial dynamics of an evolutionary game with environmental feedback and environmental-driven motion from a biased random walk on a one-dimension grid. Our derivation will follow the approach used by Alsenafi and Barbaro \cite{alsenafi2018convection,alsenafi2021multispecies}, who studied the dynamics of two populations performing biased random walks that were species to quantities that served as separate signaling species for each population. Similar approaches have been used previously to derive PDE models with a logarithmic chemotaxis term based on slightly different forms for the probabilities used in biased random walks \cite{short2008statistical,zipkin2014cops}, and we chose the approach of Alsenafi and Barbaro to obtain the linear form of the sensitivity to environmental-driven motion for our model. 

\subsection{Comparison of Discrete and Continuum Models for Eco-Evolutionary Games with Environmental-Driven Directed Motion}

Before presenting the full derivation of our PDE model, we use this Section to provide a comparison between the PDE and the discrete model we use to describe the motivation of model in terms of replication rates and individual-based rules for the spatial movement of harvesters and environmental resources. We first recall that our PDE model from Equation \eqref{eq:PDEsystemuvn} describes the change in the densities $u(t,x)$ and $v(t,x)$ of low-impact and high-aspect harvests and the profile of environmental quality $n(t,x)$ over our spatial domain, and is given by
\begin{subequations}
\begin{align}
\frac{\partial u(t,x)}{\partial t} &= D_u \frac{\partial^2 u(t,x)}{\partial x^2} - \chi_u \frac{\partial}{\partial x} \left( u(t,x) \frac{\partial n(t,x)}{\partial x} \right)+ \epsilon^{-1} u \left( f_L(u,v,n) - \kappa(u+v) \right), \\
\frac{\partial v(t,x)}{\partial t} &= D_v \frac{\partial^2 v(t,x)}{\partial x^2} - \chi_v \frac{\partial}{\partial x} \left( v(t,x) \frac{\partial n(t,x)}{\partial x} \right) + \epsilon^{-1} u \left( f_H(u,v,n) - \kappa(u+v) \right), \\
\frac{\partial n(t,x)}{\partial t} &= D_n \frac{\partial^2 n(t,x)}{\partial x^2} + \left(r - a(e_L n + e_H (1-n))\right) \left( \frac{u}{u+v} - n \right).
\end{align}
\end{subequations}

We will now describe a discrete-time, discrete-state stochastic process that can be used to derive this PDE model, as will be shown in Section \ref{sec:continuumderivation}. We consider a population of individuals with locations described by a discrete grid with step size $\Delta x$ between grid points, and we assume that birth, death, and spatial movements of individuals occur in discrete time with a time step of $\Delta t$. Following the approach of Durrett and Levin \cite{durrett1994importance}, we assume that the net per-capita birth rate of low-impact and high-impact harvesters at location $x$ and time $t$ are given by
\begin{subequations}
\begin{align}
\rho_u \left( t,x \right) &= f_L\left(u(t,x), v(t,x), n(t,x) \right) - \kappa \left( u(t,x) + v(t,x) \right)  \\
\rho_v\left(t,x\right) &= f_H\left(u(t,x), v(t,x), n(t,x) \right) - \kappa \left( u(t,x) + v(t,x) \right)
\end{align}
and that the environmental quality index $n(t,x)$ at location $x$ and time $t$ changes with rate 
\begin{equation}
\rho_n(t,x) =  \left\{ r - a(e_L n(t,x) + e_H (1 - n(t,x)))\right\} \left(\frac{u(t,x)}{u(t,x) + v(t,x)} - n(t,x)\right).
\end{equation}
\end{subequations}
As the environmental quality index was used by Tilman and coauthors to describe a normalized measure of the quantity of an environmental resource \cite{tilman2020evolutionary}, we can interpret $\rho_n(t,x)$ a birth/death or proliferation/destruction event for such a resource. 

During the time step between times $t$ and $t + \Delta t$ in our discrete-time model, we first update the densities of harvesters and the level of environmental quality at each grid point, and then we allow both harvesters and the environmental variable to possibly move to a neighboring grid-point, with environmental quality performing an unbiased walk and both low-impact and high-impact harvesters performing biased random walk based on the environmental quality of neighboring sites on the lattice.  

We assume that low-impact harvesters individuals at location $x$ after the reproduction or death events between time $t$ and $t + \Delta t$ will move to a neighboring site with probability $\gamma_v$ at time $t+\Delta t$, and then they choose to move to a neighboring site at location $x \pm \Delta x$ with probability 
\begin{subequations} 
\begin{equation} \label{eq:discrete_prob_u}
M_u\left(t,x \to x \pm \Delta x \right) = \frac{e^{\beta_u n(t,x \pm \Delta x)}}{e^{\beta_u n(t,x + \Delta x)} + e^{\beta_u n(t,x -\Delta x)}},
\end{equation}
where the parameter $\beta_u$ describes the sensitivity of directed motion for low-impact harvesters in response to gradients in environmental quality. We similarly assume that individuals at location $x$ after the demographic events between time $t$ and $t+\Delta t$ will move to a neighboring site with probability $\gamma_v$, and then they choose to move to a neighboring site at location $x \pm \Delta x$ with probability 
\begin{equation} \label{eq:discrete_prob_v}
M_v\left(t,x \to x \pm \Delta x\right) = \frac{e^{\beta_v n(t,x \pm \Delta x)}}{e^{\beta_v n(t,x + \Delta x)} + e^{\beta_v n(t,x -\Delta x)}}.
\end{equation}
\end{subequations}
Finally, we assume that the a portion of resource described by the environmental quality metric moves with probability $\gamma_n$ in a given time-step, choosing between the two neighboring lattice sites with equal probability.

While we present the full derivation of the PDE model from the stochastic process model in Section \ref{sec:continuumderivation}, we will now comment on the connection between the parameters describe spatial movement in the two models. The derivation of the PDE model is derived by simultaneously taking the limit as $\Delta t \to 0$ and $\Delta x \to 0$, resulting in a PDE description of our system that changes in continuous time and continuous space. In order to obtain the appropriate spatial partial derivatives on the righthand side, we show that we impose the following scaling relationship between $\Delta t$ and $\Delta x$ as each quantity tends to $0$:
\begin{equation}
\lim_{\substack{\Delta x \to 0 \\ \Delta t \to 0}} \frac{\left( \Delta x \right)^2}{\Delta t} = \alpha
\end{equation}
for a nonzero constant $\alpha$. Under this assumption, we can define the diffusivities in our PDE model in terms of the movement probabilities of the stochastic model as
\begin{subequations}
\begin{equation}
D_u := \frac{1}{2} \alpha \gamma_u \: \: , \: \: D_v := \frac{1}{2} \alpha \gamma_v \: \:, \: \: D_n = \frac{1}{2} \alpha \gamma_n,
\end{equation}
while we define the parameters $\chi_u$ and $\chi_v$ characterizing the strength of environmental-driven motion in terms of both the movement probabilities and environmental sensitivity parameters as 
\begin{equation}
\chi_u := \alpha \gamma_u \beta_u \: \: , \: \: \chi_v := \alpha \gamma_v \beta_v.
\end{equation}
\end{subequations}

For completeness, we summarize the relationship between the parameters of the stochastic and PDE models in Table \ref{table:parameter_relationships}. Notably, the expressions used for the PDE parameters tell us that both the diffusion coefficients $D_u$ and $D_v$ and the strengths of environmental-driven motion $\chi_u$ and $\chi_v$ depend on the respective movement probabilities $\gamma_u$ and $\gamma_v$, but only the parameters $\chi_u$ and $\chi_v$ depend on the sensitivity parameters describing $\beta_u$ and $\beta_v$ describing the bias in directed motions towards climbing gradients of environmental quality. We can therefore model a scenario in which both types of harvesters are equally like to move but low-effort harvesters are more effective than high-effort harvesters in climbing environmental gradients by setting $D_u = D_v$ (corresponding to $\gamma_u = \gamma_v$) and by considering $\chi_u > \chi_v$ (corresponding to $\beta_u > \beta_v$). 

\begin{table}[!ht]
\centering
\begin{tabular}{|c|c|c|c|c|c|}
\hline
  & $\gamma_u$ & $\gamma_v$ & $\gamma_n$ & $\beta_u$ & $\beta_v$ \\ \hline
$D_u$ & \makecell{Directly \\ Proportional} & Independent & Independent & Independent & Independent \\ \hline
$D_v$ & Independent & \makecell{Directly \\ Proportional}& Independent & Independent & Independent\\ \hline
$D_n$ & Independent & Independent & \makecell{Directly \\ Proportional}& Independent & Independent \\ \hline
$\chi_u$  & \makecell{Directly \\ Proportional} & Independent & Independent & \makecell{Directly \\ Proportional}& Independent \\ \hline
$\chi_v$  & Independent & \makecell{Directly \\ Proportional}& Independent & Independent & \makecell{Directly \\ Proportional}\\ \hline
\end{tabular}
\caption{Summary of dependence the diffusion coefficients ($D_u$, $D_v$, and $D_n$) and the strengths of environmenal-driven motion ($\chi_u$ and $\chi_v$) of the PDE model upon the movement probability ($\gamma_u$, $\gamma_v$, and $\gamma_n$) and sensitivity parameter for environmental-driven biased random walk ($\beta_u$ and $\beta_n$) of the stochastic model.}
\label{table:parameter_relationships}
\end{table}

\subsection{Deriving the PDE Model as the Continuum Limit of Biased Random-Walk Model}
\label{sec:continuumderivation}

We assume that the update of the strategic composition of the populations and the distribution of resources describing environmental quality occur in two stages during the time-step between times $t$ and $t + \Delta t$. We first assume that low-impact and high-impact harvesters reproduce or die according to at their respective per-capita reproduction rate $\rho_L(u,v,n)$ and $\rho_H(u,v,n)$, while the level of resource changes at rate $\rho_n(u,v,n)$. After reproduction and death have occurred, we assume that individuals move according to a biased random walk with probability $\gamma_u$ or $\gamma_v$, with the movement probabilities following the biased random walk according to the movement rule from Equations \eqref{eq:discrete_prob_u} and \eqref{eq:discrete_prob_v} for the low-impact and high-impact harvesters, respectively. Similarly, we assume that the resources describing environmental quality move in a given time-step with probability $\gamma_n$, moving with equal probability to the locations to the left and right.

Using these assumptions about the spatial dynamics of strategies and environmental quality, we can begin to derive our PDE model for spatial games with environmental feedback by describing the limiting behavior of our discrete model in the limit of infinitesimal time-step $\Delta t$ and infinitesimal spatial movement $\Delta x$. We start by deriving the PDE describing the change in time for the density $u(t,x)$ of the low-impact harvesters.

We can use the rates defined for reproduction and spatial movement of low-impact harvesters to find the following expression for the expected density $u\left(t+\Delta t,x\right)$ of low-impact harvesters at location $x$ at time $t + \Delta t$:
\begin{equation}
\begin{aligned}
u\left(t + \Delta t, x \right) &=  u(t,x) \left[ 1 + \epsilon^{-1} \left\{f_L\left(u(t,x),v(t,x),n(t,x) \right) - \kappa \left[u(t,x) + v(t,x) \right] \right\} \Delta t \right] \left( 1 - \gamma_u \right) \\
&+ \gamma_u \sum_{\tilde{x} \sim x} u\left(t,\tilde{x} \right) \left[ 1 + \epsilon^{-1} f_L \left( u(t,\tilde{x}), v(t,\tilde{x}), n(t,\tilde{x})\right) - \kappa \left[u(t,\tilde{x}) + v(t,\tilde{x}) \right] \right] M_u\left(t, \tilde{x} \to x \right).
\end{aligned}
\end{equation}
We can then expand this equation and use the expression for the movement probability from Equation \eqref{eq:discrete_prob_u} to write the expected change in the density of low-impact harvesters at location $x$ as
\begin{equation}
\begin{aligned}
& u\left(t + \Delta t, x \right) - u(t,x) \\ &= \epsilon^{-1} u(t,x) \left\{f_L\left(u(t,x),v(t,x),n(t,x) \right) - \kappa \left[u(t,x) + v(t,x) \right] \right\} \Delta t  \\
&+ \gamma_u  \sum_{\tilde{x} \sim x} \left\{ \frac{e^{\beta_u n(t,x)}}{\sum_{\tilde{\tilde{x}} \sim \tilde{x}} e^{\beta_u n(t,\tilde{\tilde{x}})}} u(t,\tilde{x}) - \frac{1}{2} u(t,x) \right\} \\
&+ \gamma_u \sum_{\tilde{x} \sim x} \left\{ \epsilon^{-1} \left( f_L\left(u(t,\tilde{x}),v(t,\tilde{x}),n(t,\tilde{x}) \right) - \kappa \left[u(t,\tilde{x}) + v(t,\tilde{x}) \right] \right) \Delta t  \frac{e^{\beta_u n(t,x)}}{\sum_{\tilde{\tilde{x}} \sim \tilde{x}} e^{\beta_u n(t,\tilde{\tilde{x}})}} u(t,\tilde{x}) \right\} \\ &- \gamma_u u(t,x) \epsilon^{-1} \left( f_L\left(u(t,x),v(t,x),n(t,x) \right) - \kappa \left[u(t,x) + v(t,x) \right]  \right) \left(\Delta t \right) u(t,x) 
\end{aligned}
\end{equation}
Next, we can expand the term for the net birth rate $f_L\left( u(t,\tilde{x}),v(t,\tilde{x}),n(t,\tilde{x}) \right) - \kappa \left[ u(t,\tilde{x}) + v(t,\tilde{x}) \right]$ around $x$ for the cases of $\tilde{x} = x + \Delta x$ and $\tilde{x} = x - \Delta x$ to further write that
\begin{equation}
\begin{aligned} \label{eq:udiffexpanded}
& \frac{u\left(t + \Delta t, x \right) - u(t,x)}{\Delta t} \\ &= \epsilon^{-1} \left\{f_L\left(u(t,x),v(t,x),n(t,x) \right) - \kappa \left[u(t,x) + v(t,x) \right] \right\}   \\
&+ \frac{\gamma_u}{\Delta t} \left[1 + \epsilon^{-1} \left\{f_L \left(u,v,n \right) - \kappa \left[ u + v\right] \right\} \left( \Delta t \right)  \right] \sum_{\tilde{x} \sim x} \left\{ \frac{e^{\beta_u n(t,x)}}{\sum_{\tilde{\tilde{x}} \sim \tilde{x}} e^{\beta_u n(t,\tilde{\tilde{x}})}} u(t,\tilde{x}) - \frac{1}{2} u(t,x) \right\} \\
&+ \left(\Delta x \right) \gamma_u \epsilon^{-1} \left[ \dsdel{f_L}{u} \dsdel{u}{x} + \dsdel{f_L}{v} \dsdel{v}{x} + \dsdel{f_L}{v} \dsdel{v}{x} - \kappa \left( \dsdel{u}{x} + \dsdel{v}{x} \right) +  \mc{O}\left(\Delta x\right) \right] \left[ \frac{e^{\beta_u n(t,x)}}{e^{\beta_u n(t,x)} + e^{\beta_u n(t,x + 2 \Delta x)}}  \right] \\
&- \left(\Delta x \right) \gamma_u \epsilon^{-1} \left[ \dsdel{f_L}{u} \dsdel{u}{x} + \dsdel{f_L}{v} \dsdel{v}{x} + \dsdel{f_L}{v} \dsdel{v}{x} - \kappa \left( \dsdel{u}{x} + \dsdel{v}{x} \right) +  \mc{O}\left(\Delta x\right) \right] \left[ \frac{e^{\beta_u n(t,x)}}{e^{\beta_u n(t,x)} + e^{\beta_u n(t,x - 2 \Delta x)}}  \right]%
\end{aligned}
\end{equation}
We see that the first line of the righthand side corresponds to the desired reaction term in our PDE model, and that the third and fourth line will tend to zero in the limit as $\Delta x \to 0$. To understand the behavior of the continuum limit as $\Delta x \to 0$ and $\Delta t \to 0$, we need to further understand the behavior of the second line of the righthand side, which describes the leading-order effects of the role of the biased random walk taken by low-impact harvesters towards regions of greater environmental quality. 

We show in Section \ref{sec:dispersalderivation} that the sum in the second line on the righthand side of Equation \eqref{eq:udiffexpanded} can be expanded as
\begin{equation} \label{eq:diffusiontermexpanded}
\sum_{\tilde{x} \sim x} \left( \left[ \frac{e^{\beta_u n(t,x)}}{\sum_{\tilde{\tilde{x}}\sim \tilde{x}} e^{\beta_un(t,\tilde{\tilde{x}})}} u(t,\tilde{x}) \right] - \frac{1}{2} u(t,x) \right) = \left( \Delta x \right)^2   \left[ \doubledelsame{u}{x} - 2 \beta_u \dsdel{}{x} \left(  \dsdel{n}{x} u \right) \right] + \mc{O}\left( \left( \Delta x \right)^4 \right).
\end{equation}
The derivation of this expansion follows the approach used by Alsenafi and Barbaro of a model for directed motion from a biased random walk model on a two-dimensional grid, so our derivation in Section \ref{sec:dispersalderivation} modifies this approach to derive the analogous terms describing directed motion on a one-dimensional lattice. Plugging the expansion from Equation \eqref{eq:diffusiontermexpanded} to the righthand side of Equation \eqref{eq:udiffexpanded}, we can then write that the rate of change of $u(t,x)$ satisfies
\begin{equation}
\begin{aligned} \label{eq:udiffexpandedsecond}
& \frac{u\left(t + \Delta t, x \right) - u(t,x)}{\Delta t} \\ &= \epsilon^{-1} \left\{f_L\left(u,v,n\right) - \kappa \left[u + v \right] \right\}   \\
&+ \frac{\gamma_u \left( \Delta x \right)^2}{\Delta t} \left[1 + \epsilon^{-1} \left\{f_L \left(u,v,n \right) - \kappa \left[ u + v\right] \right\} \left( \Delta t \right)  \right]   \left[ \doubledelsame{u}{x} - 2 \beta_u \dsdel{}{x} \left(  \dsdel{n}{x} u \right) \right] + \frac{1}{\Delta t} \mc{O}\left( \left( \Delta x \right)^4 \right)  \\
&+ \left(\Delta x \right) \gamma_u \left[ \dsdel{f_L}{u} \dsdel{u}{x} + \dsdel{f_L}{v} \dsdel{v}{x} + \dsdel{f_L}{v} \dsdel{v}{x} - \kappa \left( \dsdel{u}{x} + \dsdel{v}{x} \right) +  \mc{O}\left(\Delta x\right) \right] \left[ \frac{e^{\beta_u n(t,x)}}{e^{\beta_u n(t,x)} + e^{\beta_u n(t,x + 2 \Delta x)}}  \right] \\
&- \left(\Delta x \right) \gamma_u \left[ \dsdel{f_L}{u} \dsdel{u}{x} + \dsdel{f_L}{v} \dsdel{v}{x} + \dsdel{f_L}{v} \dsdel{v}{x} - \kappa \left( \dsdel{u}{x} + \dsdel{v}{x} \right) +  \mc{O}\left(\Delta x\right) \right] \left[ \frac{e^{\beta_u n(t,x)}}{e^{\beta_u n(t,x)} + e^{\beta_u n(t,x - 2 \Delta x)}}  \right]
\end{aligned}
\end{equation}
We can then take the continuum limit of this model by considering a joint limit in which $\Delta t \to 0$ and $\Delta x \to 0$, corresponding to a limit of continuous time-steps and continuous space. Because we see terms in the expansion from Equation \eqref{eq:udiffexpandedsecond} proportional to ratios of powers of $\Delta t$ and $\Delta x$, we also need to determine a scaling relation between $\Delta t$ and $\Delta x$ as both quantities tend to $0$. In order to retain a nonzero diffusive term in the continuum limit, we would like the ratio $\frac{\left(\Delta x\right)^2}{\Delta t}$ to remain nonzero and finite as $\Delta x \to 0$ and $\Delta t \to 0$. For this reason, we will consider the following scaling limit for $\Delta t$ and $\Delta x$
\begin{equation} \label{eq:Lscalingrelationship}
\lim_{\substack{\Delta x \to 0 \\ \Delta t \to 0}} \frac{\left( \Delta x \right)^2}{\Delta t} = \alpha,
\end{equation}
where $\alpha$ is a nonzero constant, and we introduce the corresponding parameters characterizing diffusivity and the strength of directed motion for low-impact harvesters as 
\begin{subequations} \label{eq:chiudefined}
\begin{align} 
D_u &:= \frac{1}{2} \gamma_u \\
\chi_u &:= \gamma_u \beta_u,
\end{align}
\end{subequations}
respectively. 
Using the scaling relationship defined by Equation \eqref{eq:Lscalingrelationship} and the parameters defined in Equation \eqref{eq:chiudefined}, we can take the limit of both sides of Equation \eqref{eq:udiffexpandedsecond} to obtain the following partial differential equation for the density of low-impact harvesters
\begin{align}
    \frac{\partial u}{\partial t}=D_u  \frac{\partial^2 u}{\partial x^2}  - \chi_u \frac{\partial }{\partial x}  \left( u \frac{\partial n}{\partial x}  \right)+\epsilon^{-1} u(f_L(u,v,n) - \kappa(u+v)).
\end{align}

Using an analogous approach, we can use our assumptions regarding birth, death, and movement events to write the expected density $v\left(t + \Delta t, x \right)$ of high-impact harvesters at location $x$ and time $t + \Delta t$ as
\begin{equation}
\begin{aligned}
v\left(t + \Delta t, x \right) &=  v(t,x) \left[ 1 + \epsilon^{-1} \left\{f_H\left(u(t,x),v(t,x),n(t,x) \right) - \kappa \left[u(t,x) + v(t,x) \right] \right\} \Delta t \right] \left( 1 - \gamma_v \right) \\
&+ \gamma_v \sum_{\tilde{x} \sim x} v\left(t,\tilde{x} \right) \left[ 1 + \epsilon^{-1} f_H \left( u(t,\tilde{x}), v(t,\tilde{x}), n(t,\tilde{x})\right) - \kappa \left[u(t,\tilde{x}) + v(t,\tilde{x}) \right] \right] M_v\left(t, \tilde{x} \to x \right),
\end{aligned}
\end{equation}
where we recall that $\gamma_v$ is the probability that a given high-impact harvester moves between time $t$ and time $t + \Delta t$ and we write the probability $M_v\left(t, \tilde{x} \to x\right)$ of a high-impact harvester moving from site $\tilde{x}$ to neighboring site $x$ in such a jump as
\begin{equation}
M_v\left(t,\tilde{x}\to x \right) = \frac{e^{\beta_v n(t,x)}}{\sum_{\tilde{\tilde{x}}\sim \tilde{x}} e^{\beta_v n(t,\tilde{x})}}.
\end{equation}
We can perform a similar expansion to derive the following PDE for the  density $v(t,x)$ for high-impact harvesters
\begin{align}
    \frac{\partial v}{\partial t}=D_v \frac{\partial^2 v}{\partial x^2}  - \chi_v \frac{\partial }{\partial x}  \left( v \frac{\partial n}{\partial x}  \right)+\epsilon^{-1} v(f_H(u,v,n) - \kappa(u+v)),
\end{align}
where we again consider the scaling limit in which $\frac{\left(\Delta x\right)^2}{\Delta t} \to \alpha$ as $\Delta x, \Delta t \to 0$, defining the diffusion coefficient $D_v$ for the high-impact harvesters by
\begin{equation}
  D_v := \frac{\gamma_v \alpha}{2}  %
\end{equation}
and defining the sensitivity of directed motion of high-impact harvesters to gradients of environmental quality as
\begin{equation}
\chi_v := 2 \beta_v D_v. 
\end{equation}
Finally, we assume that the environmental quality index $n(t,x)$ will first undergo its feedback reactions at rate
\begin{equation} \label{eq:rhonappendix}
\rho_n(t,x) = \left\{ r - a(e_L n(t,x) + e_H (1 - n(t,x)))\right\} \left(\frac{u(t,x)}{u(t,x) + v(t,x)} - n(t,x)\right)
\end{equation}
due to the strategic composition at location $x$ and time $t$, and then the resulting level of the environmental quality will disperse in the population following a step of an unbiased random walk on the grid. 
assuming that environmental quality $n$ undergoes an unbiased random walk, we derive its dynamics. The environmental quality at site $x$ at time $t$ can shift to neighboring sites with equal probability, and the probability of $n$ leaving site $x$ is given by $\gamma_n$. 

Given our assumption about the unbiased random walk of environmental resources and that the creation or destruction of the environmental resource occurs with the reaction rate $\rho_n(t,x)$ described by Equation \eqref{eq:rhonappendix}, we can write the following expression for the expected environmental quality index $n(t+\Delta t, x)$ at location $x$ at time $t+\Delta t$:
\begin{equation}
\begin{aligned}
n(t + \Delta t, x) = \left[ n(t,x) + \rho_n(t,x) \Delta t \right] \left( 1 - \gamma_n \right) + \frac{\gamma_n}{2} \sum_{\tilde{x} \sim x} \left[n(t,\tilde{x}) + \rho_n(t,\tilde{x}) \Delta t \right].
\end{aligned}
\end{equation}
We can then rearrange both sides to write that
\begin{equation} \label{eq:ndiscreterates}
\begin{aligned}
\frac{n\left(t + \Delta t, x \right) - n(t,x)}{\Delta t} &= \rho_n(t,x) + \frac{\gamma_n}{2 \Delta t} \sum_{\tilde{x} \sim x} \left( n(t,\tilde{x}) - n(t,x) \right) + \frac{\gamma_n}{2} \sum_{\tilde{x} \sim x} \left( \rho_n(t,\tilde{x}) - \rho_n(t,x) \right).
\end{aligned}
\end{equation}
We can then perform a Taylor expansion for the first sum on the righhand side by noting that
\begin{equation}
 \sum_{\tilde{x} \sim x} \left( n(t,\tilde{x}) - n(t,x) \right) = n\left(t,x+\Delta x\right) - 2 n(t,x) + n\left(t ,x-\Delta x \right) = \left(\Delta x \right)^2 \doubledelsame{n}{x}  + \mc{O}\left(  \left(\Delta x  \right)^2\right),
\end{equation}
and we can use the fact that the reaction rate for the environmental quality index $\rho_n(t,x)$ depends in a sufficiently differentiable way on $u(t,x)$, $v(t,x)$, and $n(t,x)$ to apply the chain rule and expand the second term on the righthand side of Equation \eqref{eq:ndiscreterates} as
\begin{equation}
\sum_{\tilde{x} \sim x} \left(\rho_n(t,\tilde{x}) - \rho_n(t,x)  \right) = \mc{O}\left( \left( \Delta x \right)^2 \right). 
\end{equation}
Applying these expansions to the righthand side of Equation \eqref{eq:ndiscreterates}, we obtain the following expansion for the difference quotient for $n(t,x)$ over a time-step of length $\Delta t$:
\begin{equation} \label{eq:ndifffinal}
\begin{aligned}
\frac{n\left(t+\Delta t,x\right) - n(t,x)}{\Delta t} &= \rho_n(t,x) + \frac{\gamma_n}{2} \left( \frac{\left(\Delta x\right)^2}{\Delta t} \right) \doubledelsame{n}{x} + \frac{1}{\Delta t} \mc{O}\left( \left(\Delta x\right)^4\right) + \mc{O}\left( \left(\Delta x\right)^2 \right).
\end{aligned}
\end{equation}
We can then consider the scaling $\frac{\left(\Delta x\right)^2}{\Delta t} \to \alpha$ as $\Delta t, \Delta x \to 0$ and introduce the diffusion coefficient
\begin{equation}
D_n := \frac{\gamma_n \alpha}{2}
\end{equation}
for the environmental quality index. Under these assumptions, we can take the limit as $\Delta t$ and $\Delta x$ tend to $0$ on both sides of Equation \eqref{eq:ndifffinal} and use the definition of the reaction rate $\rho_n(t,x)$ to obtain the following partial differential equation for the spatial profile of the environmental quality index:
\begin{equation}
\dsdel{n}{t} = D_n \doubledelsame{n}{x} + \left(r - a(e_L n + e_H (1 - n))\right) \left(\frac{u}{u + v} - n\right).
\end{equation}
\subsubsection{Expansion of Discrete Dispersal Term with Environmental-Driven Directed Motion}
\label{sec:dispersalderivation}

In this section, we present the derivation of the directed motion term describing environmental-driven motion for harvesters in our PDE model, starting from the discretized dispersal term from our random walk model given by
\begin{equation} \label{eq:mainsum}
\sum_{\tilde{x} \sim x} \left( M_u(t,\tilde{x} \to x) u(t,\tilde{x}) - \frac{1}{2} u(t,x) \right) = \sum_{\tilde{x} \sim x} \left( \left[ \frac{e^{\beta_u n(t,x)}}{\sum_{\tilde{\tilde{x}}\sim \tilde{x}} e^{\beta_un(t,\tilde{\tilde{x}})}} u(t,\tilde{x}) \right] - \frac{1}{2} u(t,x) \right).
\end{equation}
The general approach was previously presented by Alsenafi and Barbaro \cite{alsenafi2018convection} for the case of directed motion on a 2D lattice, but we include the similar derivative for the case of a 1D lattice for completeness. 

Noting that the neighbors of $x$ on our one-dimensional lattice are $\tilde{\tilde{x}} = \tilde{x} + \Delta x$ and $\tilde{\tilde{x}} = \tilde{x} - \Delta x$, we can perform a Taylor expansion to write the denominator of the first term in our sum as 
\begin{align} \label{eq:denominator}
    \sum_{\tilde{\tilde{x}} \sim \tilde{x}} e^{\beta_u n(\tilde{\tilde{x}},t)} = 2e^{\beta_u n(t,\tilde{x})} + \left(\Delta x\right) ^2 \frac{\partial^2 \left( e^{\beta_u n(t,\tilde{x})} \right)}{\partial x^2} + \mathcal{O}\left(\left(\Delta x\right) ^4\right)
\end{align}
as $\Delta x \to 0$. We can then further differentiate the second term in Equation \eqref{eq:denominator} to see that
\begin{equation}
\begin{aligned}
 \frac{\partial^2 \left( e^{\beta_u n(t,\tilde{x})} \right)}{\partial x^2} = \frac{\partial}{\partial x} \left(\frac{\partial\left(e^{\beta_u n(t,\tilde{x})}\right)}{\partial x}\right) 
   &=\frac{\partial}{\partial x}\left( \beta_u \frac{\partial n(t,\tilde{x})}{\partial x}e^{\beta_u n(t,\tilde{x})}\right) 
  \\  &=\left[ \left(\beta_u\frac{\partial n(t,\tilde{x})}{\partial x}\right)^2 + \beta_u \frac{\partial^2 n(t,\tilde{x})}{\partial x^2}\right]e^{\beta_u n(t,\tilde{x})},   
\end{aligned}
\end{equation}
and we can plug this expression back into Equation \eqref{eq:denominator} to see that
\begin{align}
    \sum_{\tilde{\tilde{x}} \sim \tilde{x}}e^{\beta_u n(\tilde{\tilde{x}},t)}= e^{\beta_u n(t,\tilde{x})} \left(2 + \left(\Delta x\right) ^2\left(\beta_u\frac{\partial n(t,\tilde{x})}{\partial x}\right)^2+\beta_u \frac{\partial^2 n(t,\tilde{x})}{\partial x^2} \right) +\mathcal{O}\left(\left(\Delta x\right) ^4\right).
\end{align}
We can then use this to write an expansion for the transition probability $M(t,\tilde{x} \to x)$ in the following form %
\begin{equation}
\begin{aligned}
    M_u(t,\tilde{x} \rightarrow x) &= e^{\beta_u n(t,x)} \left[\frac{e^{-\beta_u n(t,\tilde{x})}}{2 + \left(\Delta x\right) ^2 \left(\beta_u \frac{\partial n(t,\tilde{x})}{\partial x} \right)^2 + \beta_u \frac{\partial^2 n(t,\tilde{x})}{\partial x^2} + \mathcal{O}\left(\left(\Delta x\right)^4\right)} \right] \\
    &= e^{\beta_u n(t,x)} \left[\frac{e^{-\beta_u n(t,\tilde{x})}}{2 + \left(\Delta x\right) ^2 \left(\beta_u \frac{\partial n(t,\tilde{x})}{\partial x} \right)^2 + \beta_u \frac{\partial^2 n(t,\tilde{x})}{\partial x^2}} \right] +  \mathcal{O}\left(\left(\Delta x\right)^4\right).
\end{aligned}
\end{equation}
To continue our analysis of the expression on the righthand side, we introduce the following shorthand notation
\begin{align} \label{eq:Tushorthand}
    T_u(t,x)=\frac{e^{-\beta_u n(t,x)}}{2+ \left(\Delta x\right) ^2\left(\beta_u\frac{\partial n(t,x)}{\partial x} \right)^2+\beta_u \frac{\partial^2 n(t,x)}{\partial x^2} },
\end{align}
which allows us to write the probability of a dispersing low-impact harvester to move from site $\tilde{x}$ to neighboring site $x$ as
\begin{align} \label{eq:M_u_rewritten}
     M_u(t,\tilde{x} \rightarrow x) = e^{\beta_u n(t,x)} T_u(t,\tilde{x}) + \mathcal{O}\left(\left(\Delta x\right) ^4\right).
\end{align}

We can then further understand the sum on the righthand side by performing a Taylor expansion of  $u(t,\tilde{x})T_u(t,\tilde{x})$ to see that
\begin{align} \label{eq:sum_expanded}
    \sum_{\tilde{x}\sim x} u (t,\tilde{x})T_u(t,\tilde{x}) = 2u (t,x)T_u(t,x) + \left(\Delta x\right)^2 \frac{\partial^2 \left( u (t,x)T_u(t,x)\right)}{\partial x^2} + \mathcal{O}\left(\left(\Delta x\right) ^4\right).
\end{align}
Noting that there are two neighbors of the point $\tilde{x}$ on our one-dimensional lattice, we can use the expansion from Equation \eqref{eq:sum_expanded} to rewrite the sum from Equation \eqref{eq:mainsum} as
\begin{equation}
\begin{aligned}
&\sum_{\tilde{x} \sim x} \left( M_u(t,\tilde{x} \to x) u(t,\tilde{x}) - \frac{1}{2} u(t,x) \right) \\ &= \sum_{\tilde{x} \sim x} \left\{ e^{\beta_u n(t,x)} T_u(t,\tilde{x}) u(t,\tilde{x}) - \frac{1}{2} u(t,x) \right\} \\ &= e^{\beta_u n(t,x)} \left[ 2u (t,x)T_u(t,x) + \left(\Delta x\right)^2 \frac{\partial^2 \left( u (t,x)T_u(t,x)\right)}{\partial x^2} \right] + \mathcal{O}\left(\left(\Delta x\right) ^4\right) - \sum_{\tilde{x} \sim x}  \frac{1}{2} u(t,x) \\  &=  u(t,x) \left[ 2 e^{\beta_u n(t,x)} T_u(t,x) - 1 \right] + \left(\Delta x\right)^2 \frac{\partial^2 \left( u (t,x)T_u(t,x)\right)}{\partial x^2} + \mathcal{O}\left(\left(\Delta x\right) ^4\right), 
\end{aligned}
\end{equation}
and we can then %
use the definition of $T_u(t,x)$ from Equation \eqref{eq:Tushorthand} for the first term on the righthand side, and perform a Taylor expansion of the resulting expression  to write that
\begin{equation} \label{eq:biaseddiffusionexpanded}
\begin{aligned}
&\sum_{\tilde{x} \sim x} \left\{ e^{\beta_u n(t,x)} T_u(t,\tilde{x}) u(t,\tilde{x}) - \frac{1}{2} u(t,x) \right\} \\ &= u(t,x) \left[ 2 \left( \frac{1}{2 + \left(\Delta x\right)^2 \left[ \left( \beta_u \frac{\partial n}{\partial x} \right)^2 + \beta_u \frac{\partial^2 n}{\partial x^2} \right]} \right) - 1 \right] + \left(\Delta x\right)^2 \frac{\partial^2 \left( u (t,x)T_u(t,x)\right)}{\partial x^2} + \mathcal{O}\left(\left(\Delta x\right)^4\right) \\
&= u(t,x) \left[2 \left( \frac{1}{2} - \frac{1}{4} \left( \Delta x \right)^2 \left[ \left( \beta_u \dsdel{n}{x} \right)^2 + \beta_u \doubledelsame{n}{x} \right] \right) - 1 \right] + \left(\Delta x\right)^2 \frac{\partial^2 \left( u (t,x)T_u(t,x)\right)}{\partial x^2} + \mathcal{O}\left(\left(\Delta x\right)^4\right) \\
&= \left(\Delta x\right)^2 \left[ - \frac{1}{2} u \left[ \left( \beta_u \dsdel{n}{x} \right)^2 + \beta_u \doubledelsame{n}{x} \right] +   \frac{\partial^2 \left( u (t,x)T_u(t,x)\right)}{\partial x^2} \right] + \mathcal{O}\left(\left(\Delta x\right)^4\right)
\end{aligned}
\end{equation}

We can then apply the product rule to simplify the second spatial derivative of $u(t,x) T_u(t,x)$
\begin{align} \label{eq:secondderivativeproduct}
     \frac{\partial^2 (u T_u)}{\partial x^2}=T_u   \frac{\partial^2 u}{\partial x^2} +2\frac{\partial T_u }{\partial x} \frac{\partial u}{\partial x}  + u   \frac{\partial^2 T_u}{\partial x^2},
\end{align}
and we seek to obtain an expansion for the derivative $\frac{\partial^2 (u T_u)}{\partial x^2}$ by performing the following Taylor expansion of the expression for $T_u(t,x)$ given by Equation \eqref{eq:Tushorthand}:
\begin{align}
    T_u=\frac{e^{-\beta_u n}}{2}\left(1-\frac{\left(\Delta x\right) ^2}{2}\left(\left(\beta_u \frac{\partial n}{\partial x} \right)^2+\beta_u   \frac{\partial^2 n}{\partial x^2} \right)\right)+\mathcal{O}\left(\left(\Delta x\right) ^4\right).
\end{align}
We then differentiate this expanded expression with respect to $x$ to see that
\begin{align}
    \frac{\partial T_u}{\partial x} = \frac{e^{-\beta_u n}}{2}\left(-\beta_u \frac{\partial n}{\partial x}  -\frac{\left(\Delta x\right) ^2}{2}\left(-\left(\beta_u \frac{\partial n}{\partial x} \right)^3+\beta_u^2\frac{\partial n}{\partial x}    \frac{\partial^2 n}{\partial x^2}  +\beta_u \frac{\partial^3 n}{\partial x^3} \right)\right) +\mathcal{O}\left(\left(\Delta x\right) ^4\right),
\end{align}
and differentiating again allows us to see that
\begin{align}
\begin{split}
    \frac{\partial^2 T_u}{\partial x^2}  &= \frac{\partial}{\partial x} \left(\frac{\partial T_u}{\partial x} \right)\\
    &= \frac{e^{-\beta_u n}}{2}\left(\left(\left(\beta_u \frac{\partial n}{\partial x}\right)^2-\beta_u   \frac{\partial^2 n}{\partial x^2}\right) \right) \\
    &- \frac{e^{-\beta_u n}}{2} \left( \frac{\left(\Delta x\right) ^2}{2} \left(\left(\beta_u \frac{\partial n}{\partial x}\right)^4-4\beta_u^3 \left(\frac{\partial n}{\partial x}\right)^2  \frac{\partial^2 n}{\partial x^2}  +\beta_u^2 \left(\frac{\partial^2 n}{\partial x^2}\right)^2 +\beta_u \frac{\partial^4 n}{\partial x^4} \right)\right)  \\
    &+\mathcal{O}\left(\left(\Delta x\right) ^4\right).
\end{split}
\end{align}
We can then apply the expressions we found for $T_u(t,x)$, $\dsdel{T_u(t,x)}{x}$, and $\doubledelsame{T_u(t,x)}{x}$ to Equation \eqref{eq:secondderivativeproduct} to obtain the following expansion for the second-order spatial derivative for $u(t,x) T_u(t,x)$:
\begin{align}
    \begin{split}
        \frac{\partial^2 (u T_u)}{\partial x^2} &= \frac{e^{-\beta_u n}}{2}   \frac{\partial^2 u }{\partial x^2} - \frac{2\beta_u e^{-\beta_u n}}{2} \frac{\partial n}{\partial x}\frac{\partial u }{\partial x}+ \frac{e^{-\beta_u n}}{2} u \left(\left(\beta_u \frac{\partial n}{\partial x} \right)^2-\beta_u   \frac{\partial^2 n}{\partial x^2} \right) +\mathcal{O}\left(\left(\Delta x\right)^2\right) \\
    &=\frac{e^{-\beta_u n}}{2}\left(  \frac{\partial^2 u}{\partial x^2}  -2 \beta_u \frac{\partial n}{\partial x}  \frac{\partial u}{\partial x}  +u \left(\left(\beta_u \frac{\partial n}{\partial x} \right)^2-\beta_u   \frac{\partial^2 n}{\partial x^2}\right) \right) +\mathcal{O}\left(\left(\Delta x\right)^2\right).
    \end{split}
\end{align}
Finally, we can plug this expression into the righthand side of Equation \eqref{eq:biaseddiffusionexpanded} to see that our desired expansion described the effects of the biased random walk is given by
\begin{equation}
\begin{aligned}
\sum_{\tilde{x} \sim x} \left( M_u(t,\tilde{x} \to x) u(t,\tilde{x}) - \frac{1}{2} u(t,x) \right) &= \sum_{\tilde{x} \sim x} \left( \left[ \frac{e^{\beta_u n(t,x)}}{\sum_{\tilde{\tilde{x}}\sim \tilde{x}} e^{\beta_un(t,\tilde{\tilde{x}})}} u(t,\tilde{x}) \right] - \frac{1}{2} u(t,x) \right) \\ &= \left( \Delta x \right)^2   \left[ \doubledelsame{u}{x} - 2 \beta_u \dsdel{}{x} \left(  \dsdel{n}{x} u \right) \right] + \mc{O}\left( \left( \Delta x \right)^4 \right).
\end{aligned}
\end{equation}
This expansion then allows us to complete the derivation of the PDE describing the spatial density $u(t,x)$ of the low-impact harvester.

\section{Derivation and Linearization of Frequency-Dependent PDE System}

In this section, we provide additional derivations and analysis of our PDE model, showing how the description of spatial distribution of the the population in terms of the fraction of low-impact harvesters $p(t,x)$, the total number of individuals $q(t,x)$, and the environmental quality index $n(t,x)$ can help us to describe spatial pattern formation in eco-evolutionary games. We first present the derivation of the form of our PDE model expressed in terms of $p(t,x)$, $q(t,x)$, and $n(t,x)$ from the original model formulated in terms of the densities $u(t,x)$ and $v(t,x)$ for the low-impact and high-impact harvesters (Section \ref{sec:FrequencyDerivation}) and then we show how to linearize the resulting PDE system around a uniform equilibrium state (Section \ref{sec:App_linearization}). Finally, we describe some additional details required to perform a linear stability analysis to determine the possibility of spatial pattern formation in the case in which the diffusivities of the low-impact and high-impact harvesters are unequal (Section \ref{sec:app_LSA_unequal}). 

\subsection{Derivation of PDE Model in Terms of Frequencies}
\label{sec:FrequencyDerivation}

We now consider the initial description of our PDE model for spatial evolutionary games with environmental feedback with the following system of PDEs for the densities $u(t,x)$ and $v(t,x)$ of low-impact and high-impact harvesters and the spatial profile $n(t,x)$ of environmental quality index:
\begin{subequations}\label{eq:PDEsystemuvnappendix}
\begin{align}
\dsdel{u(t,x)}{t} &= D_u \doubledelsame{u(t,x)}{x} - \chi_u \dsdel{}{x} \left( u(t,x) \dsdel{n(t,x)}{x} \right)+ \epsilon^{-1} u (f_L\left(u,v,n\right) - \kappa(u+v)) \\
\dsdel{v(t,x)}{t} &= D_v \doubledelsame{v(t,x)}{x} - \chi_v \dsdel{}{x} \left( v(t,x) \dsdel{n(t,x)}{x} \right) + \epsilon^{-1} v (f_H\left(u,v,n\right) - \kappa(u+v)) \\
\dsdel{n(t,x)}{t} &= D_n \doubledelsame{n(t,x)}{x} + \left(r - a(e_L n + e_H (1-n))\right)\left(\frac{u}{u+v}-n\right).
\end{align}
\end{subequations}

We now look to achieve an equivalent description this system in terms of the spatial profiles of the fraction of low-impact harvesters and the total number of harvesters. To do this, we define the density of the fraction of low-impact harvesters by $p(t,x) := \frac{u(t,x)}{u(t,x) + v(t,x)}$ and the total density of harvesters by $q(t,x) = u(t,x) + v(t,x)$. 
By using the quotient rule, we can then express the time derivatives $\frac{\partial p}{\partial t}$ and $\frac{\partial q}{\partial t}$ of these two quantities in terms of $p$, $q$, $\frac{\partial u}{\partial t}$, and $\frac{\partial v}{\partial t}$, which allows us to see that
\begin{subequations} \label{eq:timederivatives}
\begin{align} 
    \frac{\partial p}{\partial t} &= \frac{1}{q} \frac{\partial u}{\partial t} - \frac{p}{q} \left( \frac{\partial u}{\partial t} + \frac{\partial v}{\partial t} \right), \\
    \frac{\partial q}{\partial t} &= \frac{\partial u}{\partial t} + \frac{\partial v}{\partial t}.
\end{align}
\end{subequations}
Next, we rewrite %
the righthand sides of the system of PDEs from Equation \eqref{eq:PDEsystemuvnappendix} in terms of $p$, $q$, and $n$. Using the substitutions $u = p q$ and $v = (1 - p) q$, we can then note that the derivatives $\frac{\partial u}{\partial t}$, $\frac{\partial v}{\partial t}$, and $\frac{\partial n}{\partial t}$ can also be written as
\begin{subequations}\label{eq:derivation_uvn}
    \begin{align}
\frac{\partial u}{\partial t} &= D_u \frac{\partial^2}{\partial x^2}(p q) - \chi_u \frac{\partial}{\partial x} \left( p q \frac{\partial n}{\partial x} \right) + \epsilon^{-1} p q \left(f_L(p,q,n) - \kappa q\right), \\
  \frac{\partial v}{\partial t} &= D_v \frac{\partial^2}{\partial x^2} \left( (1 - p) q \right) - \chi_v \frac{\partial}{\partial x} \left( q (1 - p) \frac{\partial n}{\partial x} \right) + \epsilon^{-1}  (1 - p) q \left(f_H(p,q,n) - \kappa q\right), \\
\frac{\partial n}{\partial t} &= D_n \frac{\partial^2 n}{\partial x^2} + \left( r - a \left(e_L n + e_H (1-n) \right) \right) \left( p - n \right).
\end{align}
\end{subequations}
Here, the functions $f_L$ and $f_H$ are payoffs of $L$-type and $H$-type individuals to find that
\begin{subequations}
\begin{align}
    f_L(p,q,n) &=  f_L(p,n)=(1 - n)(R_0 p + S_0 (1 - p)) + n (R_1 p + S_1 (1 - p)), \\
    f_H(p,q,n) &=  f_H(p,n)=(1 - n)(T_0 p + P_0 (1 - p)) + n (T_1 p + P_1 (1 - p)).
\end{align}
\end{subequations}
To simplify further, we apply the product rule to expand the spatial derivatives in Equation \eqref{eq:derivation_uvn} to see that
\begin{subequations}
\begin{align}
    \frac{\partial^2}{\partial x^2} (pq) &= 2 \frac{\partial p}{\partial x} \frac{\partial q}{\partial x} + p \frac{\partial^2 q}{\partial x^2} + q \frac{\partial^2 p}{\partial x^2}, \\
    \frac{\partial^2}{\partial x^2} \left( (1 - p) q \right) &= -2 \frac{\partial p}{\partial x} \frac{\partial q}{\partial x} + (1 - p) \frac{\partial^2 q}{\partial x^2} - q \frac{\partial^2 p}{\partial x^2}, \\
    -\chi_u \frac{\partial}{\partial x} \left( p q \frac{\partial n}{\partial x} \right) &= - \chi_u \left( q \frac{\partial p}{\partial x}  \frac{\partial n}{\partial x} + p \frac{\partial q}{\partial x} \frac{\partial n}{\partial x} + p q \frac{\partial^2 n}{\partial x^2} \right), \\
    -\chi_v \frac{\partial}{\partial x} \left( q (1 - p) \frac{\partial n}{\partial x} \right) &= -  \chi_v \left((1 - p) \frac{\partial q}{\partial x} \frac{\partial n}{\partial x} - q \frac{\partial p}{\partial x} \frac{\partial n}{\partial x} + q (1 - p) \frac{\partial^2 n}{\partial x^2} \right).
\end{align}
\end{subequations}

By substituting these results into %
the expressions for the time derivatives of $p$ and $q$ provided by Equation \eqref{eq:timederivatives}, we obtain the following system of PDEs expressed entirely in terms of the dependent variables $p$, $q$, and $n$:
\begin{subequations}
\begin{align}
\frac{\partial q(t,x)}{\partial t} &= \left(D_u p + D_v (1-p) \right) \frac{\partial^2 q}{\partial x^2} 
 + \left( D_u - D_v \right) q \frac{\partial^2 p}{\partial x^2} -q \left[(1-p) \chi_v + p \chi_u \right] \frac{\partial^2 n}{\partial x^2}\\
& + 2 \left( D_u - D_v \right) \frac{\partial q}{\partial x} \frac{\partial p}{\partial x} - q \left(\chi_u - \chi_v \right) \frac{\partial p}{\partial x} \frac{\partial n}{\partial x} - \left[ p \chi_u  + (1-p)\chi_v \right] \frac{\partial q}{\partial x} \frac{\partial p}{\partial x}  \\
&+ q \left[ p f_L(p,n) + (1-p) f_H(p,n) - \kappa q \right],  \nonumber \\
\frac{\partial p(t,x)}{\partial t} &= \frac{1}{q} \left[ p(1-p) \left( D_u - D_v \right) \right] \frac{\partial^2 q}{\partial x^2} + \left( (1-p) D_u + p D_v \right) \frac{\partial^2 p}{\partial x^2} -p\left(1-p\right)\left(\chi_u-\chi_v\right)\frac{\partial^2 n}{\partial x^2}\\
& + \frac{2}{q} \left[(1-p)D_u - p D_v \right] \frac{\partial q}{\partial x} \frac{\partial p}{\partial x}-\left[\left(1-p\right)\chi_u+p\chi_v\right]\frac{\partial p}{\partial x} \frac{\partial n}{\partial x} -\frac{1}{q}\left[p\left(1-p\right)\left(\chi_u-\chi_v\right)\right]\frac{\partial q}{\partial x} \frac{\partial n}{\partial x}\\
&+ p \left( 1 - p \right) \left[ f_L(p,n) - f_H(p,n) \right],  \nonumber \\
\frac{\partial n(t,x)}{\partial t} &= D_n \frac{\partial^2 n}{\partial x^2} + \left( r - a \left(e_L n + e_H (1-n) \right) \right) \left( p - n \right).
\end{align}
\end{subequations}

This system of PDEs in terms of $p(t,x)$, $q(t,x)$, and $n(t,x)$ is particularly amenable to study via linear stability analysis, and therefore we choose this description as the main form of the spatial model that we will use for analytical exploration throughout the paper.

\subsection{Derivation of Linearized PDE}
\label{sec:App_linearization}

In this section, we derive of the linearized system of PDEs that we use in Section \ref{sec:LSAPDE} to analyze the stability of spatially uniform equilibria for our PDE model featuring coexistence between the low-impact and high-impact harvesters. We will perform our linearization using the system of PDEs from Equation \eqref{eq:PDEsystemqpn} that describes the population in terms of spatial profiles of the total density of individuals $q(t,x)$, the fraction of cooperators $p(t,x)$, and the environmental quality index $n(t,x)$, and we will consider a uniform state $(q(x),p(x),n(x)) = (q_0,p_0,n_0)$ where the quantities $(q_0,p_0,n_0)$ correspond to an equilibrium point to our ODE model from Equation \eqref{eq:ODEafterCOV} for eco-evolutionary games with non-constant population size. To linearize our system around this equilibrium, we consider spatial profiles of the form
\begin{subequations} \label{eq:perturbations}
\begin{align}
    q(t,x) &= q_0 + \delta \tilde{q}(t,x) \\
    p(t,x) &= p_0 + \delta \tilde{p}(t,x)\\
    n(t,x) &= n_0 + \delta \tilde{n}(t,x),
\end{align}
\end{subequations}   
where $\delta$ is a small parameter and $\tilde{q}(t,x)$, $\tilde{p}(t,x)$, and $\tilde{n}(t,x)$ are spatially non-constant functions describing the deviation of the spatial quantities $q(t,x)$, $p(t,x)$, $n(t,x)$ from the uniform state. 

We can then use our expressions for $q(t,x)$, $p(t,x)$, and $n(t,x)$ from Equation \eqref{eq:perturbations} to rewrite our system of PDEs from Equation \eqref{eq:PDEsystemqpn}, using the fact that $(q(x),p(x),n(x) = (q_0,p_0,n_0)$ is a uniform equilibrium state to find that
\begin{subequations} \label{eq:PDEperturbations}
\begin{align}
\delta \frac{\partial \tilde{q}(t,x)}{\partial t} &= \delta \left[ \left(D_u p_0 +  D_v (1-p_0) \right) \frac{\partial^2 \tilde{q}}{\partial x^2} 
 + \left( D_u - D_v \right) q_0 \frac{\partial^2 \tilde{p}}{\partial x^2} -q_0 \left[(1-p_0) \chi_v + p_0 \chi_u \right] \frac{\partial^2 \tilde{n}}{\partial x^2} \right] \\
& + \delta^2 \left[ 2 \left( D_u - D_v \right) \frac{\partial \tilde{q}}{\partial x} \frac{\partial \tilde{p}}{\partial x} - q_0 \left(\chi_u - \chi_v \right) \frac{\partial \tilde{p}}{\partial x} \frac{\partial \tilde{n}}{\partial x} - \left[ p_0 \chi_u  + (1-p_0)\chi_v \right] \frac{\partial \tilde{q}}{\partial x} \frac{\partial \tilde{p}}{\partial x} \right]   \nonumber \\
&+ q_0 \left[ p_0 f_L(p_0 + \delta \tilde{p},n_0 + \delta \tilde{n}) + (1-p_0) f_H(p_0 + \delta \tilde{p},\tilde{n}) - \kappa \left( q_0 + \delta \tilde{q} \right) \right],  \nonumber \\
\delta \frac{\partial \tilde{p}(t,x)}{\partial t} &= \delta \left[ \frac{1}{q_0} \left[ p_0(1-p_0) \left( D_u - D_v \right) \right] \frac{\partial^2 \tilde{q}}{\partial x^2} + \left( (1-p_0) D_u + p_0 D_v \right) \frac{\partial^2 \tilde{p}}{\partial x^2} -p_0\left(1-p_0\right)\left(\chi_u-\chi_v\right)\frac{\partial^2 \tilde{n}}{\partial x^2} \right] \nonumber \\
& + \delta^2 \left[ \frac{2}{q_0} \left[(1-p_0)D_u - p_0 D_v \right] \frac{\partial \tilde{q}}{\partial x} \frac{\partial \tilde{p}}{\partial x}-\left[\left(1-p_0\right)\chi_u+p\chi_v\right]\frac{\partial \tilde{p}}{\partial x} \frac{\partial \tilde{n}}{\partial x} \right] \nonumber \\ &-\delta^2 \left[ \frac{1}{q_0}\left[p_0\left(1-p_0\right)\left(\chi_u-\chi_v\right)\right]\frac{\partial \tilde{q}}{\partial x} \frac{\partial \tilde{n}}{\partial x} \right] \nonumber \\
&+ p_0 \left( 1 - p_0 \right) \left[ f_L\left(p_0 + \delta \tilde{p},n_0 + \delta \tilde{n} \right) - f_H\left(p_0 + \delta \tilde{p},n_0 + \delta \tilde{n} \right) \right], \\
\delta \frac{\partial \tilde{n}(t,x)}{\partial t} &= \delta D_n \frac{\partial^2 \tilde{n}}{\partial x^2} + \left( r - a \left[e_L \left(n_0 + \delta \tilde{n} \right) + e_H \left( 1-\left(n_0 + \delta \tilde{n} \right) \right) \right] \right) \left( p_0  - n_0 + \delta \left( \tilde{p} - \tilde{n_0} \right) \right).
\end{align}
\end{subequations}
In particular, we were able to use the fact at $q_0$, $p_0$, and $n_0$ are constant to see that all partial derivatives of these quantities will vanish, allowing us interpret the system of PDEs from Equation \eqref{eq:PDEperturbations} in terms of the spatially varying functions $\tilde{q}(t,x)$, $\tilde{p}(t,x)$, and $\tilde{n}(t,x)$ describing the perturbations from the uniform steady state. 

We note that term terms containing the product of spatial derivatives $\dsdel{\tilde{p}}{x} \dsdel{\tilde{q}}{x}$, $\dsdel{\tilde{p}}{x} \dsdel{\tilde{n}}{x}$, and $\dsdel{\tilde{n}}{x} \dsdel{\tilde{q}}{x}$ are of order $\mc{O}(\delta^2)$ as $\delta \to 0$, so these terms will not appear in the resulting linearized system. In addition, we can further rewrite our system of PDEs by performing a Taylor expansion of the payoff functions $f_L\left(p_0 + \delta \tilde{p},n_0 + \delta \tilde{n} \right)$ and $f_H\left(p_0 + \delta \tilde{p},n_0 + \delta \tilde{n} \right)$ and using the fact that we are linearizing around an equilibrium $(q_0,p_0,n_0)$ of the reaction dynamics of Equation \eqref{eq:ODEafterCOV}. This allows us to write the system of Equation \eqref{eq:PDEperturbations} in terms of the following asymptotic expansion
\begin{subequations} \label{eq:PDEexpanded}
    \begin{align}
   \delta  \frac{\partial \tilde{q}}{\partial t} &= \delta \left[\left(D_u p_0 + D_v (1 - p_0) \right) \frac{\partial^2 \tilde{q}}{\partial x^2} 
 + \left( D_u - D_v \right) q_0 \frac{\partial^2 \tilde{p}}{\partial x^2}\right] \\
 &+ \delta \left[ - q_0 \left[(1 - p_0) \chi_v + p_0 \chi_u \right] \frac{\partial^2 \tilde{n}}{\partial x^2} + c_{11} \tilde{q} + c_{12} \tilde{p} + c_{13} \tilde{n} \right] + \mc{O}\left(\delta^2 \right) \nonumber, \\
    \delta \frac{\partial \tilde{p}}{\partial t} &= \delta \left[\frac{1}{q_0} \left[ p_0 (1 - p_0) \left( D_u - D_v \right) \right] \frac{\partial^2 \tilde{q}}{\partial x^2} + \left( (1 - p_0) D_u + p_0 D_v \right) \frac{\partial^2 \tilde{p}}{\partial x^2} \right]\\
    &+ \delta \left[- p_0 (1 - p_0) (\chi_u - \chi_v) \frac{\partial^2 \tilde{n}}{\partial x^2} + c_{22} \tilde{p} + c_{23} \tilde{n} \right] + \mc{O}\left(\delta^2\right) \nonumber, \\
   \delta  \frac{\partial \tilde{n}}{\partial t} &= \delta \left[ D_n \frac{\partial^2 \tilde{n}}{\partial x^2} + c_{32} \tilde{p} + c_{33} \tilde{n} \right] + \mc{O}\left(\delta^2\right)
\end{align}
\end{subequations}
as $\delta \to 0$, where the constants $c_{ij}$ for the linearized reaction terms are given by the following entries of the Jacobian matrix for the reaction dynamics of Equation \eqref{eq:ODEafterCOV}
\begin{subequations}
\begin{align}
c_{11} &= \frac{\partial}{\partial q} (\epsilon^{-1} q \left[ p f_L + (1-p) f_H - \kappa q \right]) \bigg\rvert_{q=q_0,p=p_0, n=n_0} \\ &= \epsilon^{-1}\left( p_0f_L(p_0,n_0) + (1-p_0)f_H(p_0,n_0)  - 2\kappa q_0\right) \nonumber \\
c_{12} &= \frac{\partial}{\partial p} (\epsilon^{-1}q \left[ p f_L + (1-p) f_H - \kappa q \right]) \bigg\rvert_{q=q_0,p=p_0, n=n_0}\\
    &= \epsilon^{-1} q_0 \left[ f_L(p_0,n_0) - f_H(p_0,n_0)  + p_0 \dsdel{f_L}{p}\bigg\rvert_{p_0,n_0} + (1-p_0) \dsdel{f_H}{p}\bigg\rvert_{p_0, n_0} \right] \nonumber \\
    c_{13} &= \frac{\partial}{\partial n} (\epsilon^{-1}q \left[ p f_L + (1-p) f_H - \kappa q \right]) \bigg\rvert_{q=q_0,p=p_0, n=n_0}\\
    &= \epsilon^{-1} q_0 \left[ p_0 \dsdel{f_L}{n}\bigg\rvert_{p_0, n_0}  + (1-p_0) \dsdel{f_H}{n}\bigg\rvert_{p_0, n_0} \right] \nonumber \\
     c_{22} &= \frac{\partial}{\partial p} \left(\epsilon^{-1} p \left( 1 - p \right) \left(f_L - f_H \right)\right) \bigg\rvert_{p=p_0, n=n_0} \\
    &= \epsilon^{-1} (1-2p_0) \left[f_L(p_0,n_0) - f_H(p_0,n_0) \right] + \epsilon^{-1} p_0 (1-p_0) \left[\dsdel{f_L}{p}\bigg\rvert_{p_0, n_0} - \dsdel{f_H}{p}\bigg\rvert_{p_0, n_0} \right] \nonumber \\
    c_{23} &= \frac{\partial}{\partial n} \left(\epsilon^{-1}p \left( 1 - p \right) \left(f_L - f_H \right)\right) \bigg\rvert_{p=p_0, n=n_0}\\
    &= \epsilon^{-1} p_0(1-p_0) \left[\dsdel{f_L}{n}\bigg\rvert_{p_0, n_0} - \dsdel{f_H}{n}\bigg\rvert_{p_0, n_0}  \right] \nonumber \\
     c_{32} &= \frac{\partial}{\partial p} ((r - a(e_L n + e_H (1-n)))(p-n)) \bigg\rvert_{p=p_0, n=n_0}\\
    &= r - a \left( e_L n_0 + e_H (1-n_0) \right) \nonumber \\
    c_{33} &= \frac{\partial}{\partial n} ((r - a(e_L n + e_H (1-n)))(p-n)) \bigg\rvert_{p=p_0, n=n_0}\\
    &= -a \left( e_L - e_H \right) (p_0-n_0) - \left( r - a\left( e_L n_0 + e_H \left(1-n_0 \right) \right) \right) \nonumber.
\end{align}
\end{subequations}
In particular, we note the that there are no terms $c_{21}$ and $c_{31}$ on the righthand side of the expanded system of PDEs in Equation \eqref{eq:PDEexpanded}, as the reaction dynamics for the fraction of cooperators $p$ and environmental quality $n$ are independent of the total density of harvesters $q$ in Equation \eqref{eq:ODEafterCOV}.

Finally, we can obtain the linearization of our system of PDEs around the equilibrium $(q_0,p_0,n_0)$ by dividing both sides of each PDE in Equation \eqref{eq:PDEexpanded} by $\delta$ and taking the limit as $\delta \to 0$, yielding the linear system 

\begin{subequations}
    \begin{align}
    \frac{\partial \tilde{q}}{\partial t} &= \left(D_u p_0 + D_v (1 - p_0) \right) \frac{\partial^2 \tilde{q}}{\partial x^2} 
 + \left( D_u - D_v \right) q_0 \frac{\partial^2 \tilde{p}}{\partial x^2} \\
 &- q_0 \left[(1 - p_0) \chi_v + p_0 \chi_u \right] \frac{\partial^2 \tilde{n}}{\partial x^2} + c_{11} \tilde{q} + c_{12} \tilde{p} + c_{13} \tilde{n} \nonumber \\
    \frac{\partial \tilde{p}}{\partial t} &= \frac{1}{q_0} \left[ p_0 (1 - p_0) \left( D_u - D_v \right) \right] \frac{\partial^2 \tilde{q}}{\partial x^2} + \left( (1 - p_0) D_u + p_0 D_v \right) \frac{\partial^2 \tilde{p}}{\partial x^2} \\
    &- p_0 (1 - p_0) (\chi_u - \chi_v) \frac{\partial^2 \tilde{n}}{\partial x^2} + c_{22} \tilde{p} + c_{23} \tilde{n} \nonumber \\
    \frac{\partial \tilde{n}}{\partial t} &= D_n \frac{\partial^2 \tilde{n}}{\partial x^2} + c_{32} \tilde{p} + c_{33} \tilde{n}.
\end{align}
\end{subequations}
\

\subsection{Additional Results on Linear Stability Analysis for Nonidentical Diffusion Rates}
\label{sec:app_LSA_unequal}

In this section, we describe the use of the Routh-Hurwitz conditions to derive conditions for the onset of spatial pattern formation in the case of unequal diffusivities $D_u \ne D_v$ of the low-impact and high-impact harvesters. We showed in Section \ref{sec:LSAPDE} that whether a uniform coexistence equilibrium will be stable to sinusoidal perturbations with integer wavenumber $m$ under our PDE model from Equation \eqref{eq:PDEsystemqpn} will depend on the eigenvalues of the matrix 

\begin{equation}\resizebox{\textwidth}{!}{$
    A(m) = \begin{pmatrix} \label{eq:PDE_Jaco_A_appendix}
c_{11} - \left(\frac{m \pi}{l}\right)^2 (D_u p_0 + D_v (1 - p_0)) & c_{12} - \left(\frac{m \pi}{l}\right)^2 (D_u - D_v) q_0 & c_{13} + \left(\frac{m \pi}{l}\right)^2 (\chi_u p_0 q_0 + \chi_v q_0 (1 - p_0)) \\
-q_0^{-1} \left(\frac{m \pi}{l}\right)^2 (1-p_0) p_0(D_u -D_v)  & c_{22} - \left(\frac{m \pi}{l}\right)^2 (D_u (1-p_0) + p_0 D_v) & c_{23} + \left(\frac{m \pi}{l}\right)^2 p_0(1-p_0) (\chi_u - \chi_v) \\
0 & c_{32} & -\left(\frac{m \pi}{l}\right)^2 D_n + c_{33},
\end{pmatrix}
$}
\end{equation}
where the terms of the form $c_{ij}$ are the entries of the Jacobian matrix for the ODE model with non-constant population size from Equation \eqref{eq:ODEafterCOV}. In particular, the uniform state will be unstable to perturbations with wavenumber $m$ provided that at least one eigenvalue of the matrix $A(m)$ has a positive real part. 

For our system of three PDEs from Equation \eqref{eq:PDEsystemqpn}, we know that the matrix for the linearization $A(m)$ has characteristic polynomial given by
\begin{align}
    P_{A(m)}(\lambda) = \lambda^3 - \operatorname{Tr}(A(m)) \lambda^2 + \frac{1}{2}\left( \operatorname{Tr}(A)^2 - \operatorname{Tr}(A(m)^2) \right) \lambda - \det(A(m)).
\end{align}
Using the Routh-Hurwitz criteria for our characteristic polynoomial, we have that the spatially uniform state will be stable provided that the following three conditions hold: 
\begin{subequations}
\begin{align}
\textnormal{I} &:= \operatorname{Tr}(A) < 0 \\
\textnormal{II} &:= \det(A) < 0 \\
\textnormal{III} &:= -\frac{1}{2} \left( \operatorname{Tr}(A^2) - \operatorname{Tr}^2(A) \right) \operatorname{Tr}(A) +\det(A) > 0.
\end{align}
\end{subequations}
We can observe from the linearization matrix that $\operatorname{Tr}(A(m)) < 0$ for all possible parameters, so it will only be possible to achieve instability of the uniform state through finding a parameter regime in which $\textnormal{II} > 0$ or $\textnormal{III} < 0$. 
For a given integer wavenumber $m$ and diffusivity $D_v$ of the high-impact harvesters, we define the threshold strengths of environmental-driven motion $\chi^{\textnormal{II}}_{u}\left(m; D_v \right)$ and $\chi^{\textnormal{III}}_u\left( m; D_v \right)$ to represent the minimum values $\chi_u$ such that $\textnormal{II} > 0$ and $\textnormal{III} < 0$, respectively. We can calculate these thresholds analytically, but due to their lengthy expression we have placed the expressions for these thresholds in a supplementary Mathematica notebook uploaded to Github. 

Furthermore, we can calculate the minimum $\chi_u$ value required for $\mathrm{II} > 0$ and $\mathrm{III} < 0$ for any integer wavenumber $m$, defining the corresponding threshold quantities as 
\begin{subequations}
\begin{align}
\chi^{\textnormal{II}}_{u}\left(D_v \right) &:= \min_{m \in \ZZ} \chi^{\textnormal{II}}_{u}\left( m; D_v \right) \\
\chi^{\textnormal{III}}_{u}\left(D_v \right) &:= \min_{m \in \ZZ} \chi^{\textnormal{III}}_{u}\left( m; D_v \right).
\end{align}
\end{subequations}
Because we only need either $\chi_u^* > \chi^{\textnormal{II}}_{u}\left(D_v \right)$ or $\chi_u > \chi^{\textnormal{III}}_{u}\left(D_v \right)$ in order for spatial patterns to form, we can then deduce that spatial pattern formation will occur provided that $\chi_u$ satisfies
\begin{equation}
\chi_u > \chi_u^* := \min\left(\chi^{\textnormal{II}}_{u}\left( m; D_v \right), \chi^{\textnormal{III}}_{u}\left( m; D_v \right) \right)
\end{equation}
for the case of zero-flux boundary conditions in our model of spatial eco-evolutionary games with environmental-driven directed motion and the possibility of unequal diffusivities $D_u \ne D_v$ for the low-impact and high-impact harvesters.

\begin{figure}[!htbp]
    \centering
    \includegraphics[width=0.65\linewidth]{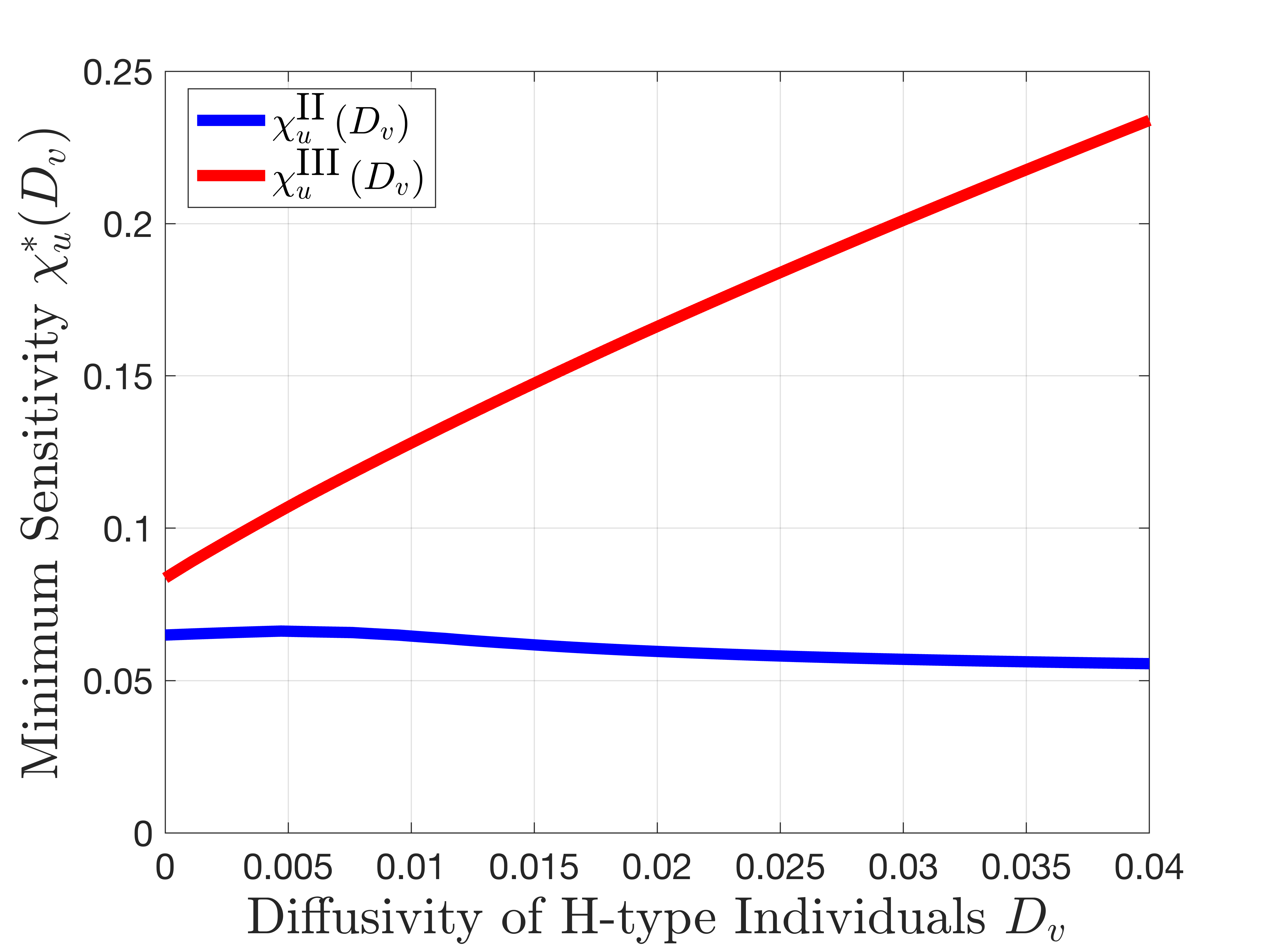}
    \caption{Comparison of threshold strengths of environmental-driven motion $\chi_u$ derived from different instability conditions from the Routh-Hurwitz criteria, plotted as a function of the diffusivity $D_v$ for high-impact harvesters. The diffusivity of the low-impact harvesters was fixed to $D_u = 0.1$, and all other game-theoretic and ecological parameters are the same as in Figure \ref{fig:chi_u_m}. }
    \label{fig:chi_u_stability}
\end{figure}

In Figure \ref{fig:chi_u_stability}, we illustrate an example the threshold conditions for the strength of environmental-driven motion $\chi_u^{\textnormal{II}}(D_v)$ and $\chi_u^{\textnormal{III}}(D_v)$ for the second or third Routh-Hurwitz conditions. Using the same game-theoretic and ecological parameters as considered for the figures in the main text, we see that the $\chi_u^{\textnormal{II}}(D_v) < \chi_u^{\textnormal{III}}(D_v)$ for all values of the high-impact harvester diffusivity $D_v$ under consideration, suggesting that the onset of spatial patterns will occur due to the strength of environmental-driven motion $\chi_u$ exceeding the threshold set by the second Routh-Hurwitz condition for this choice of game and environmental feedback. We then take the minimum of these two threshold curves for $\chi_u^{\textnormal{II}}(D_v) < \chi_u^{\textnormal{III}}(D_v)$ to obtain the threshold $\chi_u^*$ for the onset of spatial pattern formation, which we then plot as a function of $D_v$ in Figure \ref{fig:Chi_u_D_v}.

\end{document}